\documentclass[final]{lmcs} 
\pdfoutput=1

\usepackage{lastpage}
\lmcsdoi{21}{1}{19}
\lmcsheading{}{\pageref{LastPage}}{}{}%
{Feb.~27,~2023}{Feb.~26,~2025}{}


\usepackage{hyperref}

\usepackage[utf8]{inputenc}  
\usepackage[T1]{fontenc}  
   
\usepackage{xspace}
\usepackage{tikz}
\usetikzlibrary{arrows,automata,shapes,calc,positioning,intersections,backgrounds}

\tikzset{every picture/.style={>=latex}}
\tikzstyle{PlayerMin}=[draw,circle,minimum size=7mm,inner sep=1.5pt]
\tikzstyle{target}=[circle, minimum size=1mm,inner sep=-2pt]
\tikzstyle{PlayerMax}=[draw,rectangle,minimum size=7mm,inner sep=1.5pt]
\tikzstyle{proba}=[draw,circle,minimum height=0pt,inner sep=2pt,minimum width=0pt,fill=black]

\usepackage[obeyFinal]{todonotes}
\usepackage{bbm}
\usepackage{amsfonts,amssymb,amsmath}
\usepackage{wasysym}

\newtheorem{hypothesis}[thm]{Hypothesis} 

\DeclareMathOperator*{\argmin}{\arg\!\min}

\DeclareMathOperator*{\arginf}{\arg\!\inf}
\DeclareMathOperator*{\argsup}{\arg\!\sup}
\newcommand{\tuple}[1]{\langle #1 \rangle}

\newcommand{\arginfepsilon}[1][]{\ifthenelse{\equal{#1}{}}
	{{\arginf}^{\varepsilon}}{{\arginf}^{#1}}}
\newcommand{\argsupepsilon}[1][]{\ifthenelse{\equal{#1}{}}
	{{\argsup}^{\varepsilon}}{{\argsup}^{#1}}}

\newcommand{\MinPl}{\ensuremath{\mathsf{Min}}\xspace}
\newcommand{\MaxPl}{\ensuremath{\mathsf{Max}}\xspace}

\newcommand{\vertices}{\ensuremath{L}\xspace}
\newcommand{\minvertices}{\ensuremath{\vertices_{\MinPl}}\xspace}
\newcommand{\maxvertices}{\ensuremath{\vertices_{\MaxPl}}\xspace}
\newcommand{\edges}{\ensuremath{\Delta}\xspace}
\newcommand{\edge}{\ensuremath{\delta}\xspace}
\newcommand{\finalvertices}{\ensuremath{\vertices_T}\xspace}
\newcommand{\edgeweights}{\ensuremath{\weight}\xspace}
\newcommand{\game}{\ensuremath{\mathcal G}\xspace}

\newcommand{\arenaEx}{\ensuremath{\tuple{\maxvertices,\minvertices,\edges,
			\edgeweights,\finalvertices}}\xspace} 
\ProvideDocumentCommand{\gameEx}{o}{\IfNoValueTF{#1}
	{\tuple{\vertices,\edges,\edgeweights,\Payoff}}
	{\tuple{\vertices,\edges,\edgeweights,#1}}}

\newcommand{\strategy}{\ensuremath{\xi}\xspace}
\newcommand{\detminstrategy}{\ensuremath{\sigma}\xspace}
\newcommand{\detmaxstrategy}{\ensuremath{\tau}\xspace}
\newcommand{\minstrategy}{\ensuremath{\eta}\xspace}
\newcommand{\maxstrategy}{\ensuremath{\theta}\xspace}
\newcommand{\minstrategytrans}
	{\ensuremath{\minstrategy_{\scriptscriptstyle\Trans}}\xspace}
\newcommand{\minstrategydelay}
	{\ensuremath{\minstrategy_{\scriptscriptstyle\R^+}}\xspace}
\newcommand{\maxstrategytrans}
	{\ensuremath{\maxstrategy_{\scriptscriptstyle\Trans}}\xspace}
\newcommand{\maxstrategydelay}
	{\ensuremath{\maxstrategy_{\scriptscriptstyle\R^+}}\xspace}
\newcommand{\strategytrans}
	{\ensuremath{\xi_{\scriptscriptstyle\Trans}}\xspace}
\newcommand{\strategydelay}{\ensuremath{\xi_{\scriptscriptstyle\R^+}}\xspace}

\newcommand{\dist}{\ensuremath{d}\xspace}
\newcommand{\delay}{\ensuremath{t}\xspace}

\newcommand{\MDP}{\ensuremath{\game^{\minstrategy}}\xspace}
\newcommand{\MC}{\ensuremath{\game^{\minstrategy,\maxstrategy}}\xspace}

\newcommand{\outcomes}{\ensuremath{\mathsf{Play}}\xspace}

\newcommand{\Strat}{\ensuremath{\mathsf{Strat}}\xspace}
\newcommand{\StratProper}{\ensuremath{\mathsf{Strat}^\mathsf p}\xspace}
\newcommand{\dStratProper}{\ensuremath{\mathsf{dStrat}^\mathsf p}\xspace}
\newcommand{\StratMin}[1][]{\ifthenelse{\equal{#1}{}}
	{\ensuremath{\Strat_{\MinPl}}}{\ensuremath{\Strat_{\MinPl,#1}}}\xspace}
\newcommand{\StratMinProper}[1][]{\ifthenelse{\equal{#1}{}}
	{\ensuremath{\StratProper_{\MinPl}}}{\ensuremath{\StratProper_{\MinPl,#1}}}\xspace}
\newcommand{\dStratMinProper}[1][]{\ifthenelse{\equal{#1}{}}
	{\ensuremath{\dStratProper_{\MinPl}}}{\ensuremath{\dStratProper_{\MinPl,#1}}}\xspace}
\newcommand{\StratMax}[1][]{\ifthenelse{\equal{#1}{}}
	{\ensuremath{\Strat_{\MaxPl}}}{\ensuremath{\Strat_{\MaxPl,#1}}}\xspace}
\newcommand{\dStrat}{\ensuremath{\mathsf{dStrat}}\xspace}
\newcommand{\dStratMin}[1][]{\ifthenelse{\equal{#1}{}}
	{\ensuremath{\dStrat_{\MinPl}}}{\ensuremath{\dStrat_{\MinPl,#1}}}\xspace}
\newcommand{\dStratMax}[1][]{\ifthenelse{\equal{#1}{}}
	{\ensuremath{\dStrat_{\MaxPl}}}{\ensuremath{\dStrat_{\MaxPl,#1}}}\xspace}
\newcommand{\mStrat}{\ensuremath{\mathsf{mStrat}}\xspace}
\newcommand{\mStratProper}{\ensuremath{\mathsf{mStrat}^\mathsf p}\xspace}
\newcommand{\mStratMin}[1][]{\ifthenelse{\equal{#1}{}}
	{\ensuremath{\mStrat_{\MinPl}}}{\ensuremath{\mStrat_{\MinPl,#1}}}\xspace}
\newcommand{\mStratMinProper}[1][]{\ifthenelse{\equal{#1}{}}
	{\ensuremath{\mStratProper_{\MinPl}}}{\ensuremath{\mStratProper_{\MinPl,#1}}}\xspace}
\newcommand{\mStratMax}[1][]{\ifthenelse{\equal{#1}{}}
	{\ensuremath{\mStrat_{\MaxPl}}}{\ensuremath{\mStrat_{\MaxPl,#1}}}\xspace}
\newcommand{\sdStrat}{\ensuremath{\mathsf{sdStrat}}\xspace}
\newcommand{\sdStratMin}[1][]{\ifthenelse{\equal{#1}{}}
	{\ensuremath{\sdStrat_{\MinPl}}}{\ensuremath{\sdStrat_{\MinPl,#1}}}\xspace}
\newcommand{\sdStratMax}[1][]{\ifthenelse{\equal{#1}{}}
	{\ensuremath{\sdStrat_{\MaxPl}}}{\ensuremath{\sdStrat_{\MaxPl,#1}}}\xspace}
\newcommand{\sdStratProper}{\ensuremath{\mathsf{sdStrat}^\mathsf p}\xspace}
\newcommand{\sdStratMinProper}[1][]{\ifthenelse{\equal{#1}{}}
	{\ensuremath{\sdStratProper_{\MinPl}}}{\ensuremath{\sdStratProper_{\MinPl,#1}}}\xspace}

\newcommand{\Value}{\ensuremath{\mathsf{Val}}\xspace}
\newcommand{\dValue}{\ensuremath{\mathsf{dVal}}\xspace}
\newcommand{\mValue}{\ensuremath{\mathsf{mVal}}\xspace}
\newcommand{\sdValue}{\ensuremath{\mathsf{sdVal}}\xspace}
\newcommand{\uppervalue}{\ensuremath{\overline{\Value}}\xspace}
\newcommand{\lowervalue}{\ensuremath{\underline{\Value}}\xspace}
\newcommand{\duppervalue}{\ensuremath{\overline{\dValue}}\xspace}
\newcommand{\dlowervalue}{\ensuremath{\underline{\dValue}}\xspace}
\newcommand{\sduppervalue}{\ensuremath{\overline{\sdValue}}\xspace}
\newcommand{\sdlowervalue}{\ensuremath{\underline{\sdValue}}\xspace}
\newcommand{\muppervalue}{\ensuremath{\overline{\mValue}}\xspace}
\newcommand{\mlowervalue}{\ensuremath{\underline{\mValue}}\xspace}

\newcommand{\Distr}[1]{\ensuremath{\mathsf{Dist}(#1)}\xspace}
\newcommand{\support}[1]{\ensuremath{\mathop{\mathrm {supp}}(#1)}\xspace}
\newcommand{\Proba}{\ensuremath{\mathbb P}\xspace}
\newcommand{\E}{\ensuremath{\mathbb{E}}\xspace}
\newcommand{\Dirac}[1]{\ensuremath{\mathsf{Dirac}_{#1}}\xspace}

\newcommand{\R}{\ensuremath{\mathbb{R}}\xspace}
\newcommand{\Rplus}{\ensuremath{\R_{\geq 0}}\xspace}
\newcommand{\Rbar}{\ensuremath{\R_\infty}\xspace}
\newcommand{\Z}{\ensuremath{\mathbb{Z}}\xspace}

\newcommand{\N}{\ensuremath{\mathbb{N}}\xspace}

\newcommand{\A}{\ensuremath{\mathcal{A}}\xspace}
\newcommand{\Cl}{\ensuremath{\mathcal{X}}\xspace}

\newcommand{\g}{\ensuremath{\mathcal{G}}\xspace}
\newcommand{\rg}{\ensuremath{\mathcal{R}(\g)}\xspace}
\newcommand{\Cyl}{\ensuremath{\mathsf{Cyl}}\xspace}

\newcommand{\WTG}{WTG\xspace}
\newcommand{\SPG}{SPG\xspace}
\newcommand{\clockbound}{\ensuremath{M}\xspace}
\newcommand{\val}{\ensuremath{\nu}\xspace}

\newcommand{\guard}{\ensuremath{g}\xspace}
\newcommand{\Guards}{\ensuremath{\mathsf{Guards}(\Cl)}\xspace}
\newcommand{\reset}{\ensuremath{Y}\xspace}

\newcommand{\regions}[2]{\ensuremath{\mathsf{Reg}(#1,#2)}\xspace}
\newcommand{\loc}{\ensuremath{\ell}\xspace}
\newcommand{\Locs}{\ensuremath{L}\xspace}
\newcommand{\LocsMin}{\ensuremath{\Locs_{\MinPl}}\xspace}
\newcommand{\LocsMax}{\ensuremath{\Locs_{\MaxPl}}\xspace}

\newcommand{\weight}{\ensuremath{\mathsf{wt}}\xspace}
\newcommand{\LocsT}{\ensuremath{\Locs_T}\xspace}

\newcommand{\weightC}{\ensuremath{\weight_\Sigma}\xspace}
\newcommand{\Trans}{\ensuremath{\Delta}\xspace}
\newcommand{\trans}{\ensuremath{\delta}\xspace}
\newcommand{\powerset}[1]{\ensuremath{2^{#1}}\xspace}
\newcommand{\play}{\ensuremath{\rho}\xspace}
\newcommand{\playBis}{\ensuremath{\play_0}\xspace}
\newcommand{\Plays}{\ensuremath{\mathsf{Plays}}\xspace}

\newcommand{\FPlays}{\ensuremath{\mathsf{FPlays}}\xspace}
\newcommand{\FPlaysMin}{\ensuremath{\FPlays_\MinPl}\xspace}
\newcommand{\FPlaysMax}{\ensuremath{\FPlays_\MaxPl}\xspace}

\newcommand{\TPlays}{\ensuremath{\mathsf{TPlays}}\xspace}
\newcommand{\wmax}{\ensuremath{w_{\mathrm{max}}}\xspace}
\newcommand{\rgame}{\ensuremath{\mathcal{R}(\game)}\xspace}

\newcommand{\RStates}{\ensuremath{S}\xspace}

\newcommand{\rpath}{\ensuremath{\pi}\xspace} 
\newcommand{\ppath}{\ensuremath{\pi}\xspace} 
\newcommand{\ppathBis}{\ensuremath{\ppath_0}\xspace} 
\newcommand{\TPaths}{\ensuremath{\mathsf{TPaths}}\xspace} 
\newcommand{\FPaths}{\ensuremath{\mathsf{FPaths}}\xspace} 

\newcommand{\F}{\ensuremath{\mathcal F}\xspace}
\newcommand{\Hfunction}{\ensuremath{\mathcal H}\xspace}

\newcommand{\EXP}{\ensuremath{\mathsf{EXPTIME}}\xspace}
\newcommand{\PSPACE}{\ensuremath{\mathsf{PSPACE}}\xspace}

\newcommand{\fract}{\ensuremath{\mathsf{fract}}}

\newcommand{\charact}{\ensuremath{\chi}\xspace}
\newcommand{\C}{\mathcal C}

\newcommand{\extendto}[1]{\ensuremath{[#1]}\xspace}
\newcommand{\moveto}[1]{\ensuremath{\xrightarrow{#1}}\xspace}

\begin{document}

\title{Playing Stochastically in Weighted Timed Games to Emulate Memory}

\thanks{We thank the reviewers of the various versions of this article that helped us greatly improve the quality of the results and the writing. This work has been partly funded by the QuaSy project (ANR-23-CE48-0008), and the NCN grant 2019/35/B/ST6/02322.}	

\author[B.~Monmege]{Benjamin Monmege\lmcsorcid{0000-0002-4717-9955}}[a]
\author[J.~Parreaux]{Julie Parreaux\lmcsorcid{0009-0009-2744-780X}}[b]
\author[P.-A.~Reynier]{Pierre-Alain Reynier\lmcsorcid{0009-0008-4345-704X}}[a]

\address{Aix Marseille Univ, CNRS, LIS, Marseille, France}	
\email{benjamin.monmege@univ-amu.fr, pierre-alain.reynier@univ-amu.fr}  
\address{University of Warsaw, Poland}
\email{j.parreaux@uw.edu.pl}




\begin{abstract}
	Weighted timed games are two-player zero-sum games played in a timed
	automaton equipped with integer weights. We consider optimal
	reachability objectives, in which one of the players, that we call
	$\MinPl$, wants to reach a target location while minimising the
	cumulated weight. While knowing if $\MinPl$ has a strategy to
	guarantee a value lower than a given threshold is known to be
	undecidable (with two or more clocks), several conditions, one of
	them being divergence, have been given to recover
	decidability. In such weighted timed games (like in untimed weighted
	games in the presence of negative weights), $\MinPl$ may need finite
	memory to play (close to) optimally. This is thus tempting to try to
	emulate this finite memory with other strategic capabilities. In
	this work, we allow the players to use stochastic decisions, both in
	the choice of transitions and of timing delays. We give 
	a definition of the expected value in weighted timed
	games. We then show that,
	in divergent weighted timed games as well as in (untimed) weighted games (that we call shortest-path games in the following), 
	the stochastic value is indeed
	equal to the classical (deterministic) value, thus proving that
	$\MinPl$ can guarantee the same value while only using stochastic
	choices, and no memory.
\end{abstract}

\maketitle


\section{Introduction}

Game theory is now an established model in the computer-aided design
of correct-by-construction programs. Two players, the controller and
an environment, are fighting
one against the other in a zero-sum game
played on a graph of all possible configurations. A~winning strategy
for the controller results in a correct program, while the environment
is a player modelling all uncontrollable events that the program must
face. Many possible objectives have been studied in such two-player zero-sum
games played on graphs: reachability, safety, repeated reachability,
and even all possible $\omega$-regular objectives \cite{GraTho02}.

Apart from such \emph{qualitative} objectives, more
\emph{quantitative} ones are useful in order to select a particular
strategy among all the ones that are correct with respect to a
qualitative objective. Some metrics of interest, mostly studied in the
quantitative game theory literature, are mean-payoff,
discounted-payoff, or total-payoff. All these objectives have in
common that both players have strategies using no memory or randomness
to win or play optimally~\cite{GimbertZielonka-05}.

Combining quantitative and qualitative objectives, in order to select
a good strategy among the valid ones for the selected metrics, often
leads to the need of memory to play optimally. One of the simplest
combinations showing this consists in the shortest-path games (\SPG{s}) combining
a reachability objective with a total-payoff quantitative objective
(studied in~\cite{KhaBor08,BrihayeGeeraertsHaddadMonmege-17} under the name of \emph{min-cost
	reachability games}). Another case of interest is the combination of
a parity qualitative objective (modelling every possible
$\omega$-regular condition), with a mean-payoff objective (aiming for
a controller of good quality in the average long-run), where
controllers need memory, and even infinite memory, to play optimally
\cite{ChaHen05}.

It is often crucial to enable randomisation in the strategies. For
instance, Nash equilibria are only ensured to exist in matrix games
(like rock-paper-scissors) when players can play at random
\cite{Nas50}. In the context of games on graphs, a player may choose,
depending on the current history, the probability distribution on the
successors. In contrast, strategies that do not use randomisation are
called \emph{deterministic} (we sometimes say \emph{pure}).

\begin{figure}[tbp]
	\centering
	\begin{tikzpicture}[xscale=.8]
	
	\node[PlayerMax] at (0, 0)  (s0){$\loc_{\MaxPl}$}; 
	\node[PlayerMin] at (4, 0)  (s) {$\loc_{\MinPl}$}; 
	\node[target] at (2, -2) (s1) {\LARGE $\smiley$};
	
	\draw[->]
	(s)  edge [bend left=10] node[below]{$\edge_1, \mathbf{0}$} (s0) 
	(s0) edge [bend left=10] node[above] {$\edge_2, \mathbf{-1}$} (s)
	(s0) edge [auto = right] node {$\edge_3, \mathbf{-10}$} (s1)
	(s) edge [auto = left] node{$\edge_4, \mathbf{0}$} (s1)
	;
	
	\begin{scope}[xshift=6cm]
	\node[PlayerMax] at (0, 0)  (s0){$\loc_{\MaxPl}$}; 
	\node[PlayerMin] at (4, 0)  (s) {$\loc_{\MinPl}$}; 
	\node[target] at (2, -2) (s1) {\LARGE $\smiley$};
	
	\node[proba] at (3.3,-.7) (s2) {};
	
	\draw[->]
	(s)  edge (s2)
	(s0) edge [bend left=10] node[above] {$\mathbf{-1}$} (s)
	(s0) edge [auto = right] node {$\mathbf{-10}$} (s1)
	(s) edge [auto = left] (s2)
	(s2) edge node[below] {$p,\mathbf{0}$} (s0)
	(s2) edge node[below right,yshift=1mm] {$1{-}p,\mathbf{0}$} (s1)
	;
	
	\end{scope}
	\begin{scope}[xshift=12cm]
	\node[PlayerMax] at (0, 0)  (s0){$\loc_{\MaxPl}$};
	\node[PlayerMin] at (4, 0)  (s) {$\loc_{\MinPl}$}; 
	\node[target] at (2, -2) (s1) {\LARGE $\smiley$};
	
	\node[proba] at (3.3,-.7) (s2) {};
	\node[proba] at (0.7,-.7) (s3) {};
	
	\draw[->]
	(s)  edge (s2)
	(s0) edge (s3)
	(s3) edge [bend right=10,auto=right] node[near start,yshift=1mm,xshift=-3mm] 
	{$q,\mathbf{-1}$} (s)
	(s3) edge [auto = right] node[yshift=1mm] {$1{-}q,\mathbf{-10}$} (s1)
	(s) edge [auto = left] (s2)
	(s2) edge[auto=right] node[very near end] {$p,\mathbf{0}$} (s0)
	(s2) edge node[below right,yshift=1mm] {$1{-}p,\mathbf{1}$} (s1)	
	;
	
	\end{scope}
	\end{tikzpicture}
	\caption{On the left, an SPG, where $\MinPl$ requires
		memory to play optimally. In the middle, the Markov Decision
		Process obtained when letting $\MinPl$ play at random, with a
		parametric probability $p\in(0,1)$. On the right, the Markov Chain
		obtained when $\MaxPl$ plays along a memoryless randomised
		strategy, with a parametric probability $q\in[0,1]$.}
	\label{fig:SP1}
\end{figure}

An example of SPG is
depicted on the left of \figurename~\ref{fig:SP1}. The objective of
$\MinPl$ is to reach vertex $\text{\Large\smiley}$, while minimising
the cumulated weight. Let us consider the vertex $\loc_\MinPl$ as
initial. Player $\MinPl$ could reach directly $\text{\Large\smiley}$,
thus leading to a weight of $0$. But $\MinPl$ can also choose to go to
$\loc_\MaxPl$, in which case $\MaxPl$ either jumps directly
to~$\text{\Large\smiley}$ (leading to a beneficial payoff $-10$), or
comes back to $\loc_\MinPl$, but having already capitalised a total
payoff $-1$. We can continue this way \emph{ad libitum} until $\MinPl$
is satisfied (at least 10 times) and jumps to
$\text{\Large\smiley}$. This guarantees a value at most $-10$ for
$\MinPl$ when starting in $\loc_\MinPl$. Reciprocally, $\MaxPl$ can
guarantee a value at least $-10$ by directly jumping into
$\text{\Large\smiley}$ when playing for the first time. Thus,
the optimal value is $-10$ when starting from~$\loc_\MinPl$
or~$\loc_\MaxPl$. However, $\MinPl$ cannot achieve this optimal value by
playing \emph{without memory} (we sometimes say \emph{positionally}),
since it either results in a total-payoff $0$ (directly going to the
target) or $\MaxPl$ has the opportunity to keep $\MinPl$ in the
negative cycle for ever, thus never reaching the target. Therefore,
$\MinPl$ needs memory to play optimally. He can do so by playing a
\emph{switching strategy}, looping in the negative cycle long enough
so that no matter how he reaches the target finally, the value he gets
is lower than the optimal value. This strategy uses
pseudo-polynomial memory with respect to the weights of the game
graph. It can be described by two memoryless strategies $A$ and $B$, following $A$ for a given number of terms (to loop in negative cycles long enough) before switching to $B$ (to ensure reaching $\text{\Large\smiley}$).

In this example, such a switching strategy can be \emph{mimicked}
using randomisation only (and no memory), $\MinPl$ deciding to go to
$\loc_\MaxPl$ with high probability $p<1$ and to go to the target vertex
with the remaining low probability $1-p>0$ (we enforce this
probability to be positive, in order to reach the target with
probability $1$, no matter how the opponent is playing). The resulting
\emph{Markov Decision Process (MDP)} is depicted in the middle of
\figurename~\ref{fig:SP1}. The shortest path problem in such MDPs has
been thoroughly studied in~\cite{BertsekasTsitsiklis91}, where it is
proved that $\MaxPl$ does not require memory to play
optimally. Denoting by~$q$ the probability that $\MaxPl$ jumps in
$\loc_\MinPl$ in its memoryless strategy, we obtain the \emph{Markov
	chain (MC)} on the right of \figurename~\ref{fig:SP1}. We can
compute (see Example~\ref{example:for-intro}) the expected value in
this MC, as well as the best strategy for both players: in the
overall, the optimal value remains $-10$, even if $\MinPl$ no longer
has an optimal strategy. He rather has an $\varepsilon$-optimal
strategy, consisting in choosing $p=1-\varepsilon/10$ that ensures a
value at most $-10+\varepsilon$.

A first aim of this article is thus to study the trade-off between memory and
randomisation in strategies of \SPG{s}. The study is only
interesting in the presence of both positive and negative weights,
since both players have optimal memoryless deterministic strategies
when the graph contains only non-negative weights~\cite{KhaBor08}. The
trade-off between memory and randomisation has already been
investigated in many classes of games where memory is required to win
or play optimally. This is for instance the case for qualitative games
like Street or Müller games thoroughly studied (with and without
randomness in the arena) in \cite{ChatterjeeAlfaroHenzinger-04}. The study has been
extended to timed games \cite{ChatterjeeHenzingerPrabhu-08} where the goal is to use
as little information as possible about the precise values of
real-time clocks. Memory or randomness is also crucial in
multi-dimensional objectives \cite{ChatterjeeRR14}: for instance, in
mean-payoff parity games, if there exists a deterministic
finite-memory winning strategy, then there exists a randomised
memoryless almost-sure winning strategy.
We show in this article that deterministic memory and
memoryless randomisation provide the same power to $\MinPl$ in \SPG{s}.

Another aim of this article is to extend this study to the case of weighted games on timed automata. 
\emph{Timed automata}~\cite{AlurDill-94} extend
finite-state automata with timing constraints, providing an
automata-theoretic framework to model and verify real-time
systems. 
While this has led to the development of mature verification tools,
the design of programs verifying some real-time specifications remains
a notoriously difficult problem. In this context, \emph{timed
	games} have been explored, with an objective of reaching a target as quickly as possible: they are decidable~\cite{AsarinMaler-99}, and
\EXP-complete~\cite{JurdzinskiTrivedi-07}. As in the untimed setting of \SPG{s}, timed games have been extended with quantitative objectives, in the form of \emph{weighted or (priced) timed games} (\WTG{s} for
short)~\cite{BouyerCassezFleuryLarsen-04,AlurBernadskyMadhusudan-04}.

While solving the optimal reachability problem on weighted timed
automata has been shown to be \PSPACE-complete~\cite{BouyerBrihayeBruyereRaskin-07}
(i.e.~the same complexity as the non-weighted version), \WTG{s} are
known to be undecidable~\cite{BrihayeBruyereRaskin-05}. Many restrictions have then been considered in
order to regain decidability, the first and most interesting one being
the class of strictly non-Zeno cost with only non-negative weights (in
transitions and locations)~\cite{BouyerCassezFleuryLarsen-04}: this
hypothesis requires that every execution of the timed automaton that
follows a cycle of the region automaton has a weight far from 0 (in
interval $[1,+\infty)$, for instance). This setting has been extended
in the presence of negative weights in transitions and locations
\cite{BusattoGastonMonmegeReynier-17}: in the so-called
\emph{divergent \WTG{s}}, each execution that follows a cycle of the
region automaton has a weight in $(-\infty,-1]\cup[1,+\infty)$. A
triply-exponential-time algorithm allows one to compute the values and
almost-optimal strategies, while deciding the divergence of a \WTG is
\PSPACE-complete. 

When studying optimal reachability objectives with both positive and
negative weights, it is known that strategies of player $\MinPl$
require memory to play optimally (see~\cite{BrihayeGeeraertsHaddadMonmege-17} 
for the case of
finite games).  More precisely, the memory needed is pseudo-polynomial
(i.e.~polynomial if constants are encoded in unary).  For
\WTG{s}, the memory needed even becomes exponential.
An important challenge is thus to find ways to avoid 
using such complex strategies, e.g.~by proposing
alternative classes of strategies that are more easily amenable
to implementation. 

We thus use the same ideas as developed before for \SPG{s}, introducing 
stochasticity in strategies. 
A first important challenge is to analyse how to play stochastically
in \WTG{s}. To our knowledge, this has not been studied before.
Starting from a notion of stochastic behaviours in a timed automaton
considered
in~\cite{BertrandBouyerBrihayeMenetBaierGroserJurdzinski-14} (for the
one-player setting), we propose a new class of stochastic
strategies. Compared
with~\cite{BertrandBouyerBrihayeMenetBaierGroserJurdzinski-14}, our
class is larger in the sense that we allow Dirac distributions for
delays, which subsumes the setting of deterministic
strategies. However, in order to ensure that strategies yield a
well-defined probability distribution on sets of executions, we need
measurability properties stronger than the one considered
in~\cite{BertrandBouyerBrihayeMenetBaierGroserJurdzinski-14} (we
actually provide an example showing that their hypothesis was not
strong enough).

Then, we turn our attention towards the expected cumulated weight of
the set of plays conforming to a pair of stochastic strategies.  We
first prove that under the previous measurability hypotheses, this
expectation is well-defined when restricting to the set of plays
following a finite sequence of transitions. In order to have the
convergence of the global expectation, we identify another property of
strategies of $\MinPl$, which intuitively ensures that the set of
target locations is reached quickly enough. This allows us to define a
notion of stochastic value (resp.~memoryless stochastic value) of the
game, i.e.~the best value $\MinPl$ can achieve using stochastic
strategies (resp.~memoryless stochastic strategies), when $\MaxPl$
uses stochastic strategies (resp.~memoryless stochastic strategies)
too.

Our main contribution is to show that in divergent \WTG{s} and in \SPG{s}, the stochastic value and the (stochastic) memoryless value are determined 
(i.e.~it does not matter which player chooses first their strategy), and the 
stochastic, memoryless and deterministic values are equal: when 
allowing players to play with memory, no players benefit from using also 
randomisation; moreover $\MinPl$ can emulate memory using randomisation, and vice versa.

\paragraph{Outline} We will recall the model of \WTG{s} in Section~\ref{sec:prelim}. We present in Section~\ref{sec:stochasticWTG} the notion of stochastic strategy and ground mathematically the associated definitions of expected weight and stochastic value. We adapt the notion of deterministic value to restrict deterministic strategies to be \emph{smooth} (in order for them to be considered as particular stochastic strategies). We relate in Section~\ref{sec:Max_best-response} the new deterministic value and the stochastic value in all \WTG{s}: as a technical tool, also used afterwards, we show that the player $\MaxPl$ always has a deterministic best-response against any strategy of player $\MinPl$ in all \WTG{s}. We then recall in Section~\ref{sec:divergent} the notion of divergence in \WTG{s}, and show new results on the form of almost-optimal (deterministic and smooth) strategies for both players. Section~\ref{sec:trading} then contains the proof of the possible trade-off between memory and randomisation in strategies in divergent \WTG{s}. We then focus on the special case of \SPG{s} (i.e.~\WTG{s} without clocks) in Section~\ref{sec:SPG}, where we are also able to obtain the previous trade-off, as well as characterising the existence of an optimal deterministic (or memoryless) strategy.

This article is an extended version of the two conference articles \cite{MonmegeParreauxReynier-20,MonmegeParreauxReynier-ICALP21}, showing respectively the results in untimed, and timed games. With respect to these works, we have tried to uniformise the notations and techniques, and obtained the results about the stochastic value and the determinacy results shown in this article.

\section{Weighted timed games}
\label{sec:prelim}

We let \Cl be a finite set of variables called clocks. A valuation is a mapping
$\val\colon \Cl\to \Rplus$ to the set $\Rplus$ of non-negative real numbers.
For a valuation \val, a delay $t\in\Rplus$ and a subset $Y\subseteq \Cl$ of
clocks, we define the valuation $\val+t$ as $(\val+t)(x)=\val(x)+t$, for all
$x\in \Cl$, and the valuation $\val[Y:=0]$ as $(\val[Y:=0])(x)=0$ if $x\in Y$,
and $(\val[Y:=0])(x)=\val(x)$ otherwise. A (non-diagonal) guard on clocks
of~\Cl is a conjunction of atomic constraints of the form $x\bowtie c$, where
${\bowtie}\in\{{\leq},<,=,>,{\geq}\}$ and $c\in \N$. A valuation $\val$
satisfies an atomic constraint $x\bowtie c$ if $\val(x)\bowtie c$. The
satisfaction relation is extended to all guards~$g$ naturally, and denoted by
$\val\models g$. We let $\Guards$ denote the set of guards over \Cl. For all
$a\in\Rplus$, $\lfloor a\rfloor\in\N$ denotes the integral part of $a$, and
$\fract(a)\in[0,1)$ its fractional part, such that $a=\lfloor
a\rfloor+\fract(a)$.

\begin{figure}
	\centering
	\scalebox{0.8}{
		\begin{tikzpicture}[node distance=3cm,auto,->,>=latex,every label/.style={font=\scriptsize}]
		
		\node[PlayerMax](1){\makebox[0mm][c]{$\mathbf{-2}$}};
		\node()[below of=1,node distance=6mm]{$\loc_1$};
		
		\node[PlayerMin](2)[below right
		of=1]{\makebox[0mm][c]{$\mathbf{2}$}};
		\node()[below of=2,node distance=6mm]{$\loc_2$};
		
		\node[target](3)[above right
		of=2]{\LARGE $\smiley$};
		\node()[below of=3,node distance=6mm]{$\loc_3$};
		
		\node[PlayerMax](4)[below right
		of=3]{\makebox[0mm][c]{$\mathbf{-1}$}};
		\node()[below of=4,node distance=6mm]{$\loc_4$};
		
		\node[PlayerMin](5)[above right
		of=4]{\makebox[0mm][c]{$\mathbf{-2}$}};
		\node()[below of=5,node distance=6mm]{$\loc_5$};
		
		\path
		(2) edge node[below left,xshift=2mm,yshift=2mm]{$
			\begin{array}{c}
			x\leq2 \\ x:=0 \\ \mathbf{0}
			\end{array}
			$} (1);
		
		\path
		(2) edge node[above left,xshift=4mm,yshift=-4mm]{
			$\begin{array}{c}
			1\leq x\leq3 \\ \mathbf{1}
			\end{array}$} (3);
		
		\path
		(2) edge node[below]{$x\leq3,\;  x:=0,\; \mathbf{0}$} (4);
		
		\path
		(4) edge node[above right,xshift=-4mm,yshift=-4mm]{$
			\begin{array}{c}
			2\leq x\leq3 \\ \mathbf{3}
			\end{array}$} (3);
		
		\path
		(4) edge node[below right,xshift=-2mm,yshift=2mm]{$
			\begin{array}{c}
			x\leq3\\ \mathbf{0}
			\end{array}$} (5);
		
		\path
		(1) edge node[above]{$x\leq3,\; \mathbf{0}$} (3);
		
		\path
		(5) edge node[above]{$x\leq3, \; \mathbf{0}$} (3);
		
		\path
		(1) edge [loop above] node {$
			\begin{array}{c}
			x\leq 1\\ x:=0,\; \mathbf{3}
			\end{array}$}  (1);
		
		\path
		(5) edge [loop above] node {$
			\begin{array}{c}
			1<x\leq 3 \\ x:= 0, \;\mathbf{1}
			\end{array}$} (5);
		
		\end{tikzpicture}}
	\raisebox{1mm}{\scalebox{.77}{
			\begin{tikzpicture}[node distance=3cm,auto,>=latex,
			every label/.style={font=\scriptsize}]
			
			\draw[->] (0,0) -- (0,4);
			\draw[->] (0,0) -- (4,0);
			
			\draw[dashed] (1,0) -- (1,4);
			\draw[dashed] (2,0) -- (2,4);
			\draw[dashed] (3,0) -- (3,4);
			
			\draw[dashed] (0,1) -- (4,1);
			\draw[dashed] (0,2) -- (4,2);
			\draw[dashed] (0,3) -- (4,3);
			
			\draw[dashed] (.666,0) 
				node[below,xshift=-1mm]{$2/3$}-- (.666,1.666);
			
			\node at (4,-.5) {$x$};
			\node at (-.5,4) {$\dValue$};
			
			\node at (0,-.5) {$0$};
			\node at (1,-.5) {$1$};
			\node at (2,-.5) {$2$};
			\node at (3,-.5) {$3$};
			
			\node at (-.5,0) {$0$};
			\node at (-.5,1) {$1$};
			\node at (-.5,2) {$2$};
			\node at (-.5,3) {$3$};
			
			\draw[red,very thick] (0,3) -- (1,1) -- (3,1);
			
			\draw[blue,very thick] (0,1) -- (2,3) -- (3,3);
			
			\node[rectangle,fill=white,fill opacity=.9,text opacity=1] at 
				(2.4,3.4) {$\loc_2 \rightarrow \loc_4 \rightarrow \loc_3$};
			
			\node[rectangle,fill=white,fill opacity=.9,text opacity=1]  at 
				(2.2,1.4) {$\loc_2 \rightarrow \loc_3$};
			
			\end{tikzpicture}}}
	\caption{On the left, a (one-clock) weighted timed game. 
		On the right, the value function of location $\loc_2$, with respect to the value of clock $x$, is obtained as the minimum of two curves.}\label{fig:wtg}
\end{figure}

\begin{defi}
	A \emph{weighted timed game} (\WTG) is a tuple
	$\game=\langle\LocsMin,\LocsMax, \LocsT, \Trans, \weight\rangle$
	where $\LocsMin$, $\LocsMax$, $\LocsT$ are finite disjoint subsets
	of \MinPl locations, \MaxPl locations, and target locations,
	respectively (we let $\Locs=\LocsMin\uplus\LocsMax\uplus\LocsT$),
	$\Trans\subseteq \Locs\times\Guards\times \powerset \Cl \times
	\Locs$ is a finite set of transitions,
	$\weight\colon \Trans\uplus\Locs \to \Z$ is the weight function.
\end{defi}

An example of \WTG{s} is depicted on the left of Figure~\ref{fig:wtg}. Locations belonging to
\MinPl (resp.~\MaxPl) are depicted by circles (resp.~squares). The
target location is $\loc_3$. Weights of transitions and locations are depicted in bold font.

The semantics of a \WTG $\game$ is defined in terms of a game played
on an infinite transition system whose vertices are configurations of
the \WTG. A configuration is a pair $(\loc,\val)$ with a location \loc and
a valuation \val of the clocks. Configurations are split into players
according to the location. A configuration is final if its location is
a target location of $\LocsT$. The alphabet of the transition system
is given by $\Trans\times\Rplus$: a pair $(\trans,t)$ encodes the
delay $t$ that a player wants to spend in the current location, before
firing transition $\trans$. For every delay $t\in\Rplus$, transition
$\trans=(\loc,g,Y,\loc')\in \Trans$ and valuation~$\val$, there is an
edge $\edge = (\loc,\val)\moveto{\trans,\delay}(\loc',\val')$ if
$\val+t\models g$ and $\val'=(\val+t)[Y:=0]$. 
Without loss of generality, we suppose the absence of deadlocks except
on target locations, i.e.~for each location
$\loc\in \Locs\backslash\LocsT$ and valuation $\val$, there exists
$(\loc,g,Y,\loc')\in \Trans$ and a delay $t \in \Rplus$ such that 
$\val + t \models g$, and no transitions start in~\LocsT. This is generally not true on the examples we draw in this article (the example seen above) because we will bound the clocks later, but the automaton can always be completed to allow for a useless move reaching a sink vertex. 
A \emph{finite play} is a finite sequence of consecutive edges
$\play=(\loc_0,\val_0)\moveto{\trans_0,\delay_0}(\loc_1,\val_1)
\moveto{\trans_1,\delay_1}\cdots (\loc_k,\val_k)$. We denote by
$|\play|$ the length $k$ of \play. The concatenation of two finite 
plays $\play_1$ and $\play_2$, such that $\play_1$ ends in the same 
configuration as $\play_2$ starts, is denoted by $\play_1\play_2$. We 
denote by $I(\play,\trans)$ the interval\footnote{It is an interval since guards 
are conjunctions of inequality constraints on clocks. More precisely, for 
each constraint given by an inequality over the valuation of one clock 
adequate delays are described by an (possibly empty) interval. Then, 
$I(\play,\trans)$ is defined by the intersection of these intervals.} 
of delays $t$ such that the play \play can be extended with the edge 
$\moveto{\trans,\delay}$. This extended play is denoted by $\play \extendto{\trans,\delay}$.  
In particular, we sometimes denote a play \play by 
$(\loc_0,\val_0)\extendto{\trans_0,\delay_0} \cdots 
\extendto{\trans_{k-1},\delay_{k-1}}$,  
since intermediate locations and
valuations are uniquely defined by the initial configuration and the
sequence of transitions and delays. 
We let $\FPlays$ be the set of all finite plays, 
whereas $\FPlaysMin$ (resp. $\FPlaysMax$ and $\TPlays$) denote the finite plays
that end in a configuration of $\MinPl$ (resp. $\MaxPl$ and $\LocsT$). 
We let $\TPlays_\play$ (resp. $\TPlays^n_\play$) be the subset 
of $\TPlays$ that start in the last configuration of the finite play \play
(resp. containing $n$ transitions without taking into account the size of \play). A \emph{play} is then a maximal sequence of 
consecutive edges (it is either infinite or it reaches \LocsT).

We call \emph{path} a finite or infinite sequence \ppath of
transitions of~$\game$. As for finite plays, the concatenation of two finite 
paths $\ppath_1$ and $\ppath_2$, such that $\ppath_1$ ends in the same 
location as $\ppath_2$ starts, is denoted by $\ppath_1\ppath_2$. Each 
play~\play of $\game$ is associated with a unique path \ppath (by 
projecting away everything but the transitions): we say that \play 
\emph{follows} the path \ppath. A \emph{target path} is a finite path 
ending in the target set $\LocsT$. We denote by $\TPaths$ the set of 
target paths. We let $\TPaths_\play$ (resp. $\TPaths^n_\play$) the 
subset of target paths that start from the last location of the 
finite play \play (resp. containing $n$ transitions without taking 
into account the size of \play). A path is said to be \emph{maximal} if
it is infinite or if it is a target path.

A \emph{deterministic strategy} for $\MinPl$ (resp.~$\MaxPl$) is a mapping
$\detminstrategy\colon \FPlaysMin \to \Trans\times\Rplus$
(resp.~$\detmaxstrategy\colon \FPlaysMax \to \Trans\times
\Rplus$) such that for all finite plays $\play\in\FPlaysMin$
(resp.~$\play\in\FPlaysMax$) ending in (non-target) 
configuration $(\loc,\val)$, there exists an edge
$(\loc,\val)\xrightarrow{\detminstrategy(\play)}(\loc',\val')$. 
We let $\dStratMin$ and $\dStratMax$ denote the set of
deterministic strategies in $\game$ for players $\MinPl$ and $\MaxPl$, respectively. 
A play or finite play
$\play = (\loc_0,\val_0)\extendto{\trans_0, \delay_0}
\extendto{\trans_1, \delay_1}\cdots$ conforms to a
deterministic strategy $\detminstrategy$ of $\MinPl$ (resp.~$\MaxPl$)
if for all $k$ such that $(\loc_k,\val_k)$ belongs to $\MinPl$
(resp.~$\MaxPl$), we have that
$(\trans_k, t_{k}) =
\detminstrategy\big((\loc_0,\val_0)\extendto{\trans_0, \delay_0} \cdots 
\extendto{\trans_{k-1}, \trans_{k-1}}\big)$. For all deterministic 
strategies \detminstrategy and \detmaxstrategy of players \MinPl and 
\MaxPl, respectively, and for all configurations~$(\loc_0,\val_0)$, we 
let $\outcomes((\loc_0,\val_0),\detminstrategy,\detmaxstrategy)$ be
the outcome of $\detminstrategy$ and $\detmaxstrategy$, defined as the
unique maximal play conforming to $\detminstrategy$ and
$\detmaxstrategy$ and starting in~$(\loc_0,\val_0)$.

The objective of \MinPl is to reach a target configuration, while
minimising the cumulated weight up to the target. Hence, we
associate with every finite play
$\play=(\loc_0,\val_0) \moveto{\trans_0, \delay_0} (\loc_1,\val_1)
\moveto{\trans_1, \delay_1}\cdots (\loc_k,\val_k)$ its cumulated
weight, taking into account both discrete and continuous costs:
\begin{displaymath}
\weightC(\play)=\sum_{i=0}^{k-1} [t_i\times \weight(\loc_i) +
\weight(\trans_i)]\,.
\end{displaymath} 
Then, the weight of a maximal play \play,
denoted by $\weight(\play)$, is defined by $+\infty$ if \play is
infinite, i.e. it does not reach $\LocsT$, and $\weightC(\play)$ if 
it ends in $(\loc_T,\val)$ with $\loc_T\in \LocsT$.

A deterministic strategy $\detminstrategy \in\dStratMin$
guarantees a certain value, against all possible strategies of the
opponent: for all locations $\loc$ and valuations~$\val$, we let
\begin{displaymath}
\dValue^\detminstrategy_{\loc,\val}=
\sup_{\detmaxstrategy\in\dStratMax}
\weight(\outcomes((\loc,\val),\detminstrategy,\detmaxstrategy))\,.
\end{displaymath} 
Then, for all locations $\loc$ and valuations~$\val$, we define the 
\emph{deterministic value} from $(\loc, \val)$ under the point of view 
of \MinPl by 
\begin{displaymath}
\duppervalue_{\loc,\val} = \inf_{\detminstrategy\in\dStratMin} \,
\sup_{\detmaxstrategy\in\dStratMax}
\weight(\outcomes((\loc,\val),\detminstrategy,\detmaxstrategy))\,.
\end{displaymath} 
Similarly, we define the deterministic value from $(\loc, \val)$ under 
the point of view of \MaxPl by 
\begin{displaymath}
\dlowervalue_{\loc,\val} =
\sup_{\detmaxstrategy\in\dStratMax}\inf_{\detminstrategy\in\dStratMin}
\weight(\outcomes((\loc,\val),\detminstrategy,\detmaxstrategy))\,.
\end{displaymath}
Since \WTG{s} are known to be 
determined~\cite{BrihayeGeeraertHaddadLefaucheuxMonmege-15}, i.e.~ 
$\dlowervalue_{\loc, \val} = \duppervalue_{\loc, \val}$, we note by 
$\dValue_{\loc, \val}$ the deterministic value, for all 
configurations $(\loc, \val)$. 

In the example of Figure~\ref{fig:wtg}, location $\loc_1$ (resp.~$\loc_5$) has
deterministic value $+\infty$ (resp.~$-\infty$). As a
consequence, the value in~$\loc_4$ is determined by the edge to
$\loc_3$, and depicted in blue on the right of Figure~\ref{fig:wtg}. 
In location $\loc_2$,
the deterministic value associated with the transition to $\loc_3$ is 
depicted in	red, and the deterministic value in $\loc_2$ is obtained as the
minimum of these two curves.

Finally, we say that a deterministic 
strategy \detminstrategy of \MinPl is 
\emph{$\varepsilon$-optimal wrt the deterministic value} if 
$\dValue^\detminstrategy_{\loc,\val} \leq \dValue_{\loc,\val} + \varepsilon$ 
for all $(\loc,\val)$. It is said optimal if this holds for $\varepsilon=0$.

Seminal works in weighted timed games
\cite{AlurBernadskyMadhusudan-04,BouyerCassezFleuryLarsen-04} have assumed that
clocks are \emph{bounded}. This is known to be without loss of generality for
(weighted) timed
automata~\cite[Theorem~2]{BehrmannFehnkerHuneLarsenPetterssonRomijnVaandrager-01}:
it suffices to replace transitions with unbounded delays with self-loop
transitions periodically resetting the clocks. We do not know if it is the case
for the weighted timed games defined above. Indeed, the technique of
\cite{BehrmannFehnkerHuneLarsenPetterssonRomijnVaandrager-01} cannot be
directly applied. This would give too much power to player $\MaxPl$ that would
then be allowed to loop in a location where an unbounded delay could originally
be taken before going to the target. In \cite{BouyerCassezFleuryLarsen-04}, the
situation is simpler since the game is \emph{concurrent}, and thus \MinPl
always has a chance to move outside of such a situation. Trying to detect and
avoid such situations in our turn-based case seems difficult in the presence of
negative weights, since the opportunities of $\MaxPl$ crucially depend on the
configurations of value $-\infty$ that $\MinPl$ could control afterwards: the
problem of detecting such configurations is undecidable
\cite[Prop.~9.2]{Bus19}, which is an additional evidence to
motivate the decision to focus only on bounded weighted timed games. We thus
suppose from now on that all clocks are bounded by a constant
$\clockbound\in\N$, i.e.~every transition of the \WTG is equipped with a guard
$g$ such that $\val\models g$ implies $\val(x) < \clockbound$ for all clocks
$x\in \Cl$. 

We denote by $\wmax^\Locs$ (resp.~$\wmax^\Trans$, $\wmax^e$) the maximal weight
in absolute values of locations (resp.~of transitions, edges) of $\game$,
i.e.~$\wmax^\Locs = \max_{\loc\in \Locs} |\weight(\loc)|$ (resp.~$\wmax^\Trans
= \max_{\trans \in \Trans} |\weight(\trans)|$, $\wmax^e = M\wmax^\Locs +
\wmax^\Trans$).

In the following, we rely on the crucial notion of regions, as introduced in
the seminal work on timed automata \cite{AlurDill-94}. Formally, with respect
to the set $\Cl$ of clocks and the upper bound $\clockbound$ on the valuation
of clocks, we denote by $\regions \Cl \clockbound$ the set of regions bounded
by $\clockbound$. The regions partition the set $[0,\clockbound)^\Cl$ of
valuations. Each such region is characterised by a pair $(\iota, \beta)$ where
$\iota\colon \Cl\to [0,\clockbound)\cap \N$ and $\beta$ is a partition
of~$\Cl$ into subsets $\beta_0\uplus \beta_1\uplus \cdots \uplus \beta_m$
(with $m\geq 0$), where $\beta_0$ can be empty but $\beta_i\neq\emptyset$ for
$1\leq i\leq m$. A valuation $\val$ of $[0,\clockbound)^\Cl$ belongs to the
region characterised by $(\iota, \beta)$ if
\begin{itemize}
\item for all $x\in\Cl$,
$\iota(x) = \lfloor \val(x)\rfloor$;
\item for all $x\in \beta_0$, $\fract(\val(x))=0$;
\item for all $0\leq i\leq m$, for all $x,y\in \beta_i$,
$\fract(\val(x))=\fract(\val(y))$;
\item for all $0\leq i< j\leq m$, for all $x\in \beta_i$
and all $y\in \beta_j$, $\fract(\val(x))
< \fract(\val(y))$.
\end{itemize}
\noindent 
The set of valuations contained in a region $r$ characterised by $(\iota,
\beta)$ with $\beta = \beta_0\uplus \beta_1\uplus \cdots \uplus \beta_m$ can be
described by the guard $g_0\land g_1\land\cdots \land g_m$ with
\[g_0 = \bigwedge_{x\in \beta_0} \big(x=\iota(x)\big), \qquad
g_1 = \bigwedge_{x,y\in \beta_1} \big(0 < x-\iota(x) = y-\iota(y)< 1\big)\]
and for $i\in\{2,\ldots,m\}$,
\[g_i = \bigwedge_{x,y\in \beta_i} \big(z-\iota(z) < x-\iota(x) = y-\iota(y) <
1 \big)\] where $z$ is any clock of $\beta_{i-1}$.

If $r$ is a region, then the time successors of valuations in $r$ form a finite
union of regions. Formally, 
a region $r'$ is said to be a time successor of the region $r$ if
there exist $\val\in r$, $\val'\in r'$, and $\delay>0$ such that
$\val'=\val+\delay$. Moreover, the reset $r[\reset:=0]$ of $\reset\subseteq\Cl$, containing all valuations of $r$ where clocks of $\reset$ have been reset, is also a region. 

A game $\game$ can be populated with the region information, without
loss of generality, as described formally in
\cite{BusattoGastonMonmegeReynier-17}. Letting $\regions \Cl \clockbound$ be 
the set of regions for the set of clocks \Cl bounded by \clockbound, the 
\emph{region automaton, \emph{or} region game}, $\rgame$ is thus the \WTG with 
locations $\RStates = \Locs\times \regions \Cl \clockbound$ and all transitions
$((\loc,r),\guard'',\reset,(\loc',r'))$ with
$(\loc,\guard,\reset,\loc')\in \Trans$ such that the model of the guard
$\guard''$ (i.e.~all valuations $\val$ such that
$\val\models \guard''$) is a region $r''$, time successor of $r$ such
that $r''$ satisfies the guard $\guard$, and $r'$ is the region
obtained from $r''$ by resetting all clocks of $\reset$. Distribution
of locations to players, final locations, and weights are inherited
from $\game$. We call \emph{region path} a finite or infinite sequence
of transitions in this automaton, and we again denote by $\rpath$ such
paths. A play~\play in $\game$ is projected on a region path
$\rpath$, with a similar definition as the projection on paths:
we again say that \play \emph{follows} the region path $\rpath$. It
is important to notice that, even if $\rpath$ is a \emph{cycle}
(i.e.~starts and ends in the same location of the region game), there
may exist plays following it in $\game$ that are not cycles, due to
the fact that regions are sets of valuations. 

As shown in previous works
\cite{BouyerCassezFleuryLarsen-04,BusattoGastonMonmegeReynier-17},
knowing whether $\dValue_{\loc,\val}=+\infty$ for a certain
configuration $(\loc, \val)$ is a purely qualitative problem that can 
be decided easily by using the region game: indeed, 
$\dValue_{\loc,\val} = +\infty$
if and only if \MinPl has no strategies that guarantee that a
target of $\LocsT$ is reached. This is thus a reachability objective, where weights
are useless. Moreover, \MaxPl has a strategy that guarantees that no
plays reach the target $\LocsT$ from any configuration $(\loc,\val)$
such that $\dValue_{\loc,\val}=+\infty$. In this situation, considering
stochastic choices is not interesting. \textbf{We thus rule out this
	case by supposing in the following that no configurations of $\game$
	have a deterministic value $+\infty$}: such configurations can be removed 
in the region game by strengthening the guards on transitions.

\section{Playing stochastically in \WTG{s}}
\label{sec:stochasticWTG}

Our first contribution consists in allowing both players to use
stochastic choices in their strategies. From a game theory point of
view, this seems natural: it is necessary to find Nash 
equilibria in finite strategic games \cite{Nas50}. From a controller 
synthesis point of view, we claim that using stochastic choices is 
natural too in \WTG{s}, especially because player
\MinPl may require exponential memory to play optimally in
\WTG{s}. This is already the case even without clocks (such games are
then sometimes called \emph{shortest-path games}) as in 
\figurename{~\ref{fig:SP1}}. We aim at proving that the memory required by 
\MinPl could be traded for stochastic choices instead (and vice versa). Before
doing so, we must introduce stochastic strategies in the context of
\WTG{s}, which has never been explored until now, as far
as we are aware of. We will however strongly rely on a recent line of
works aiming at studying \emph{stochastic timed automata}
\cite{BertrandBouyerBrihayeMenetBaierGroserJurdzinski-14,
	BouyerBrihayeCarlierMenet-16,BertrandBouyerBrihayeCarlier-18,
	BouyerBrihayeRandourRiviereVandenhove-20},
thus extending the results in the context of two-player games (instead
of model-checking) and with weights, which indeed represents the main
challenge in order to give a meaning to the expected payoff.

Naturally, deterministic strategies for $\MinPl$ are extended to more
general stochastic strategies. We let $\Distr S$ be the set
of all probability distributions over a set $S$ (equipped with an
underlying $\sigma$-algebra). Stochastic strategies are then mappings
$\minstrategy\colon \FPlaysMin \to \Distr{\Trans\times\Rplus}$ where
each finite play is associated to a probability distribution over the
set of pairs of transition and delay. 
Since $\Trans$ is a finite set, we choose to decouple the distribution on pairs of
$\Trans\times \Rplus$ by first selecting a transition and then delay,
whereas authors of
\cite{BertrandBouyerBrihayeMenetBaierGroserJurdzinski-14} consider
independent choices, the one on transitions being described by some
weights on transitions (depending on the current region).
Thus, the strategy $\minstrategy$ can be decoupled as first choosing a transition via
$\minstrategytrans\colon \FPlaysMin \to \Distr{\Trans}$, and then,
knowing the chosen transition, choosing a delay via
$\minstrategydelay\colon \FPlaysMin\times \Trans \to \Distr{\Rplus}$,
the support of the distribution $\minstrategydelay(\play, \trans)$, 
denoted by $\support{\minstrategydelay(\play, \trans)}$, 
being included in the interval $I(\play,\trans)$ of valid delays. 
Similar definitions hold for $\MaxPl$ whose
stochastic strategies will be denoted by~\maxstrategy.

Notice that deterministic strategies are a special case of stochastic 
strategies, where the distributions are chosen to be Dirac distributions. 
Another useful restriction over strategies is the non-use of memory: a
strategy $\minstrategy$ is \emph{memoryless} if for all finite plays
$\play, \play'$ ending in the same configuration, we have that
$\minstrategy (\play)=\minstrategy (\play')$. A similar definition
holds for $\MaxPl$. Finally, we let  
$\FPlays^{\minstrategy, \maxstrategy}_{\play}$ be the set of plays 
reaching \LocsT from \play (without containing \play) conforming to 
\minstrategy and~\maxstrategy.

\subsection{Probability measure on plays.}
\label{subsec:proba}
We fix a strategy \minstrategy and \maxstrategy for each player, and an initial
configuration $(\loc_0,\val_0)$. Our goal is to define a probability measure on
plays. To do so, and following the contribution
of~\cite{BertrandBouyerBrihayeMenetBaierGroserJurdzinski-14} for stochastic
timed automata, the set of plays of a WTG $\game$ starting from
$(\loc_0,\val_0)$ and conforming to $\minstrategy$ and $\maxstrategy$ that we
denote by $\Plays^{\minstrategy, \maxstrategy}_{\loc_0, \val_0}$ can be
equipped with a structure of $\sigma$-algebra generated by \emph{cylinders},
that are all subsets of plays that start with a finite prefix following the
same finite path \ppath (remember that paths are sequences of transitions, with
no information on the delayed time) with some Lebesgue-measurable constraints
on the delays taken along \ppath. The idea is thus to define a probability
measure $\Proba^{\minstrategy,\maxstrategy}_{\loc_0,\val_0}$ on the algebra
generated by such cylinders (i.e.~the closure of cylinders by finite union and
complement) which extends uniquely as a probability measure over the whole
$\sigma$-algebra (i.e.~the closure of cylinders by countable union and
complement), by Carathéodory's extension theorem. First, we formally define the
notion of cylinders, generalising the single configuration $(\loc_0,\val_0)$ by
a finite play $\playBis$ to later take into account the possible use of memory
in strategies of both players. 
\begin{defi}
	Let \playBis be a finite play, $\ppath$ be a finite path, and $\C$ be a
	Lebesgue-measurable subset of $\Rplus^{|\ppath|}$, the \emph{cylinder
	$\Cyl_{\playBis}(\ppath, \C)$} (denoted by $\Cyl_{\playBis}(\ppath)$ when
	$\C = \Rplus^{|\ppath|}$) is the set of maximal plays $\play$ that start in
	the last configuration of $\playBis$ and such that the maximal play
	$\playBis\play$ \emph{satisfies $\ppath,\C$} (denoted by $\playBis\play
	\models \ppath, \C$), i.e.~the prefix of length~$|\ppath|$ of the maximal
	play $\playBis\play$ follows \ppath and its sequence of delays belongs
	to~$\C$. 
\end{defi} 


We define the probability of the cylinder $\Cyl_{\playBis}(\ppath, \C)$ under
the strategies \minstrategy of $\MinPl$ and \maxstrategy of $\MaxPl$, and
denote it by $\Proba^{\minstrategy,\maxstrategy}_{\playBis}(\ppath, \C)$
(instead of the longer
$\Proba^{\minstrategy,\maxstrategy}(\Cyl_{\playBis}(\ppath, \C))$). If the
cylinder is empty (because the finite play $\playBis$ does not follow a prefix
of $\ppath$ or contains delays not consistent with the constraint $\C$), we let
its probability be 0. Otherwise, the probability of the non-empty cylinder is
defined by induction on the length of $\playBis$. If $|\playBis|=|\ppath|$, we
let $\Proba^{\minstrategy,\maxstrategy}_{\playBis}(\ppath, \C) = 1$. Otherwise,
letting $\trans$ be the $(|\playBis|+1)$-th transition of \ppath, and \strategy
be the strategy \minstrategy or \maxstrategy according if $\playBis \in
\FPlaysMin$ or $\playBis \in \FPlaysMax$, we let 
\begin{equation}
	\label{eq:proba-cylinder}
	\Proba^{\minstrategy,\maxstrategy}_{\playBis}(\ppath, \C) = 
	\int_{I(\playBis, \trans)} \strategytrans(\playBis)(\trans) \times
	\Proba^{\minstrategy,\maxstrategy}_{\playBis \extendto{\trans,\delay}}(\ppath, \C) 
	\;\mathrm d\strategydelay(\playBis,\trans)(\delay) 
\end{equation} 
As in the previous definition, when $\C = \Rplus^{|\ppath|}$, we denote by
$\Proba^{\minstrategy,\maxstrategy}_{\playBis}(\ppath)$ the probability of
$\Cyl_{\playBis}(\ppath)$.

For modelling purposes, the authors
of~\cite{BertrandBouyerBrihayeMenetBaierGroserJurdzinski-14} enforce that
probability distributions on delays in $\minstrategydelay$ do not forbid any
delays of the interval $I(\playBis,\trans)$ of possible delays, thus ruling out
singular distributions like Dirac ones that would consider taking a single
possible delay (like deterministic strategies do). More formally, they require
$\minstrategydelay(\playBis,\trans)$ to be absolutely continuous (i.e.~equivalent
to the Lebesgue measure) on interval $I(\playBis,\trans)$. We claim that even with
this assumption, the previous definition of the probability of a cylinder may
not be well-founded, as demonstrated in the following example.

\begin{figure}
	\centering
	\begin{tikzpicture}[xscale=.8,every label/.style={font=\scriptsize}]
	\node[PlayerMin, label={below:$\loc_0$}] at (0, 0)  (s0) {};
	\node[PlayerMin, label={below:$\loc_1$}] at (3, 0)  (s2) {};
	\node[target] at (6, 0) (s3) {\LARGE \smiley}; 
	\node[PlayerMin, label={below:$\loc_2$}] at (9, 0) (s4) {};
	
	\draw[->]
	(s0) edge node[above] {
		$\begin{array}{c}
		\trans_1 \\ x < 1 
		\end{array}$} (s2)
	(s2) edge node[above] {
		$\begin{array}{c}
		\trans_2 \\ x < 2 
		\end{array}$} (s3)
	(s2) edge[bend right=15] node[below] {
		$\begin{array}{c}
		\trans_3 \\ x < 1 
		\end{array}$} (s4)
	(s4) edge node[above] {
		$\begin{array}{c}
		\trans_4 \\ x \leq 2 
		\end{array}$} (s3)
	;
	\end{tikzpicture}
	\caption{A (one-player) \WTG with a single clock $x$ and all
		weights equal to~0.}
	\label{Fig:simpleWTA}
\end{figure}

\begin{exa}
	\label{ex:defproba}
	We consider the \WTG of \figurename{~\ref{Fig:simpleWTA}}, where only \MinPl
	plays and all weights equal $0$ (so we can see it as a stochastic timed 
	automaton), and the memoryless strategy \minstrategy (partially) defined as
	follows. We let $A$ be a non-Lebesgue-measurable subset of
	$[0,1)$. We denote $\overline{A}$ the complement set $[0, 1) \setminus A$ 
	and $\charact_A$ the characteristic function of the set $A$. We start by
	defining the delays to match as closely as possible the setting of
	\cite{BertrandBouyerBrihayeMenetBaierGroserJurdzinski-14} here. For
	delays in $\loc_0$ and $\loc_2$, we consider uniform probability
	distributions on the appropriate intervals. In $\loc_1$, for all
	$\delay_1 \in [0,1)$, we let $\minstrategydelay((\loc_1, \delay_1), \trans_2)$
	be the uniform distribution on $[0, 2-\delay_1]$ if $\delay_1 \in A$, and the
	truncated exponential distribution on $[0, 2-\delay_1]$ with parameter
	$\lambda=1$ otherwise. For the choice of transitions from $\loc_1$ (the only 
	place where there is a choice), for all $\delay_1\in [0,1)$, we let
	\begin{displaymath}
	\minstrategytrans(\loc_1,\delay_1)(\trans_2) = 
	\begin{cases}
	(3-\delay_1)/(4-2\delay_1) = f(\delay_1) & \text{if } \delay_1\in A  \\
	(1+e^{-(1-\delay_1)}-2e^{-(2-\delay_1)})/(2-2e^{-(2-\delay_1)}) = g(\delay_1)  
	& \text{if } \delay_1\in \overline A
	\end{cases}
	\end{displaymath} 
	In the setting of
	\cite{BertrandBouyerBrihayeMenetBaierGroserJurdzinski-14}, the
	strategy in $\loc_1$ can be obtained by first choosing a delay
	similarly as ours, and then choosing transition $\trans_2$ with
	either probability $1/2$ or $1$ depending on whether transition
	$\trans_3$ is fireable after letting the chosen delay elapse:
	authors of \cite{BertrandBouyerBrihayeMenetBaierGroserJurdzinski-14}
	would describe this by putting weight $1$ on both transitions
	$\trans_2$ and $\trans_3$ in the stochastic timed automaton. The
	intricate formulas above for the transitions are thus simply a way
	to mimic their setting in ours.  Let us try to compute the
	probability of the cylinder $\Cyl_{\loc_0,0}(\trans_1 \trans_2)$: its 
	probability with respect to $\Proba^\minstrategy_{\loc_0,0}$ is
	\begin{displaymath}
	\Proba^\minstrategy_{\loc_0, 0}(\trans_1 \trans_2) = 
	\int_{I((\loc_0, 0), \trans_1)} \minstrategytrans(\loc_0, 0)(\trans_1) 
	\times \Proba^\minstrategy_{(\loc_0,0) \extendto{\trans_1,\delay_1}}(\trans_2)
	\; \mathrm d\minstrategydelay((\loc_0, 0), \trans_1)(\delay_1)
	\end{displaymath}
	Moreover, since $\trans_1$ is the only transition from $\loc_0$, then 
	$\minstrategytrans(\loc_0, 0)(\trans_1)=1$ and 
	\begin{displaymath}
	\Proba^\minstrategy_{\loc_0,0}(\trans_1 \trans_2) = 
	\int_0^1 \Proba^\minstrategy_{(\loc_0,0)\extendto{\trans_1,\delay_1}}(\trans_2)
	\;\mathrm d\delay_1
	\end{displaymath}
	that requires $h \colon \delay_1 \mapsto \Proba^\minstrategy_{(\loc_0,0) 
		\extendto{\trans_1,\delay_1}}(\trans_2)$ to be a measurable 
	function on $[0,1]$ to be well-defined. For $\delay_1 \in[0,1) \cap A$,
	\begin{displaymath}
	h(\delay_1)=\int_{I((\loc_1, \delay_1), \trans_1)} 
	\minstrategytrans(\loc_1,\delay_1)(\trans_2) \times 1 
	\;\mathrm d\minstrategydelay((\loc_1, \delay_1), \trans_2)(\delay_2) 
	= \int_{0}^{2-\delay_1} f(\delay_1) \frac{\mathrm d\delay_2}{2-\delay_1} 
	= f(\delay_1).
	\end{displaymath}
	Similarly, if $\delay_1\in[0,1)\cap \overline{A}$, $h(\delay_1)=g(\delay_1)$. 
	Functions $f$ and $g$ are measurable and never match over $[0,1)$. Thus, 
	if $h$ is measurable, so would be $(h-g)/(f-\delay)$ that is equal to the
	characteristic function of $A$: this contradicts the
	non-measurability of $A$. And thus, it is not possible to define the
	probability $\Proba^\minstrategy_{\loc_0,0}(\trans_1 \trans_2)$.
\end{exa}

From this example, we see the importance to moreover enforce that the
distributions $\minstrategytrans(\playBis)$ and
$\minstrategydelay(\playBis,\trans)$ are ``measurable wrt the sequence of
delays along the play \playBis''. This is easy to define for the transition
part. For delays, since we want deterministic strategies to be a subset of
stochastic strategies, we must be able to choose delays by using Dirac
distributions, and by extension discrete distributions (that are not absolutely
continuous, as \cite{BertrandBouyerBrihayeMenetBaierGroserJurdzinski-14}
requires). We restrict ourselves to discrete distributions whose
\emph{cumulative distribution functions} (CDF) are a finite linear combination
of translated Heaviside functions. Precisely, we let the Heaviside function $H$
denote the mapping from $\R$ to $[0,1]$ such that $H(t)=0$ if $t<0$ and
$H(t)=1$ otherwise. Recall that it is the CDF of the Dirac distribution
choosing $t=0$. This results in the following hypothesis:
\begin{hypothesis}[smooth strategies]\label{hyp:measurability}
	A strategy $\strategy$ is said to be \emph{smooth} if 
	for all finite plays $\playBis=(\loc_0,\val_0) \extendto{\trans_0,\delay_0} 
	\cdots \allowbreak\extendto{\trans_{k-1},\delay_{k-1}}$ and transitions \trans,
	\begin{enumerate}
		\item\label{item:hyp-measurability_1} the mapping 
		$(t_0,\ldots,t_{k-1}) \mapsto \strategytrans(\playBis)(\trans)$ 
		is Lebesgue-measurable; and 
		
		\item\label{item:hyp-measurability_2} the probability distribution
		$\strategydelay(\playBis,\trans)$ (of the random variable $t$) is
		described by a CDF of the form 
		\begin{displaymath}
		\delay \mapsto G(\playBis,\trans)(t) + \sum_{i=0}^{n} \alpha_i(\playBis,\trans) 
		H(\delay-a_i(\playBis,\trans))
		\end{displaymath} 
		where $G(\playBis,\trans)$ is an absolutely continuous function, $\alpha_i(\playBis,\trans) \in [0,1]$ and $a_i(\playBis,\trans) \in \Rplus$, for all $i \in \{0,\ldots,n\}$. 
	\end{enumerate}
		
	\noindent Moreover, the mappings 
		\begin{displaymath}
		(t_0,\ldots,t_{k-1},t)\mapsto G(\playBis,\trans)(t) \text{,} \quad 
		(t_0,\ldots,t_{k-1})\mapsto \alpha_i(\playBis,\trans) \text{, \quad and} \quad
		(t_0,\ldots,t_{k-1})\mapsto a_i(\playBis,\trans)
		\end{displaymath}
		must be Lebesgue-measurable.
\end{hypothesis}
\noindent 
By default, all stochastic strategies considered in the rest of this article are smooth. We thus let $\StratMin$ and
$\StratMax$ be the sets of smooth (stochastic) strategies 
of each player, as well as
$\mStratMin$ and $\mStratMax$ be their respective subsets of memoryless
strategies. 

However, not all deterministic strategies are smooth (Example~\ref{ex:defproba} can be modified to build non-smooth deterministic strategies), and we will thus recall the smoothness of them when we need it: we let $\sdStratMin$ and $\sdStratMax$ be the respective 
subsets of smooth deterministic strategies. 

\begin{rem}
	In divergent WTG{s} (that will be in studied in Section~\ref{sec:divergent}), we will show in Corollary~\ref{cor:sdVal-div} that there exists, for both players, $\varepsilon$-optimal smooth deterministic strategies with respect to the deterministic value. 
\end{rem}

Under the smooth hypothesis, we prove that the integral in~\eqref{eq:proba-cylinder} is always well-defined:

\begin{lem}
	\label{lem:proba_defi}
	For all finite plays $\playBis = (\loc_0,\val_0) \extendto{\trans_0,\delay_0} 
	\cdots \extendto{\trans_{k-1},\delay_{k-1}}$, finite paths 
	$\ppath$ starting in~$\loc_0$ (such that $|\ppath| \geq k$), and Lebesgue-measurable sets 
	$\mathcal C$ of $\Rplus^{|\ppath|}$, 
	\begin{enumerate}
		\item\label{item:proba_defi-measurability} 
		the mapping $(t_0,\ldots,t_{k-1})\mapsto 
		\Proba^{\minstrategy,\maxstrategy}_{\playBis}(\ppath, \C)$ is 
		Lebesgue-measurable;
		\item\label{item:proba_defi-bound} 
		$0 \leq \Proba^{\minstrategy,\maxstrategy}_{\playBis}(\ppath, \C) \leq 1$.
	\end{enumerate}
\end{lem}
\begin{proof}
	We reason by induction on the length of \playBis. If $|\playBis| =
	|\ppath|$, then (by its definition), the mapping $(t_0,\ldots,t_{k-1})
	\mapsto \Proba^{\minstrategy,\maxstrategy}_{\playBis}(\ppath, \C)$ is equal
	to $1$ if and only if $\ppath = \trans_0\trans_1 \cdots \trans_{k-1}$ and
	$(t_0,\ldots,t_{k-1}) \in \C$ (otherwise it is equal to $0$). In
	particular, in case $\ppath = \trans_0\trans_1 \cdots \trans_{k-1}$, this
	mapping is equal to the characteristic function of $\C$ (regarding to
	delays) which is Lebesgue-measurable by definition of a cylinder. Thus,
	$(t_0,\ldots,t_{k-1}) \mapsto
	\Proba^{\minstrategy,\maxstrategy}_{\playBis}(\ppath, \C)$ is
	Lebesgue-measurable. Otherwise, we suppose that the property holds for all
	finite plays of length $k+1\leq |\ppath|$ and we consider a finite play
	\playBis of length $k$. Letting \strategy be the strategy \minstrategy or
	\maxstrategy according as $\playBis \in \FPlaysMin$, or $\playBis \in
	\FPlaysMax$, then by using the notations of~\eqref{eq:proba-cylinder} and
	Hypothesis~\ref{hyp:measurability}.\eqref{item:hyp-measurability_2}, we can
	decompose $\Proba^{\minstrategy,\maxstrategy}_{\playBis}(\ppath, \C)$ as  
	$\strategytrans(\playBis)(\trans)\,(A + B)$ where 
	\begin{equation}
	\label{eq:proba-defi-A}
	A = \int_{I(\playBis, \trans) \cap \C^k}
	\Proba^{\minstrategy,\maxstrategy}_{\playBis\extendto{\trans,\delay}}(\ppath, \C) 
	\times G(\playBis, \trans)(\delay) \;\mathrm d\delay 
	\end{equation}
	and 
	\begin{equation}
	\label{eq:proba-defi-B}
	B = \sum_{i=0}^\infty \alpha_i(\playBis, \trans)
	\times \Proba^{\minstrategy,\maxstrategy}_{\playBis
		\extendto{\trans,a_i(\playBis,\trans)}}(\ppath, \C) \,.
	\end{equation}
	
	We start by studying~\eqref{eq:proba-defi-A}. In particular, by induction
	hypothesis applied to the finite play~$\playBis \extendto{\trans,\delay}$,
	we know that the mapping $(\delay_0,\ldots,\delay_{k-1},\delay) \mapsto
	\Proba^{\minstrategy,\maxstrategy}_{\playBis\extendto{\trans,\delay}}(\ppath,
	\C)$ is Lebesgue-measurable. Therefore, the mapping
	$(\delay_0,\ldots,\delay_{k-1},\delay) \mapsto
	\Proba^{\minstrategy,\maxstrategy}_{\playBis\extendto{\trans,
	\delay}}(\ppath, \C) \times G(\playBis, \trans)(\delay)$ is
	Lebesgue-measurable. Since it is also non-negative and upper-bounded by
	$1$, and the interval $I(\playBis, \trans)$ is bounded, the integral
	exists. Moreover, by Fubini Theorem\footnote{The dependency on \playBis in
	the interval $I(\playBis, \trans) \cap \C^k$ of integration can be replaced
	by the full space $[0,M]$ since clocks are bounded by $M$, and the
	characteristic function of $\delay \in I(\playBis, \trans) \cap \C^k$ which
	is measurable	wrt~$(\delay_0,\ldots,\delay_{k-1}, \delay)$.}, the
	resulting integral is Lebesgue-measurable wrt the delays in \play,
	i.e.~\eqref{eq:proba-defi-A} is Lebesgue-measurable wrt $(\delay_0, \ldots,
	\delay_{k-1})$.  
	
	We focus on~\eqref{eq:proba-defi-B}. By composition of the
	Lebesgue-measurable mappings $(\delay_0,\ldots,\delay_{k-1},\delay) \allowbreak\mapsto
	\Proba^{\minstrategy,\maxstrategy}_{\playBis\extendto{\trans,\delay}}(\ppath,
	\C)$ (by induction hypothesis applied on $\playBis\extendto{\trans,
	\delay}$) and $(\delay_0, \ldots, \delay_{k-1}) \mapsto a_i(\playBis,
	\trans)$ (by
	Hypothesis~\ref{hyp:measurability}.\eqref{item:hyp-measurability_2}), the
	mapping $(\delay_0,\ldots,\delay_{k-1}) \mapsto
	\Proba^{\minstrategy,\maxstrategy}_{\playBis
	\extendto{\trans,a_i(\play,\trans)}}(\ppath, \C)$ is Lebesgue-measurable.
	Then, thanks to
	Hypothesis~\ref{hyp:measurability}.\eqref{item:hyp-measurability_2}, the
	countable sum of~\eqref{eq:proba-defi-B} absolutely converges. Thus, by
	countable sum of Lebesgue-measurable functions,~\eqref{eq:proba-defi-B} is
	Lebesgue-measurable wrt $(\delay_0, \ldots, \delay_{k-1})$. 
	
	Finally, by using 
	Hypothesis~\ref{hyp:measurability}.\eqref{item:hyp-measurability_1},
	$(\delay_0,\ldots,\delay_{k-1}) \mapsto \strategytrans(\playBis)(\trans)$ 
	is Lebesgue-measurable. Thus, $(\delay_0,\ldots,\delay_{k-1}) \mapsto	
	\Proba^{\minstrategy,\maxstrategy}_{\playBis}(\ppath, \C)$ is 
	Lebesgue-measurable as a linear combination of Lebesgue-measurable 
	functions. Moreover, by using the fact that
	$\strategytrans(\playBis)$ is a probability distribution and the
	induction hypothesis bounding the probabilities between $0$ and $1$, 
	we obtain, from the definition, 
	\begin{displaymath}
	0 \leq 
	\Proba^{\minstrategy,\maxstrategy}_{\playBis}(\ppath, \C) 
	\leq \int_{I(\play, \trans)}\mathrm d\strategydelay(\playBis, \trans)(\delay) 
	= 1
	\end{displaymath}
	since $\strategydelay(\playBis, \trans)$ is a probability measure.
\end{proof}
\noindent 
Under the smoothness hypothesis, we want to prove that
$\Proba^{\minstrategy,\maxstrategy}_{\playBis}$ can be extended as a probability
distribution over the set $\Plays^{\minstrategy, \maxstrategy}_{\playBis}$ of
maximal plays starting in the last configuration of \playBis, and conforming to 
\minstrategy and \maxstrategy, equipped with the $\sigma$-algebra $\Sigma_{\playBis}$ 
generated by the cylinders $\Cyl_{\playBis}(\ppath, \C)$, following a similar proof
as~\cite[Proposition 3.2]{BertrandBouyerBrihayeMenetBaierGroserJurdzinski-14}. 
The complete proof is given in Appendix~\ref{app:measure}. 

\begin{restatable}{prop}{propProbaDistribution}
	\label{prop:proba_distribution}
	If $\minstrategy$ and $\maxstrategy$ are smooth strategies, 
	then, for all finite plays~\playBis, 
	there exists a probability measure 
	$\Proba^{\minstrategy,\maxstrategy}_{\playBis}$ over 
	$(\Plays^{\minstrategy, \maxstrategy}_{\playBis}, \Sigma_{\playBis})$.
\end{restatable}

\subsection{Expected payoff of plays}
\label{subsec:expected}

We now define the expected weight over the maximal plays starting from the last
configuration of a finite play $\playBis$. We denoted it by
$\E^{\minstrategy,\maxstrategy}_{\playBis}$ instead of
$\E^{\minstrategy,\maxstrategy}_{\playBis}(\weight)$, since we only consider
the expectation of the weight~\weight:
\begin{equation}
\label{eq:expectation-definition}
\E^{\minstrategy,\maxstrategy}_{\playBis} =
\int_\play \weight(\play) \; 
\mathrm d\Proba^{\minstrategy,\maxstrategy}_{\playBis}(\play)
\end{equation}
This definition of the expectation makes sense by the following result: 
\begin{lem}
The mapping $\play \mapsto 
\weight(\play)$ is a $\Proba^{\minstrategy,\maxstrategy}_{\playBis}$-measurable 
function.
\end{lem}
\begin{proof} By using the results
of~\cite[Section~3.2]{BouyerBrihayeBruyereRaskin-07}, we know that given a
finite path \ppath ending in a target location of \LocsT and an open interval
of possible weights $I_\weight$ (that is a basic Lebesgue-measurable set of
$\R$), the set of maximal plays following \ppath and with a weight in
$I_\weight$ is defined by constraints over the delays taken along the play.
More precisely, this set of plays can be defined by a set of delays
$(\delay_i)_{0 \leq i \leq |\ppath|-1}$ that satisfy a linear program where the
constraints are over partial sums $\sum_{i=j}^k \delay_i$. Thus, the set of
maximal plays with a weight in $I_\weight$ is a (countable) union of cylinders,
defined by finite maximal paths reaching the target and constraints on delays
described by a linear program (and thus a Lebesgue-measurable set of
$\Rplus^{|\ppath|}$). 
\end{proof}
\noindent 
Even if the integral $\E^{\minstrategy, \maxstrategy}_{\playBis}$ always
exists, it may be infinite. In particular, it is finite only if the probability
to reach a target location is equal to~$1$, since otherwise there is a non-zero
probability to not reach the target location, the weight of all such plays
being $+\infty$, leading to an infinite expectation. We thus now require that 
$\Proba^{\minstrategy,\maxstrategy}_{\playBis}(\TPlays_{\playBis}) = 1$
(i.e.~the probability to follow an infinite path is $0$) that is possible since
$\TPlays_{\playBis}$ is a
$\Proba^{\minstrategy,\maxstrategy}_{\playBis}$-measurable set as a countable
union of cylinders generated by all finite maximal paths ending in \LocsT.
However, this is not a sufficient condition to ensure that the expected weight
is finite. 

\begin{figure}
	\centering
	\begin{tikzpicture}[xscale=.8,every label/.style={font=\scriptsize}]
	\node[PlayerMin, label=above:$\loc$] at (0, 0) (s0) {$\mathbf{0}$};
	\node[target] at (4, 0) (s1) {\LARGE $\smiley$};
	
	\draw[->]
	(s0) edge[loop left] node[left]{$\trans_0, x \leq 1, \mathbf{1}$} (s0)
	(s0) edge node[above]{$\trans_1, x \leq 1, \mathbf{1}$} (s1)
	;
	\end{tikzpicture}
	\caption{A \WTG where $\MinPl$ has a strategy
		reaching the target with probability 1 but with an
		expected weight equal to $+\infty$.}
	\label{Fig:cyclicWTA}
\end{figure}

\begin{exa}
	\label{ex:infinitesum}
	We consider the WTG of \figurename{~\ref{Fig:cyclicWTA}}, where only \MinPl
	plays, and the memoryless strategy \minstrategy defined as follows. For all 
	$t\leq 1$, $\minstrategytrans(\ell, t)(\trans_0) = t$,
	$\minstrategytrans(\ell, t)(\trans_1) = 1-t$, and, for all $i\geq 2$
	and $\trans\in\{\trans_0,\trans_1\}$,
	$\minstrategydelay((\ell, 1 - \frac 1 i), \trans)$ is the Dirac
	distribution selecting the delay $t=\frac 1 i-\frac 1 {i+1}$. Notice
	that we can extend the delay distribution (continuously for instance)
	so that this strategy is smooth.
	For all $i\geq 1$, there is a unique play conforming to $\minstrategy$
	of length~$i$ that reaches the target from configuration
	$(\loc, \frac{1}{2})$: it has a weight $i$ and a probability
	$\frac{1}{i+1}\prod_{j=2}^{i} (1 - \frac 1 j) = \frac {1}{i(i+1)}$. In
	particular,
	$\Proba^\minstrategy_{\loc, \frac{1}{2}}(\TPlays_{\loc, \frac{1}{2}})
	= \sum_{i\geq 1} \frac 1 {i(i+1)} = 1$. Moreover, by the previous 
	computation, we can deduce that 
	\begin{displaymath}
	\E^{\minstrategy,\maxstrategy}_{\loc, \frac{1}{2}} =
	\int_\play \weight(\play) 
	\; \mathrm d\Proba^{\minstrategy,\maxstrategy}_{\loc, \frac{1}{2}}(\play) = 
	\sum_i \int_\play \weight(\play)\,\charact_{\TPlays^i_{\loc, \frac{1}{2}}}(\play)
	\; \mathrm d\Proba^{\minstrategy,\maxstrategy}_{\loc, \frac{1}{2}}(\play) 
	\end{displaymath}
	where $\TPlays^i_{\loc, \frac{1}{2}}$ is a subset of 
	$\TPlays_{\loc, \frac{1}{2}}$ that contains all plays of length $i$. The 
	expected weight would thus be 
	$\E^{\minstrategy,\maxstrategy}_{\loc, \frac{1}{2}} = 
	\sum_{i \geq 1} i \times \frac{1}{i(i+1)} = \sum_{i \geq 1} \frac{1}{i+1}$, which 
	is a diverging series. This example can easily be adapted to only consider 
	continuous distributions on the delays (instead of Dirac~ones).
\end{exa}

Even smooth deterministic strategies of $\MinPl$ that guarantee that a target location is reached may not ensure that the expected payoff is finite (see Appendix~\ref{ex:det-proper}). We thus need a stronger hypothesis to ensure that the expectation is finite in all cases. We adopt
here an asymmetrical point of view, relying only on a hypothesis on the
strategy $\minstrategy$ of $\MinPl$. Our choice is grounded in our
controller synthesis view, $\MinPl$ being the controller desiring to
reach a target location with minimum expected weight, while $\MaxPl$
is an uncontrollable~environment.

On top of ensuring that the strategy of $\MinPl$ reaches the target set of locations with probability $1$, we ask the strategies of $\MinPl$ to fulfil the following hypothesis, by separating two cases. For deterministic strategies, we simply ask that the expected payoff is finite, as we indeed want. For non-deterministic strategies, we ask a more demanding condition (and prove afterwards that it is sufficient), which will be helpful in our later proofs: not only \MinPl must reach the target almost
surely, but must do it \emph{quickly enough} so that the
expectation converges.
\begin{hypothesis}[proper strategies]\label{hyp:proper}
	A strategy $\minstrategy \in \StratMin$ of \MinPl is said to be \emph{proper} if: 
	\begin{itemize}
		\item either it is a smooth deterministic strategy and
		for all finite plays \playBis and strategies
		$\maxstrategy \in \StratMax$, $\Proba^{\minstrategy,\maxstrategy}_{\playBis}(\TPlays_{\playBis})=1$ and $\E^{\minstrategy,\maxstrategy}_{\playBis}$ is finite; 
		\item or there exist $m\in \N$ and $\alpha\in (0,1)$ such that
		for all finite plays \playBis and strategies
		$\maxstrategy \in \StratMax$,
		\begin{displaymath}
		\Proba^{\minstrategy,\maxstrategy}_{\playBis}
		(\bigcup_{n \leq m}\TPlays^n_{\playBis}) \geq \alpha \,.
		\end{displaymath}
	\end{itemize}
\end{hypothesis}
\noindent 
We let $\StratMinProper$ be the set of proper (smooth) strategies 
	of \MinPl. We let $\mStratMinProper$ be the set of proper (smooth) memoryless strategies of \MinPl. Finally, we let $\sdStratMinProper$ be the set of proper smooth deterministic strategies 
	of \MinPl. 

In the non-deterministic case, Hypothesis~\ref{hyp:proper} formalises the idea that \MinPl guarantees that the play reaches a 
target quickly enough since we can exponentially bound the probability 
of each cylinder according to its length. The complete proof of the following 
result is in Appendix~\ref{app:proof_lem-proper-bound}. 

\begin{restatable}{lem}{lemProperBound}
	\label{lem:proper_bound}
	Let $\minstrategy \in \StratMinProper$ be a strategy of \MinPl satisfying the second condition of 
	Hypothesis~\ref{hyp:proper} for the bounds $\alpha$ and $m$. Let
	$\maxstrategy \in \StratMax$ be a strategy of \MaxPl and $\playBis \in
	\FPlays$ be a finite play following a finite path $\ppathBis$. For all $n$,
	we have
	\begin{displaymath}
	\sum_{\ppath \in \TPaths^n_{\playBis}} 
	\Proba^{\minstrategy,\maxstrategy}_{\playBis}(\ppathBis\ppath)
	\leq (1-\alpha)^{\lfloor n/m\rfloor}.
	\end{displaymath}
\end{restatable}
\noindent 
In particular, this hypothesis is indeed a sufficient condition for the expected payoff to be finite:

\begin{prop}
	\label{prop:proper}
	All proper strategies of \MinPl
	are such that for all finite plays \playBis and strategies
	$\maxstrategy \in \StratMax$, $\E^{\minstrategy,\maxstrategy}_{\playBis}$ is finite.
\end{prop}
\noindent 
The proof is trivial, by definition, in the case of smooth deterministic strategies. For the other case,
the main idea of the proof is to decompose the expectation $\E^{\minstrategy,
\maxstrategy}_{\playBis}$ according to the size of maximal plays such that we
can bound their weight by an affine function and their probabilities by the
exponential function seen in Lemma~\ref{lem:proper_bound}. A natural choice is
to decompose the expectation with cylinders (as for probability). Thus, we need
to define the expected weight of a cylinder. We will not need to consider
specific constraints on the delays and  thus only consider cylinders of the
form $\Cyl_{\playBis}(\ppath)$ (with the set of constraints
$\Rplus^{|\ppath|}$).

\begin{defi}\label{def:expectation}
	Let \playBis be a finite play and \ppath be a finite path. If the cylinder $\Cyl_{\playBis}(\ppath)$ is non empty, the \emph{expected weight 
		$\E^{\minstrategy,\maxstrategy}_{\playBis}(\ppath)$} 
	of plays that are in $\Cyl_{\playBis}(\ppath)$ is defined by induction on the 
	length of \playBis. If $|\playBis| = |\ppath|$, we let
	$\E^{\minstrategy,\maxstrategy}_{\playBis}(\ppath) = 0$, meaning that there are no more steps to take into account in the expected weight. 
	Otherwise, letting $\trans=(\loc,g,Y,\loc')$ be the $(|\playBis|+1)$-th
	transition of \ppath, and \strategy be the strategy \minstrategy or
	\maxstrategy depending if $\playBis \in \FPlaysMin$ or $\playBis \in
	\FPlaysMax$, we let 
	\begin{displaymath}
	\E^{\minstrategy,\maxstrategy}_{\playBis}(\ppath) = 
	\displaystyle\int_{I(\playBis, \trans)} \strategytrans(\playBis)(\trans)
	\Big[\big(\delay\,\weight(\loc) + \weight(\trans)\big)
	\Proba^{\minstrategy,\maxstrategy}_{\playBis 	
		\extendto{\trans,\delay}}(\ppath) +
	\E^{\minstrategy,\maxstrategy}_{\playBis 	
		\extendto{\trans,\delay}}(\ppath)\Big]
	\;\mathrm d \strategydelay(\playBis,\trans)(\delay) \,.
	\end{displaymath}
	For the sake of completeness, if the cylinder $\Cyl_{\playBis}(\ppath)$ is empty, we let $\E^{\minstrategy,\maxstrategy}_{\playBis}(\ppath)=0$.
\end{defi}

By adapting the proof of Lemma~\ref{lem:proba_defi}, 
the smooth hypothesis is sufficient to show that all 
expectations $\E^{\minstrategy,\maxstrategy}_{\playBis}(\ppath)$ are finite:
\begin{lem}
	\label{lem:expectation-integral}
	For all finite plays $\playBis = (\loc_0, \val_0)\extendto{\trans_0, \delay_0} 
	\cdots \extendto{\trans_{k-1}, \delay_{k-1}}$, we let $\ppathBis=\trans_0\trans_1\cdots\trans_{k-1}$. For all finite 
	paths $\ppath$ starting in the last location of $\ppathBis$, 
	$\minstrategy \in \StratMin$ and $\maxstrategy \in \StratMax$:
	\begin{enumerate}
		\item\label{item:expectation-integral-cv}
		$\E^{\minstrategy,\maxstrategy}_{\playBis}(\ppathBis\ppath)$ exists;
		\item\label{item:expectation-integral-bound}
		$|\E^{\minstrategy,\maxstrategy}_{\playBis}(\ppathBis\ppath)| \leq 
		\Proba^{\minstrategy,\maxstrategy}_{\playBis}(\ppathBis\ppath)
		\,|\ppath|\wmax^e$ (so
		$\E^{\minstrategy,\maxstrategy}_{\playBis}(\ppathBis\ppath)$ is finite);
		\item\label{item:expectation-integral-measurable} 
		the mapping $(\delay_0,\ldots,\delay_{k-1}) \mapsto 
		\E^{\minstrategy,\maxstrategy}_{\playBis}(\ppathBis\ppath)$ is 
		Lebesgue-measurable.
	\end{enumerate}
\end{lem}
\begin{proof}
	We start by proving~\eqref{item:expectation-integral-cv} 
	and~\eqref{item:expectation-integral-bound} by induction on the length $k$ of 
	\playBis. Since \playBis follows \ppathBis, the result is 
	immediate when $k = |\ppathBis\ppath|$ (i.e.~$|\ppath| = 0$). Otherwise, 
	we suppose that the property holds for 
	all finite plays of length $k+1$ and we consider a finite play \playBis of 
	length $k$. We let \strategy be the strategy \minstrategy or \maxstrategy 
	according if $\playBis \in \FPlaysMin$ or $\playBis \in \FPlaysMax$, 
	$\trans= (\loc, \guard, \reset, \loc')$ be a transition with $\loc$ the last location of $\playBis$ and $\ppath'$ be a finite path starting in $\loc'$. We also let $\ppath = \trans \ppath'$.
	By definition of the expectation, we have 
	\begin{displaymath} 
	\E^{\minstrategy,\maxstrategy}_{\playBis}(\ppathBis\ppath) = 
	\displaystyle\int_{I(\playBis, \trans)} \strategytrans(\playBis)(\trans)
	\Big[(\delay\,\weight(\loc) + \weight(\trans))
	\Proba^{\minstrategy,\maxstrategy}_{\playBis 
		\extendto{\trans,\delay}}(\ppathBis\ppath) +
	\E^{\minstrategy,\maxstrategy}_{\playBis 
		\extendto{\trans,\delay}}(\ppathBis\ppath)\Big]
	\;\mathrm d \strategydelay(\playBis,\trans)(\delay) \, .
	\end{displaymath}
	By induction hypothesis for all finite plays 
	$\playBis \extendto{\trans,\delay}$, we know that 
	$\E^{\minstrategy,\maxstrategy}_{\playBis \extendto{\trans,\delay}}(\ppathBis\ppath)$ 
	exists and is a measurable function of $\delay$. Therefore, 
	the mapping 
	\begin{displaymath}
	\delay \mapsto (\delay\,\weight(\loc) + \weight(\trans)) 
	\Proba^{\minstrategy,\maxstrategy}_{\playBis 
		\extendto{\trans,\delay}}(\ppathBis\ppath) + 
	\E^{\minstrategy,\maxstrategy}_{\playBis \extendto{\trans,\delay}}(\ppathBis\ppath)
	\end{displaymath}
	is measurable (by Lemma~\ref{lem:proba_defi}). Thus, since  
	$(\delay_0, \ldots, \delay_{k-1}) \mapsto \strategytrans(\play)(\trans)$ 
	is Lebesgue-measurable (by 
	Hypothesis~\ref{hyp:measurability}.\eqref{item:hyp-measurability_1}),
	$\E^{\minstrategy,\maxstrategy}_{\playBis}(\ppathBis\ppath)$ exists, 
	i.e.~\eqref{item:expectation-integral-cv} holds. Moreover, by 
	triangular inequality, we~have 
	\begin{displaymath} 
	|\E^{\minstrategy,\maxstrategy}_{\playBis}(\ppathBis\ppath)| \leq 
	\displaystyle\int_{I(\playBis, \trans)} \strategytrans(\playBis)(\trans)
	\Big[|\delay\,\weight(\loc) + \weight(\trans)|
	\Proba^{\minstrategy,\maxstrategy}_{\playBis 
		\extendto{\trans,\delay}}(\ppathBis\ppath) +
	|\E^{\minstrategy,\maxstrategy}_{\playBis 
		\extendto{\trans,\delay}}(\ppathBis\ppath)|\Big]
	\;\mathrm d \strategydelay(\playBis,\trans)(\delay) \,.
	\end{displaymath}
	By definition of $\wmax^e$ and induction hypothesis applied to 
	$\playBis \extendto{\trans, \delay}$,
	\begin{displaymath}
	|\delay\,\weight(\loc) + \weight(\trans)| \leq \wmax^e
	\qquad \text{and} \qquad 
	|\E^{\minstrategy,\maxstrategy}_{\playBis 
		\extendto{\trans,\delay}}(\ppathBis\ppath)|
	\leq \Proba^{\minstrategy,\maxstrategy}_{\playBis 
		\extendto{\trans,\delay}}(\ppathBis\ppath) 
	\,|\ppath'|\, \wmax^e\,.
	\end{displaymath}
	 By properties of the integral, 
	we deduce that 
	\begin{displaymath}
	|\E^{\minstrategy,\maxstrategy}_{\playBis}(\ppathBis\ppath)| \leq 
	\underbrace{(|\ppath'|+1)}_{=|\ppath|}\wmax^e
	\underbrace{\int_{I(\playBis, \trans)} \strategytrans(\playBis)(\trans)
		\Proba^{\minstrategy,\maxstrategy}_{\playBis 
			\extendto{\trans,\delay}}(\ppathBis\ppath) 
		\;\mathrm d\strategydelay(\playBis,\trans)(\delay)}_{= \,
		\Proba^{\minstrategy,\maxstrategy}_{\playBis}(\ppathBis\ppath)} 
	\end{displaymath}  
	and~\eqref{item:expectation-integral-bound} holds.
	
	To conclude the proof, we remark that by~\eqref{item:expectation-integral-bound} 
	and Fubini theorem, $\E^{\minstrategy,\maxstrategy}_{\playBis}(\ppathBis\ppath)$
	is Lebesgue-measurable wrt the delays in \playBis, 
	i.e.~$(\delay_0,\ldots,\delay_{k-1}) \mapsto 
	\E^{\minstrategy,\maxstrategy}_{\playBis}(\ppathBis\ppath)$ is 
	Lebesgue-measurable. In particular,~\eqref{item:expectation-integral-measurable}~holds.
\end{proof}
\noindent 
Now, we give the link between the expected weight and the expected weight 
restricted to a cylinder. In particular, we prove that the expected weight can 
be decomposed over~cylinders.

\begin{lem}
	\label{lem:expectation-split}
	Let $\minstrategy \in \StratMin$, $\maxstrategy \in \StratMax$, and
	\playBis be a finite play following the path \ppathBis, then 
	\begin{displaymath}
	\E^{\minstrategy,\maxstrategy}_{\playBis} = 
	\sum_{\ppath \in \TPaths_{\playBis}}
	\E^{\minstrategy,\maxstrategy}_{\playBis}(\ppathBis\ppath) \,.
	\end{displaymath}
\end{lem}
\begin{proof}
	By property of the integral, we can split the integral into cylinders 
	and obtain that\footnote{The definition of $A$ could correspond to that of 
		$\E^{\minstrategy,\maxstrategy}_{\playBis}(\ppath)$. However, in the 
		following, we need the inductive formula given by Definition~\ref{def:expectation}.} 
	\begin{displaymath}
	\E^{\minstrategy,\maxstrategy}_{\playBis} = 
	\sum_{\ppath \in \TPaths_{\playBis}} 
	\underbrace{\int_\play \weight(\play) \, 
	\charact_{\Cyl_{\playBis}(\ppathBis\ppath)}(\play)\; 
	\mathrm d\Proba^{\minstrategy,\maxstrategy}_{\playBis}(\play)}_{= \, A}  \,.
	\end{displaymath}
	
	To conclude the proof, we prove by induction on the length of \playBis 
	that $A = \E^{\minstrategy,\maxstrategy}_{\playBis}(\ppathBis\ppath)$.
	If $|\playBis| = |\ppathBis\ppath|$, since $\ppath \in \TPaths_{\playBis}$, 
	then $\weight(\play) = 0$ for all $\play \in \Cyl_{\playBis}(\ppathBis\ppath)$. 
	Thus, the property holds in this case. Otherwise, we suppose that the property 
	holds for all finite plays of length $k+1$ and we consider a finite play  
	\playBis of length $k$, $\trans=(\loc, \guard, \reset, \loc') \in \Trans$ the 
	first transition of \ppath, and we let \strategy be a strategy \minstrategy or 
	\maxstrategy according if $\playBis$ ends in a configuration of \MinPl or \MaxPl.
	In particular, we obtain that $A$ is equal to 
	\begin{displaymath}
	\int_{I(\playBis,\trans)} \strategytrans(\playBis)(\trans) 
	\Big(\int_{\play} (\delay\,\weight(\loc) + \weight(\trans) + \weight(\play)) \, 
	\charact_{\Cyl_{\playBis \extendto{\trans,\delay}}(\ppathBis\ppath)}(\play) 
	\;\mathrm d\Proba^{\minstrategy,\maxstrategy}_{\playBis 
		\extendto{\trans,\delay}}(\play)\Big) 
	\mathrm d\strategydelay(\playBis,\trans)(\delay) \,.
	\end{displaymath}
	In particular, by linearity of the integral, $A$ can be rewritten~as
	\begin{multline*}
	\int_{I(\playBis,\trans)} \strategytrans(\playBis)(\trans)
	\Big[\big(\delay\,\weight(\loc)+\weight(\trans)\big) 
	\underbrace{\int_{\play} 
		\charact_{\Cyl_{\playBis \extendto{\trans,\delay}}(\ppathBis\ppath)}(\play) 
		\mathrm d\Proba^{\minstrategy,\maxstrategy}_{\playBis 
			\extendto{\trans,\delay}}(\play)}_{= \, \Proba^{\minstrategy, 
			\maxstrategy}_{\playBis \extendto{\trans,\delay}}(\ppathBis\ppath)} + \\
	\int_{\play} \weight(\play) \, 
		\charact_{\Cyl_{\playBis \extendto{\trans,\delay}}(\ppathBis\ppath)}(\play) \;
	\mathrm d\Proba^{\minstrategy,\maxstrategy}_{\playBis 
		\extendto{\trans,\delay}}(\play) \Big]
	\mathrm d\minstrategydelay(\playBis,\trans)(\delay) \,.
	\end{multline*}
	By hypothesis of induction applying on $\playBis 
	\extendto{\trans,\delay}$, we have  
	\begin{displaymath}
	\int_{\play} \weight(\play) \, 
	\charact_{\Cyl_{\playBis \extendto{\trans,\delay}}(\ppathBis\ppath)}(\play) \;
	\mathrm d\Proba^{\minstrategy,\maxstrategy}_{\playBis \extendto{\trans,\delay}}
	(\play) = \E^{\minstrategy,\maxstrategy}_{\playBis 
		\extendto{\trans,\delay}}(\ppathBis\ppath)
	\end{displaymath}
	so that $A$ can be rewritten~as 
	\begin{displaymath}
	A = \int_{I(\playBis,\trans)} \strategytrans(\playBis)(\trans)
	\Big[\big(\delay\,\weight(\loc)+\weight(\trans)\big) 
	\Proba^{\minstrategy, \maxstrategy}_{\playBis 
		\extendto{\trans,\delay}}(\ppathBis\ppath) +
	\E^{\minstrategy,\maxstrategy}_{\playBis 
		\extendto{\trans,\delay}}(\ppathBis\ppath) \Big]
	\mathrm d\minstrategydelay(\playBis,\trans)(\delay)
	\end{displaymath}
	that is the definition of 
	$\E^{\minstrategy,\maxstrategy}_{\playBis}(\ppathBis\ppath)$.
\end{proof}

\noindent 
Finally, we can conclude the proof of Proposition~\ref{prop:proper}.
\begin{proof}[Proof of Proposition~\ref{prop:proper}]
	Let $\minstrategy \in \StratMin$ be a strategy satisfying the second condition of
	Hypothesis~\ref{hyp:proper} with the bounds $\alpha$ and $m$. Let $\maxstrategy \in \StratMax$, and
	$\playBis \in \FPlays$ following a path~\ppathBis.
	
	We first show that
	$\Proba^{\minstrategy,\maxstrategy}_{\playBis}(\TPlays_{\playBis})=1$. The probability not to reach 
	the target can be obtained as the limit
	\begin{displaymath}
	\lim_{n \to \infty} \sum_{\ppath \in \TPaths_{\playBis}^n} 
	\Proba^{\minstrategy,\maxstrategy}_{\playBis}(\ppathBis\ppath)
	\end{displaymath} 
	By
	Lemma~\ref{lem:proper_bound}, this limit is 0, so that the probability to reach the target is indeed 1.
	
	We then show that the infinite sum
	$\E^{\minstrategy,\maxstrategy}_{\playBis} = \sum_{\ppath \in
	\TPaths_{\playBis}}
	\E^{\minstrategy,\maxstrategy}_{\playBis}(\ppathBis\ppath)$ (given by
	Lemma~\ref{lem:expectation-split}) converges. Notice that this sum can be
	decomposed as
	\begin{equation}
	\label{eq:expectation-paths-length-n}
	\E^{\minstrategy,\maxstrategy}_{\playBis} = \sum_{n=0}^\infty \;
	\sum_{\ppath \in \TPaths_{\playBis}^n}
	\E^{\minstrategy,\maxstrategy}_{\playBis}(\ppathBis\ppath)
	\end{equation} 
	Since $\TPaths_{\playBis}^n$ is a finite set, only the first sum must be
	shown to be converging. We prove that it is absolutely converging by
	bounding $\sum_{\ppath \in \TPaths_{\playBis}^n}
	|\E^{\minstrategy,\maxstrategy}_{\playBis}(\ppathBis\ppath)|$ with respect
	to $n$. By
	Lemmas~\ref{lem:expectation-integral}.(\ref{item:expectation-integral-bound})
	and~\ref{lem:proper_bound}, it is bounded by $n\, (1 -
	\alpha)^{\lfloor\frac{n}{m}\rfloor}\, \wmax^e$, which is the general term
	of a convergent series since $0<\alpha <1$. 
\end{proof}


\noindent 
Now that we have associated an expected payoff with each convenient pair
of strategies, we are able to mimic the classical definitions of
\emph{values} to stochastic strategies. 
Let $\loc$ be a location and $\val$ be a valuation. For all
$\minstrategy\in\StratMinProper$ and $\maxstrategy\in\StratMax$,
we~let 
\begin{displaymath}
\Value^\minstrategy_{\loc,\val}=
\sup_{\maxstrategy\in\StratMax}
\E^{\minstrategy,\maxstrategy}_{\loc,\val} \qquad \text{ and } \qquad 
\Value^\maxstrategy_{\loc,\val}=
\inf_{\minstrategy\in\StratMinProper}
\E^{\minstrategy,\maxstrategy}_{\loc,\val}\;.
\end{displaymath}
This definition can be generalised by replacing configurations
$(\loc,\val)$ by finite plays \playBis: we let
$\Value^\minstrategy_{\playBis} = \sup_{\maxstrategy\in\StratMax}
\E^{\minstrategy,\maxstrategy}_{\playBis}$  and $\Value^\maxstrategy_{\playBis} = \sup_{\minstrategy\in\StratMinProper}
\E^{\minstrategy,\maxstrategy}_{\playBis}$ be the generalised versions.
Then, we let $\uppervalue_{\loc,\val}$ and $\lowervalue_{\loc,\val}$ be the upper- and lower-values of $\game$ in
$(\loc,\val)$, defined as the best expected payoff \MinPl and \MaxPl can hope for, respectively:
\begin{displaymath}
\uppervalue_{\loc,\val} =
\inf_{\minstrategy\in\StratMinProper}\Value^\minstrategy_{\loc,\val} \qquad \text{and} \qquad 
\lowervalue_{\loc, \val} = \sup_{\maxstrategy \in \StratMax} 
\Value^\maxstrategy_{\loc,\val}\,.
\end{displaymath}
We also define the \emph{memoryless values}
$\mValue^\minstrategy_{\loc, \val}$, $\mValue^\maxstrategy_{\loc, \val}$, $\muppervalue_{\loc, \val}$, and $\mlowervalue_{\loc, \val}$, where all strategies are moreover taken memoryless.
We finally define the \emph{smooth deterministic values}
$\sdValue^\minstrategy_{\loc, \val}$, $\sdValue^\maxstrategy_{\loc, \val}$, $\sduppervalue_{\loc, \val}$, and $\sdlowervalue_{\loc, \val}$, where all strategies are moreover taken (smooth) deterministic.
As usual, for all configurations $(\loc, \val)$, we always have
\[\lowervalue_{\loc, \val} \leq \uppervalue_{\loc, \val}\,,\qquad 
\mlowervalue_{\loc, \val} \leq \muppervalue_{\loc, \val}\qquad \text{ and }\qquad 
\sdlowervalue_{\loc, \val} \leq \sduppervalue_{\loc, \val}\,.\]

We have introduced the (stochastic) values to generalize the deterministic one.
However, to ensure the existence of mathematical objects like the measure of probabilities or 
the expectation (against all strategies for \MaxPl), we need to constraint deterministic strategies and 
defined new deterministic values. In particular, we do not have tools to prove that smooth deterministic 
values are determined, and are equal to the deterministic value. An idea to prove it will be 
proving the existence of an $\varepsilon$-optimal proper smooth deterministic for the deterministic 
value, intuitively it consists on proving some smooth properties on the deterministic value function 
as well as the capability of \MinPl to ensure the deterministic value by enough quickly going to a 
target location. Moreover, unlike in the case of the deterministic value, we do not know that 
\WTG{s} are determined under smooth deterministic strategies. 

Our main contribution, presented in details in
Sections~\ref{sec:trading} and~\ref{sec:SPG}, is to compare the memoryless 
(stochastic) values, the (smooth) deterministic values and the stochastic values, showing
their equality for two fragments of WTGs.

\subsection{Discrete strategies}
\label{subsec:discrete-time_strat}
Between deterministic and stochastic strategies, we define a new class of
strategies that we call \emph{discrete strategies}. Intuitively, discrete
strategies generalise deterministic strategies by only allowing discrete
probabilities for both transitions \emph{and} delays. We will later use such
particular strategies and thus study them here.

\begin{defi}
	\label{def:discrete-strat}
	A smooth strategy \strategy 
	is \emph{discrete} if for all $\play \in \FPlays$ and $\trans 
	\in \Trans$, $\strategydelay(\play,\trans)$ is a piecewise-constant 
	function,~i.e. there exist $\alpha_0, \ldots, \alpha_n\in (0,1]$, and 
	$a_0, \ldots, a_n\in\Rplus$ such that, for all $t \in I(\play, \trans)$,   
	\begin{displaymath}
	\strategydelay(\play,\trans)(\delay) = \sum_{i=0}^{n} 
	\alpha_i(\play,\trans) H(\delay-a_i(\play,\trans))
	\end{displaymath}
\end{defi}

Under a discrete strategy, the current player chooses a delay on a finite  
set of possible choices (i.e. pairs of transitions and delays). In a certain 
way, discrete strategies discretize the time in \WTG. However, even if a 
discrete strategy for \MinPl only uses Dirac distributions for delays, and 
verifies that the probability to reach the target (against all strategies of 
\MaxPl) is $1$, the expectation may not exist. Thus, to ensure the existence 
of the expectation, strategies of \MinPl must moreover be proper (i.e. satisfy 
Hypothesis~\ref{hyp:proper}).

We consider $\minstrategy \in \StratMinProper$ and $\maxstrategy \in \StratMax$
two discrete strategies (with \minstrategy
being moreover proper), and we will rewrite the expectation
they induce. Intuitively, since \minstrategy and \maxstrategy only use discrete
distributions for transitions and delays, a single play reaching a target and
conforming to both strategies forms a $\Proba^{\minstrategy,
\maxstrategy}_{\playBis}$-measurable set. Moreover, when we consider a finite
path conforming to both discrete strategies, the number of finite plays
following it is countable. Thus, we prove that $\Proba^{\minstrategy,
\maxstrategy}_{\playBis}$ is a discrete measure and we obtain a simpler formula
for the expectation. 

\begin{lem}
	\label{lem:expectation_discrete}
	Let $\minstrategy \in \StratMinProper$ and $\maxstrategy \in \StratMax$ 
	be two discrete strategies. For all finite plays~\playBis, we have 
	\begin{displaymath}
	\E^{\minstrategy, \maxstrategy}_{\playBis} = 
	\sum_{\play \in \TPlays^{\minstrategy, \maxstrategy}_{\playBis}} 
	\weight(\play) \times 
	\Proba^{\minstrategy, \maxstrategy}_{\playBis}(\play)
	\end{displaymath}
	where $\TPlays^{\minstrategy, \maxstrategy}_{\playBis}$ is the 
	$\Proba^{\minstrategy,\maxstrategy}_{\playBis}$-measurable set of plays 
	starting in the last configuration of \playBis, conforming to \minstrategy 
	and \maxstrategy and ending in the target.
\end{lem}
\begin{proof}
	Since \minstrategy is proper, the probability measure of the set of 
	plays that do not reach the target is $0$. By definition of the expectation, 
	we thus~have 
	\begin{displaymath}
	\E^{\minstrategy,\maxstrategy}_{\playBis} =
	\int_\play \weight(\play) \; 
	\mathrm d\Proba^{\minstrategy,\maxstrategy}_{\playBis}(\play) = 
	\int_\play \weight(\play) \, 
	\charact_{\TPlays^{\minstrategy, \maxstrategy}_{\playBis}}(\play) \; 
	\mathrm d\Proba^{\minstrategy,\maxstrategy}_{\playBis}(\play) 
	\end{displaymath}
	We conclude by showing that $\TPlays^{\minstrategy,
	\maxstrategy}_{\playBis}$ is countable. To do so, notice that the number of
	finite maximal paths from the last location of \playBis is countable (as we
	restrict ourselves to plays that reach the target). Moreover, since the
	distributions on delays have a finite support, for each finite path, the
	number of possible delays consistent with strategies is finite. Thus, along
	a finite path, at each step, the current player can extend the play with a
	finite number of choices. In total, $\TPlays^{\minstrategy,
	\maxstrategy}_{\playBis}$ only contains a countable number of plays.
\end{proof}

\section{Probabilities are useless in the presence of memory}
\label{sec:Max_best-response}

Our first result is to show that 
the combination of stochastic choices and memory 
in strategies does not bring more power than just the memory, i.e.~that the 
stochastic values are bounded by the smooth deterministic ones.

\begin{thm}
\label{thm:detVal=Val}
In all \WTG{s}, for all locations $\loc$ and valuations $\val$, 
\begin{displaymath}
	\sdlowervalue_{\loc,\val} \leq \lowervalue_{\loc,\val} \leq \uppervalue_{\loc,\val} 
	\leq \sduppervalue_{\loc, \val}\,. 
\end{displaymath}
\end{thm}
\noindent 
The rest of this section is devoted to the proof of this result. 
The main ingredient for the proof is the fact that when \MinPl plays with a proper
strategy, \MaxPl always has a \emph{best response} strategy that is (smooth)
deterministic:
\begin{lem}
	\label{lem:best-response}
	Let $\minstrategy\in\StratMinProper$ and
	$\varepsilon>0$. There exists a smooth deterministic strategy
	$\detmaxstrategy\in\sdStratMax$
	such that for all finite plays \play,
	$\E^{\minstrategy,\detmaxstrategy}_{\play} \geq
	\Value^\minstrategy_{\play}-\varepsilon$. If
	$\minstrategy\in\mStratMinProper$ is memoryless, then there
	exists a smooth deterministic strategy $\detmaxstrategy\in\sdStratMax$
	such that for all configurations $(\loc, \val)$,
	$\E^{\minstrategy,\detmaxstrategy}_{\loc, \val} \geq
	\mValue^\minstrategy_{\loc, \val}-\varepsilon$.  
\end{lem}
\noindent 
A useful tool for the proof is a Bellman-like fixpoint 
equation fulfilled by the expected payoff. To state it, we pack all
expectations following a pair $(\minstrategy,\maxstrategy)$ of
strategies in a single mapping
$\E^{\minstrategy,\maxstrategy}\colon \play \in\FPlays\mapsto
\E^{\minstrategy,\maxstrategy}_\play\in\R$. The Bellman-like fixpoint operator
$\Hfunction$ is a function relating mappings $X \colon \FPlays \to \R$ such 
that $\Hfunction^{\minstrategy, \maxstrategy} \colon 
(\FPlays \to \R) \to (\FPlays\to \R)$ is partially defined for 
$X \colon \FPlays \to \R$, and $\play\in\FPlays$ by
\begin{equation}
\label{eq:Hfunction}
\begin{cases}
0 & \text{if } \play \text{ ends in } \LocsT \\
\sum_\trans \int_{I(\play,\trans)} \minstrategytrans(\play)(\trans) 
\big(\delay\,\weight(\loc) + \weight(\trans) + 
X(\play \extendto{\trans, \delay})\big)
\;\mathrm d \minstrategydelay(\play,\trans)(\delay) & 
\text{if } \play\in\FPlaysMin \\
\sum_\trans \int_{I(\play,\trans)} \maxstrategytrans(\play)(\trans)
\big(\delay\,\weight(\loc) + \weight(\trans) + 
X(\play \extendto{\trans, \delay})\big)
\;\mathrm d \maxstrategydelay(\play,\trans)(\delay) & 
\text{if } \play\in\FPlaysMax
\end{cases}
\end{equation}
This operator is only partially defined since it makes sense only for functions
$X$ that are measurable wrt delays along the finite play. In the following, we
prove that, when $\minstrategy$ is not-deterministic but satisfies the second item of Hypothesis~\ref{hyp:proper} (this is the only case where we will need a non-trivial proof for Lemma~\ref{lem:best-response}), $\E^{\minstrategy, \maxstrategy}$ 
is a function for which
$\Hfunction^{\minstrategy, \maxstrategy}$ is well defined, and it is a fixpoint
of $\Hfunction^{\minstrategy, \maxstrategy}$.

\begin{lem}
	\label{lem:Bellman}
	Let $\minstrategy\in\StratMinProper\setminus\sdStratMinProper$
	and $\maxstrategy \in \StratMax$. The mapping 
	$\E^{\minstrategy, \maxstrategy} \colon \play\in\FPlays \mapsto 
	\E^{\minstrategy,\maxstrategy}_\play$ is a fixpoint of the operator 
	$\Hfunction^{\minstrategy, \maxstrategy} \colon 
	(\FPlays \to \R) \to (\FPlays\to \R)$.
\end{lem}
\begin{proof}
	Let \minstrategy and \maxstrategy be two strategies, and
	$\playBis \in \FPlays$.
	If \playBis ends in \LocsT, then $\Hfunction^{\minstrategy,
	\maxstrategy}(\E^{\minstrategy, \maxstrategy})(\playBis) = 0 =
	\E^{\minstrategy, \maxstrategy}_{\playBis}$.
	
	Otherwise, we let \strategy be the strategy \minstrategy or \maxstrategy
	depending on whether $\playBis \in \FPlaysMin$ or $\playBis \in \FPlaysMax$
	and $\ppathBis$ be the path that \playBis follows. By
	Lemma~\ref{lem:expectation-split}, $\E^{\minstrategy,
	\maxstrategy}_{\playBis}= \sum_{\rpath \in \TPaths_{\playBis}}
	\E^{\minstrategy, \maxstrategy}_{\playBis}(\ppathBis\rpath)$. In
	particular, by decomposing the path (that is possible, since \playBis does
	not end in \LocsT), we have $\E^{\minstrategy, \maxstrategy}_{\playBis} =
	\sum_\trans \sum_{\ppath \mid \trans\ppath \in \TPaths_{\playBis}}
	\E^{\minstrategy,\maxstrategy}_{\playBis}(\ppathBis\trans\ppath)$. Thus, by
	definition of the expectation of a path, $\E^{\minstrategy,
	\maxstrategy}_{\playBis}$ is equal to 
	\begin{displaymath}
	\sum_{\trans} \sum_{\ppath \mid \trans\ppath \in \TPaths_{\playBis}} 
	\int_{I(\playBis,\trans)} \strategytrans(\playBis)(\trans)
	\big[(\delay\,\weight(\loc) + \weight(\trans)) 
	\Proba^{\minstrategy,\maxstrategy}_{\playBis
		\extendto{\trans, \delay}}(\ppathBis\trans\ppath) + 
	\E^{\minstrategy,\maxstrategy}_{\playBis
		\extendto{\trans, \delay}}(\ppathBis\trans\ppath)\big] 
	\;\mathrm d \strategydelay(\playBis,\trans)(\delay). 
	\end{displaymath}
	Since \minstrategy is not smooth determinstic, yet proper, it satisfies the second item of Hypothesis~\ref{hyp:proper} for bounds 
	$\alpha$ and $m$. Then by Lemma~\ref{lem:proper_bound} and Lemma~\ref{lem:expectation-integral}.(\ref{item:expectation-integral-bound}) we obtain that
	\begin{displaymath}
	\E^{\minstrategy,\maxstrategy}_{\playBis 
		\extendto{\trans, \delay}}(\ppathBis\trans\ppath) \leq 
	(1-\alpha)^{\lfloor |\ppath|/m\rfloor} \,|\ppath|\, \wmax^e 
	\end{displaymath}
	and 
	\begin{displaymath}
	(\delay\,\weight(\loc) + \weight(\trans)) 
	\Proba^{\minstrategy,\maxstrategy}_{\playBis 
		\extendto{\trans, \delay}}(\ppathBis\trans\ppath) \leq 
	(1-\alpha)^{\lfloor |\ppath|/m\rfloor} \wmax^e
	\end{displaymath}
	In particular, $
	(\delay\,\weight(\loc) + \weight(\trans)) 
	\Proba^{\minstrategy,\maxstrategy}_{\playBis 
		\extendto{\trans, \delay}}(\ppathBis\trans\ppath) + 
	\E^{\minstrategy,\maxstrategy}_{\playBis 
		\extendto{\trans, \delay}}(\ppathBis\trans\ppath)$ is bounded by 
	$(1-\alpha)^{\lfloor |\ppath|/m\rfloor} \,(|\ppath|+1)\, \wmax^e$. Thus, by dominated convergence theorem, again by Lemma~\ref{lem:expectation-split}, and by
	definition of $\Hfunction^{\minstrategy, \maxstrategy}$, 
	we obtain
	\begin{displaymath}
	\E^{\minstrategy, \maxstrategy}_{\playBis}
	= \sum_{\trans} \int_{I(\playBis,\trans)}
	\strategytrans(\playBis)(\trans)
	\big(\delay\,\weight(\loc) + \weight(\trans) + 
	\E^{\minstrategy,\maxstrategy}_{\playBis \extendto{\trans, \delay}}\big)
	\;\mathrm d\strategydelay(\playBis,\trans)(\delay)  \\
	= \Hfunction^{\minstrategy, \maxstrategy} 
	(\E^{\minstrategy,\maxstrategy})(\playBis).
	\end{displaymath}
	Thus, $\E^{\minstrategy,\maxstrategy}$ is a fixpoint of
	$\Hfunction^{\minstrategy,\maxstrategy}$.
\end{proof}
\noindent 
With this fixpoint equation in mind, we are equipped to prove
Lemma~\ref{lem:best-response}: \MaxPl has a deterministic best-response 
strategy against a proper strategy of \MinPl.  
\begin{proof}[Proof of Lemma~\ref{lem:best-response}]
	To do so, we first consider a (stochastic) strategy~$\maxstrategy$ of
	$\MaxPl$ such that for all finite plays \playBis,
	$\E^{\minstrategy,\maxstrategy}_{\playBis} \geq
	\Value^{\minstrategy}_{\playBis}-\varepsilon/2$, for a fixed
	$\varepsilon>0$, which exists by definition of the (stochastic) value.
	We show the existence of a smooth 
	deterministic strategy $\detmaxstrategy \in \sdStratMax$ 
	for $\MaxPl$ such that for all finite plays \playBis, we have
	$\E^{\minstrategy,\detmaxstrategy}_{\playBis}\geq
	\E^{\minstrategy,\maxstrategy}_{\playBis} -\varepsilon/2^{|\play|}$, which
	allows us to conclude that
	$\E^{\minstrategy,\detmaxstrategy}_{\playBis} \geq
	\Value^{\minstrategy}_{\playBis}-\varepsilon$. 
	
	To do that, we distinguish the cases where $\minstrategy$ is deterministic or is not. 
	First, suppose $\minstrategy \in \sdStratMinProper$. 
	By contradiction, we suppose that 
	for all $\detmaxstrategy \in \sdStratMax$, 
	$\weight(\outcomes(\playBis, \minstrategy, \detmaxstrategy)) \leq 
	\E^{\minstrategy,\maxstrategy}_{\playBis} -\varepsilon/2^{|\playBis|}$.  
	Thus, by definition of the expectation,	
	\begin{displaymath}
		\E^{\minstrategy,\maxstrategy}_{\playBis} = 
		\int_{\play} \weight(\play) \,\mathrm d \Proba^{\minstrategy, \maxstrategy}_{\playBis} 
		\leq \E^{\minstrategy,\maxstrategy}_{\playBis} -\varepsilon/2^{|\playBis|}
	\end{displaymath}
	We obtain a contradiction. 

	Now, we suppose that $\minstrategy$ is not in $\sdStratMinProper$.
	In this case, we will use the $\argsupepsilon$ operator defined for all mappings
	$f \colon A \to \R$ and $B \subseteq A$ by
	$\argsupepsilon_B f = \{a \in B \mid f(a) \geq \sup_B f -
	\varepsilon\}$.
	For all $\play\in\FPlaysMax$ ending in $(\loc, \val)$, we let
	\begin{equation*}
	\label{eq:best-response_Max}
	\detmaxstrategy(\play) \in \argsupepsilon[\varepsilon/2^{|\play|+1}]_{(\loc, \val) 
		\moveto{\trans, \delay}\,(\loc', \val')} 
	\big(\delay\,\weight(\loc) + \weight(\trans) +
	\E^{\minstrategy, \maxstrategy}_{\play \extendto{\trans, \delay}}\big)
	\end{equation*}
	We can use the Kuratowski and Ryll-Nardzewski measurable 
	selection theorem \cite{Kuratowski-RyllNardzewski-65} to ensure that $\detmaxstrategy$ can be chosen in $\sdStratMax$. This theorem states that if $\psi$ is a mapping from a measurable space taking values in the set of non empty closed subsets of a Borel set (for us, pairs $(\trans,\delay)$ of transitions and real delays) that is \textit{weakly measurable}, then $\psi$ has a selection, i.e.~a measurable mapping $\tau$ such that for all elements $\omega$ of the measurable space, $\tau(\omega)\in \psi(\omega)$. Weak measurability amounts to saying that for every 
	open subset $U$, we have that $\{\omega\mid \psi(\omega)\cap U \neq \emptyset\}$ is measurable. 
	
 Indeed, the mapping $\psi$ defined for all plays $\play$ by 
	\[\psi(\play) = \{(\trans,\delay) \in \R \mid \delay \, \weight(\loc) 
	+ \weight(\trans) + 
	\E^{\minstrategy, \maxstrategy}_{\play\extendto{\trans, \delay}} \geq K - 
	\varepsilon \}\] where $K$ is the supremum of the function $t\mapsto \delay \, \weight(\loc) 
	+ \weight(\trans) + 
	\E^{\minstrategy, \maxstrategy}_{\play\extendto{\trans, \delay}}$ can be shown to be weakly measurable:  
	for all open set $U$ of $\R$, the set $\{\play \mid \psi(\play) \cap U \neq \emptyset \}$
	can be generated by a countable union of cylinders, since we have a countable number of paths and 
	delays are constrained by $U$ (which is measurable) and a supremum 
	(which is also measurable). The selection theorem thus allows us to define a measurable strategy $\detmaxstrategy\in \sdStratMax$.
	
	We also define, for all $i \in \N$, the (smooth) strategy
	$\strategy_i \in \StratMax$ consisting in playing $\detmaxstrategy$ during 
	the first $i$ steps of the play, and $\maxstrategy$ for the remaining ones: since $\detmaxstrategy$ 
	and $\maxstrategy$ are smooth, the combination $\strategy_i$ does too. 
	When $i$ increases, strategy $\strategy_i$ looks more and more like
	$\detmaxstrategy$.
	In particular, for all finite plays $\playBis$, $\E^{\minstrategy, \strategy_{i}}_{\playBis}$ tends to
	$\E^{\minstrategy, \detmaxstrategy}_{\playBis}$ when $i$ tends to
	$+\infty$. To prove it formally, we decompose the sum given for the
	expectation in Lemma~\ref{lem:expectation-split} according to the length of
	the path, we have 
	\begin{displaymath}
	|\E^{\minstrategy,\strategy_i}_{\playBis} -
	\E^{\minstrategy,\detmaxstrategy}_{\playBis} |
	= \Big|\sum_{j = 0}^\infty \sum_{\ppath \in \TPaths^j_\play}
	\E^{\minstrategy,\strategy_i}_{\playBis}(\ppathBis\ppath)
	- \sum_{j=0}^\infty \sum_{\ppath \in \TPaths^j_\play}
	\E^{\minstrategy,\detmaxstrategy}_{\playBis}(\ppathBis\ppath) \Big|
	\end{displaymath}
	where \ppathBis is the path followed by \playBis. 
	Letting $k_i = \max(0, i-|\playBis|)$ be the number of steps after 
	\playBis where $\strategy_i$ and \detmaxstrategy are equal, and noticing that 
	$\strategy_i$ and $\detmaxstrategy$ behave the same until the step $k_i$ from \playBis, we deduce that 
	\begin{align*}
	|\E^{\minstrategy,\strategy_{i}}_{\playBis} -
	\E^{\minstrategy,\detmaxstrategy}_{\playBis} |
	&= \Big|\sum_{j=k_i+1}^\infty \sum_{\ppath \in \TPaths^j_{\playBis}}
	\E^{\minstrategy,\strategy_i}_{\playBis}(\ppathBis\ppath)
	- \sum_{j=k_i+1}^\infty \sum_{\ppath \in \TPaths^j_{\playBis}}
	\E^{\minstrategy,\detmaxstrategy}_{\playBis}(\ppathBis\ppath) \Big|
	\\
	&\leq \sum_{j=k_i+1}^\infty \sum_{\ppath \in \TPaths^j_{\playBis}} 
	(|\E^{\minstrategy,\strategy_i}_{\playBis}(\ppathBis\ppath)|+
	|\E^{\minstrategy,\detmaxstrategy}_{\playBis}(\ppathBis\ppath)|).
	\end{align*}
	Thus, by Lemma~\ref{lem:expectation-integral}.(\ref{item:expectation-integral-bound}) and Lemma~\ref{lem:proper_bound}, we have (with $\alpha$ and $m$ given by the property of
	being proper)
	\begin{displaymath}
	|\E^{\minstrategy,\strategy_{i}}_{\playBis} -
	\E^{\minstrategy,\detmaxstrategy}_{\playBis} |
	\leq \sum_{j=k_i+1}^\infty 2 \, j \, \wmax^e (1-\alpha)^{\lfloor j/m\rfloor}.
	\end{displaymath}
	Moreover, since $x\mapsto (1-\alpha)^x$ is decreasing, we obtain that 
	\begin{displaymath}
	|\E^{\minstrategy,\strategy_{i}}_{\playBis} -
	\E^{\minstrategy,\detmaxstrategy}_{\playBis}|
	\leq \sum_{j=k_i+1}^\infty 2 \,j\, \wmax^e (1-\alpha)^{j/m + 1}.
	\end{displaymath}
	Noticing that the formal power series $\sum x \beta^{x-1}$ is the formal derivative of the formal power series $\sum \beta^x$, whose sum we can compute, we conclude that 
	\begin{displaymath}
	|\E^{\minstrategy,\strategy_{i}}_{\playBis} -
	\E^{\minstrategy,\detmaxstrategy}_{\playBis} |
	\leq 2\, \wmax^e\, \frac{(1-\alpha)^{(k_i+2)/m}
		\big(k_i(1-(1-\alpha)^{1/m})+1\big)}{(1-(1-\alpha)^{1/m})^2} 
	\end{displaymath}
	that tends to 0 when $i$ (and thus $k_i$) tends to $+\infty$. 
	
	Independently, we relate strategies $\strategy_i$ and $\maxstrategy$, by showing that 
	\begin{equation}
	\label{eq:best-response-Max_2}
	\forall i\in \N\quad \forall \playBis\in \FPlays \qquad
	\E^{\minstrategy, \strategy_i}_{\playBis} \geq \E^{\minstrategy,
		\maxstrategy}_{\playBis} - \frac{\varepsilon}{2 ^{|\playBis|}} 
	\end{equation}
	When taking the limit when $i$ tends to $+\infty$, we
	obtain that for all finite plays \playBis,
	$\E^{\minstrategy, \detmaxstrategy}_{\playBis} \geq
	\E^{\minstrategy,\maxstrategy}_{\playBis} - \frac{\varepsilon}{2
		^{|\playBis|}}$.  In particular,
	$\E^{\minstrategy, \detmaxstrategy}_{\playBis} \geq
	\Value^\minstrategy_{\playBis} - \varepsilon$. In the case where \minstrategy is 
	memoryless, since memoryless strategies of \MaxPl are stochastic 
	strategies of \MaxPl, we have, for all configurations $(\loc, \val)$,
	$\Value^\minstrategy_{\loc, \val} \geq \mValue^\minstrategy_{\loc, \val}$. 
	Thus, by applying the previous result in case where \play is only a 
	configuration, we deduce that 
	$\E^{\minstrategy, \detmaxstrategy}_{\loc, \val} \geq
	\Value^\minstrategy_{\loc, \val} - \varepsilon \geq
	\mValue^\minstrategy_{\loc, \val} - \varepsilon$.
	
	\bigskip 
	To conclude the proof, we need to show~\eqref{eq:best-response-Max_2}. 
	First, we notice that this inequality trivially holds when $|\playBis|\geq i$,
	since then strategies $\strategy_i$ and $\maxstrategy$ coincide for the
	rest of the play. It remains to show that the property holds for all
	$i$ and plays \playBis such that $|\playBis| \leq i$. We proceed by
	induction on $i-|\playBis|$. If $i-|\playBis|=0$, we have said before that the 
	inequality trivially holds. Otherwise, we distinguish three cases according to 
	the last location of \playBis. If \playBis ends in $\LocsT$, we have 
	$\E^{\minstrategy, \strategy_i}_{\playBis} = 0 
	\geq 0 - \frac{\varepsilon}{2^{|\playBis|}} = 
	\E^{\minstrategy, \maxstrategy}_{\playBis} - \frac{\varepsilon}{2^{|\playBis|}}$.
	
	If $\playBis \in \FPlaysMin$, by Lemma~\ref{lem:Bellman}, 
	\begin{displaymath}
	\E^{\minstrategy, \strategy_i}_{\playBis} = \Hfunction^{\minstrategy, \strategy_i}(\E^{\minstrategy, \strategy_i})(\playBis)
	= \sum_\trans \int_{I(\playBis, \trans)} \minstrategytrans(\playBis)(\trans)
	\big[\delay\,\weight(\loc) + \weight(\trans)
	+ \E^{\minstrategy, \strategy_i}_{\playBis \extendto{\trans,\delay}} \big]
	\;\mathrm d \minstrategydelay(\playBis,\trans)(\delay)
	\end{displaymath}
	By induction hypothesis, we deduce that 
	\begin{displaymath}
	\E^{\minstrategy, \strategy_i}_{\playBis} 
	\geq \sum_\trans \int_{I(\playBis, \trans)} \minstrategytrans(\playBis)(\trans)
	\big(\delay\,\weight(\loc) + \weight(\trans) + 
	\E^{\minstrategy,\maxstrategy}_{\playBis \extendto{\trans,\delay}}
	- \frac\varepsilon{2^{|\playBis|+1}}\big)
	\;\mathrm d \minstrategydelay(\playBis,\trans)(\delay).
	\end{displaymath}
	Thus, by linearity of the integral, we have 
	\begin{multline*}
	\E^{\minstrategy, \strategy_i}_{\playBis} 
	\geq \sum_\trans \int_{I(\playBis,\trans)} \minstrategytrans(\playBis)(\trans)
	\big(\delay \,\weight(\loc) + \weight(\trans)  + 
	\E^{\minstrategy,\maxstrategy}_{\playBis \extendto{{\trans,\delay}}}\big) 
	\;\mathrm d \minstrategydelay(\playBis,\trans)(\delay) \\
	- \frac\varepsilon{2^{|\playBis|+1}}
	\sum_\trans \int_{I(\playBis,\trans)} \minstrategytrans(\playBis)(\trans)
	\;\mathrm d \minstrategydelay(\playBis,\trans)(\delay).
	\end{multline*}
	Since $\minstrategy$ is a distribution, we obtain that 
	\[\E^{\minstrategy, \strategy_i}_{\playBis} 
	\geq \Hfunction^{\minstrategy,\maxstrategy}
	(\E^{\minstrategy,\maxstrategy})(\playBis)
	- \frac\varepsilon{2^{|\playBis| +1}}\times 1 \geq \Hfunction^{\minstrategy,\maxstrategy}
	(\E^{\minstrategy,\maxstrategy})(\playBis)
	- \frac\varepsilon{2^{|\playBis|}}\,.\]
	
	If $\playBis \in \FPlaysMax$, Lemma~\ref{lem:expectation-split} gives again
	\begin{displaymath}
	\E^{\minstrategy, \strategy_i}_{\playBis} 
	= \sum_\trans \sum_{\ppath \mid \trans \ppath \in \TPaths_{\playBis}  } 
	\E^{\minstrategy,\strategy_i}_{\playBis} (\ppathBis\trans\rpath)\,.
	\end{displaymath}
	Since $|\playBis|\leq i$, the next step for $\MaxPl$ is dictated by
	the strategy $\detmaxstrategy$, that is deterministic. Letting
	$(\trans_0, t_0) = \detmaxstrategy(\playBis)$, we thus have
	\begin{displaymath}
	\E^{\minstrategy, \strategy_i}_{\playBis} 
	= (\delay_0\,\weight(\loc) + \weight(\trans_0)) 
	\sum_{\ppath \mid \trans_0 \ppath \in \TPaths_{\playBis} } 
	\Proba^{\minstrategy,\strategy_i}_{\playBis
		\extendto{\trans_0,\delay_0}}(\ppathBis\trans_0\ppath)
	+ \sum_{\ppath \mid \trans_0 \ppath \in \TPaths_{\playBis} } 
	\E^{\minstrategy, \strategy_i}_{\playBis
		\extendto{\trans_0,\delay_0}}(\ppathBis\trans_0\ppath)
	\end{displaymath}
	Moreover, since $\minstrategy$ is proper, we know that  
	$\sum_{\ppath \mid \trans_0 \ppath \in \TPaths_{\playBis} } \Proba^{\minstrategy,\strategy_i}_{\playBis \extendto{\trans_0,\delay_0}} 
	(\TPaths_{\playBis \extendto{\trans_0,\delay_0}}) = 1$, and we deduce that 
	\begin{displaymath}
	\E^{\minstrategy, \strategy_i}_{\playBis} 
	= \delay_0\,\weight(\loc) + \weight(\trans_0) + 
	\E^{\minstrategy, \strategy_i}_{\playBis \extendto{\trans_0,\delay_0}} 
	\end{displaymath}
	By the induction hypothesis, we  
	obtain that 
	\begin{displaymath}
	\E^{\minstrategy, \strategy_i}_{\playBis} 
	\geq \delay_0\,\weight(\loc) + \weight(\trans_0) + 
	\E^{\minstrategy,\maxstrategy}_{\playBis \extendto{{\trans_0,\delay_0}}}
	- \frac\varepsilon{2^{|\playBis|+1}} 
	\end{displaymath}
	Thus, by definition of $\detmaxstrategy$ and letting $(\loc, \val)$ the last 
	configuration of \playBis, we get
	\begin{displaymath}
	\E^{\minstrategy, \strategy_i}_{\playBis} 
	\geq \sup_{(\loc, \val) \xrightarrow{\trans, \delay}(\loc', \val')} 
	\big(t\,\weight(\loc) + \weight(\trans) + 
	\E^{\minstrategy,\maxstrategy}_{\playBis \extendto{\trans,\delay}}\big)
	- \frac{\varepsilon}{2^{|\playBis|+1}} - \frac{\varepsilon}{2^{|\playBis|+1}}.
	\end{displaymath}
	Since $\maxstrategy$ is a distribution, we have 
	\begin{align*}
	\E^{\minstrategy, \strategy_i}_{\playBis} 
	&\geq \sum_\trans  \int_{I(\playBis,\trans)} 
	\maxstrategytrans(\playBis)(\trans)
	\big(t\,\weight(\loc) + \weight(\trans) 
	+ \E^{\minstrategy,\maxstrategy}_{\playBis \extendto{\trans,\delay}} \big)
	\;\mathrm d \maxstrategydelay(\playBis,\trans)(t) 
	- \frac\varepsilon{2^{|\playBis|}} \\&= \Hfunction^{\minstrategy,\maxstrategy}(\E^{\minstrategy,\maxstrategy})
	(\playBis) - \frac{\varepsilon}{2^{|\playBis|}}.
	\end{align*}
	We conclude in all cases, by 
	Lemma~\ref{lem:Bellman}:
	\begin{displaymath}
	\E^{\minstrategy, \strategy_i}_{\playBis} 
	\geq \Hfunction^{\minstrategy,\maxstrategy}(\E^{\minstrategy,\maxstrategy})
	(\playBis) - \frac{\varepsilon}{2^{|\playBis|}}
	= \E^{\minstrategy,\maxstrategy}_{\playBis}  - \frac{\varepsilon}{2^{|\playBis|}}.
	\qedhere
	\end{displaymath}
\end{proof}
\noindent 
We are now able to come back to the proof of Theorem~\ref{thm:detVal=Val}, separated into two inequalities. The first one shows that the stochastic values are at most equal to the smooth deterministic upper-value: 
\begin{lem}
	\label{lem:divergent-valinequalites}
	In all \WTG{s} \game, for all locations $\loc$ and valuations $\val$,
	$\uppervalue_{\loc,\val} \leq \sduppervalue_{\loc,\val}$. 
\end{lem}
\begin{proof}
	Let $\detminstrategy$ be a proper smooth deterministic strategy for \MinPl. 
	Let $\varepsilon>0$. By Lemma~\ref{lem:best-response}, there exists
	$\detmaxstrategy\in \sdStratMax$ such that
	$\weight(\outcomes((\loc, \val), \detminstrategy, \detmaxstrategy)) =
	\E^{\detminstrategy, \detmaxstrategy}_{\loc, \val}\geq
	\Value^{\detminstrategy}_{\loc,\val}-\varepsilon$.  Therefore
	\begin{displaymath}
	\sup_{\detmaxstrategy \in \sdStratMax} 
	\weight(\outcomes((\loc, \val), \detminstrategy, \detmaxstrategy))
	\geq \Value^{\detminstrategy}_{\loc,\val}-\varepsilon \,.
	\end{displaymath}
	Since this holds for all $\varepsilon$, we have
	\begin{displaymath}
	\sup_{\detmaxstrategy \in \sdStratMax} 
	\weight(\outcomes((\loc, \val), \detminstrategy, \detmaxstrategy))
	\geq \Value^{\detminstrategy}_{\loc,\val} \,.
	\end{displaymath}
	By considering the infimum over all proper smooth deterministic strategies, 
	we obtain
	\begin{displaymath}
	\sduppervalue_{\loc,\val} = 
	\inf_{\detminstrategy \in \sdStratMinProper} \sup_{\detmaxstrategy \in \sdStratMax} 
	\weight(\outcomes((\loc, \val), \detminstrategy, \detmaxstrategy)) \geq 
	\inf_{\detminstrategy \in \sdStratMinProper} \Value^{\detminstrategy}_{\loc,\val} \,.
	\end{displaymath} 
	Since $\sdStratMinProper \subseteq \StratMinProper$, the infimum over all proper
	strategies of $\MinPl$ is at most the infimum over deterministic
	strategies:
	\begin{displaymath}
	\sduppervalue_{\loc,\val} \geq \inf_{\minstrategy
		\in \StratMinProper} \Value^{\minstrategy}_{\loc,\val} =
	\uppervalue_{\loc,\val} \,. \qedhere
	\end{displaymath}
\end{proof}
\noindent 
The second one shows that smooth deterministic lower-value is at most equal to the stochastic values. 
\begin{lem}
	\label{lem:detVal<lowerVal}
	In all \WTG{s} \game, for all locations $\loc$ and valuations $\val$,
	$\sdlowervalue_{\loc,\val} \leq \lowervalue_{\loc,\val}$.
\end{lem}
\begin{proof} 
	Let $\varepsilon > 0$. If $\sdlowervalue_{\loc,\val} = -\infty$, the result is trivial. 
	Otherwise, by definition, \MaxPl has a deterministic $\varepsilon$-optimal 
	strategy \detmaxstrategy for the smooth deterministic lower-value. 
	Against any proper smooth deterministic strategy of $\MinPl$, from $(\loc,\val)$, 
	it guarantees a weight at least $\sdlowervalue_{\loc, \val} - \varepsilon$. 

Let \minstrategy be any proper strategy of \MinPl. Every play
conforming to $\minstrategy$ and $\detmaxstrategy$ has a weight at least
$\sdlowervalue_{\loc, \val} - \varepsilon$ (since it is conforming to
$\detmaxstrategy$), so that, by definition of the expectation,
\[\E^{\minstrategy, \detmaxstrategy}_{\loc, \val} = \int_{\play}
\weight(\play)\mathrm{d}\Proba_{\loc,\val}^{\minstrategy,\detmaxstrategy}(\play)
\geq \sdlowervalue_{\loc, \val} - \varepsilon\,\] Since this inequality holds for all
proper strategies of \MinPl, we deduce that 
	\begin{displaymath}
	\inf_{\minstrategy \in \StratMinProper} 
	\E^{\minstrategy, \detmaxstrategy}_{\loc, \val} \geq \sdlowervalue_{\loc, \val} - 
	\varepsilon\,.
	\end{displaymath} 
	By taking the supremum over all strategies of
	$\MaxPl$ (that contain the smooth deterministic strategy
	\detmaxstrategy), we have
	\begin{displaymath}
	\sup_{\maxstrategy \in \StratMax} \inf_{\minstrategy \in \StratMinProper} 
	\E^{\minstrategy, \maxstrategy}_{\loc, \val} \geq \sdlowervalue_{\loc, \val} - 
	\varepsilon \,.
	\end{displaymath} 
	In particular, $\lowervalue_{\loc, \val} \geq \sdlowervalue_{\loc,\val} - 
	\varepsilon$. 
	To conclude, we remark that this inequality holds for all 
	$\varepsilon > 0$, and thus also when $\varepsilon$ is set to $0$. 
\end{proof}
\noindent 
Combined with the fact that $\lowervalue_{\loc,\val}\leq \uppervalue_{\loc,\val}$, 
we indeed proved the equalities of Theorem~\ref{thm:detVal=Val}.

It remains to study the memoryless values, and compare them with the (smooth) deterministic value(s).
Unfortunately, we cannot hope for a similar result as the one above, showing the equality of 
the memoryless values with the deterministic one. Indeed, the following example shows that \WTG{s} 
are not determined with respect to the memoryless value.

\begin{figure}[tbp]
	\centering
	\begin{tikzpicture}[xscale=.8,every node/.style={font=\footnotesize}, 
	every label/.style={font=\scriptsize}]
	\node[PlayerMax, label={above:$\loc_1$}] at (0, 0) (s1) {$\mathbf{1}$};
	\node[PlayerMin, label={above:$\loc_0$}] at (5, 0)  (s0) {$\mathbf{-1}$};
	\node[target] at (2.5, -2) (s3) {\LARGE \smiley}; 
	
	\draw[->]
	(s0) edge[bend right=20] node[above] {$\begin{array}{c}
		\trans_1: 0 < x < 1 \\ x := 0
		\end{array}$} (s1)
	(s1) edge node[below] {$\begin{array}{c}
		\trans_2: 0 < x < 1 \\ x := 0
		\end{array}$} (s0)
	(s1) edge node[left, near end] {$\begin{array}{c}
		\trans_3  \\ x = 0 \\ \mathbf{-10} 
		\end{array}$} (s3)
	(s0) edge node[below right] {$\trans_4 : x = 0$} (s3)
	;
	\end{tikzpicture}
	\caption{A one-clock \WTG where $\muppervalue_{\loc_0, 0} > \mlowervalue_{\loc_0, 0}$.}
	\label{fig:cex_det-mVal}
\end{figure}

\begin{exa}\label{ex:mVal}
We consider the one-clock \WTG depicted in \figurename{~\ref{fig:cex_det-mVal}}, and show that $\muppervalue_{\loc_0, 0} = 0 > -9 \geq \mlowervalue_{\loc_0, 0}$.

To do so, consider
a proper memoryless strategy  $\minstrategy$ of \MinPl. First, we remark that there exists $p \in [0, 1)$ such that 
$\minstrategytrans(\loc_0, 0) = p \times \Dirac{\trans_1} + 
(1-p) \times \Dirac{\trans_4}$. Moreover, since $\loc_0$ and $\loc_1$ are 
always reached after a reset, i.e. with the valuation $0$, $\minstrategy$ always 
chooses the same delay in $\loc_0$ that does not impact the next choice of 
the memoryless strategy of \MaxPl. Formally, let $\delay \in (0, 1)$ be the expected delay in $\loc_0$ (from the valuation $0$) 
according to $\minstrategydelay(\loc_0, 0)(\trans_1)$, and we define the discrete 
strategy $\minstrategy'$~by 
\begin{displaymath}
\minstrategy'(\loc_0, 0) = 
p\times\Dirac{\trans_1, \delay} + (1-p)\times\Dirac{\trans_4, 0} 
\end{displaymath}
For all memoryless strategies $\maxstrategy$ of \MaxPl, 
$\E^{\minstrategy, \maxstrategy}_{\loc_0, 0} = 
\E^{\minstrategy', \maxstrategy}_{\loc_0, 0}$.
Now, we consider the smooth deterministic memoryless strategy $\detmaxstrategy$ of \MaxPl 
defined by 
$\detmaxstrategy(\loc_1, 0) = (\trans_2,(1+\delay)/2)$. Then 
\begin{displaymath}
\E^{\minstrategy', \detmaxstrategy}_{\loc_0, 0} = 
\sum_{i=0}^{+\infty} \frac{1-\delay}{2}ip^i 
= \frac{1-\delay}{2}\frac{p}{(1-p)^2} \geq 0
\end{displaymath}
and this expectation is null when $p = 0$, thus 
$\muppervalue_{\loc_0, 0} = 0$.

By a similar reasoning, all (smooth) memoryless strategies $\maxstrategy$ of \MaxPl are equivalent
to a discrete strategy $\maxstrategy'$ described by the probability $q = 
\maxstrategytrans(\loc_1, 0) \in [0, 1)$ and the expected delay $\delay \in (0, 1)$ 
spent in $\loc_1$ (from the valuation $0$) 
according to $\maxstrategydelay(\loc_1, 0)(\trans_2)$: we have
\begin{displaymath}
\maxstrategy'(\loc_1, 0) = 
q\times\Dirac{\trans_2, \delay} + (1-q)\times\Dirac{\trans_3, 0} \,.
\end{displaymath}
We define a proper memoryless strategy $\minstrategy$ for \MinPl by letting
\begin{displaymath}
\minstrategy(\loc_0, 0) = 
p\times\Dirac{\trans_1, (1+\delay)/2} + (1-p)\times\Dirac{\trans_4, 0} \,.
\end{displaymath}
By Lemma~\ref{lem:Bellman}, we have 
$\E^{\minstrategy, \maxstrategy'}_{\loc_0, 0} = 
(\E^{\minstrategy, \maxstrategy'}_{\loc_1, 0}-(1+\delay)/2) \times p$ and 
$\E^{\minstrategy, \maxstrategy'}_{\loc_1, 0} = -10\times (1-q) + 
(\delay + \E^{\minstrategy, \maxstrategy'}_{\loc_0, 0}) \times q$. 
In particular, we deduce that 
\[\E^{\minstrategy, \maxstrategy'}_{\loc_0, 0} = \left(-10(1-q) + 
(\E^{\minstrategy, \maxstrategy'}_{\loc_0, 0}+\delay)q - \frac{1+\delay}{2}\right)p \]
which resolves into 
\begin{displaymath}
\E^{\minstrategy, \maxstrategy'}_{\loc_0, 0} 
=\left(- 10(1-q) - \frac{\delay(1-2q) + 1}{2}\right)\frac p{1-pq}  \,.
\end{displaymath}
First, we suppose that $q=1$, i.e. $\maxstrategy'$ is a deterministic 
strategy such that $\maxstrategy'(\loc_1, 0) = (\trans_2, \delay)$. In this 
case, we have
\begin{displaymath}
\E^{\minstrategy, \maxstrategy'}_{\loc_0, 0} 
= -\frac{1-\delay}{2}\frac p{1-p}  \,.
\end{displaymath}
Now, as we control the strategy of \MinPl, 
we can fix $p = \frac{18}{19-\delay} < 1$ (since $\delay < 1$), and 
we obtain 
\begin{displaymath}
\E^{\minstrategy, \maxstrategy'}_{\loc_0, 0} 
= -\frac{1-\delay}{2}\frac {18}{(19-\delay)(1-\frac{18}{19-\delay})} 
= -\frac{1-\delay}{2}\frac {18}{1-\delay}
= -9 \,.
\end{displaymath}
Otherwise, we suppose that $0 \leq q < 1$, we deduce that $\delay(1-2q) + 1 \geq 0$ 
(since $\delay \in (0,1]$). Thus, 
\begin{displaymath}
\E^{\minstrategy, \maxstrategy'}_{\loc_0, 0} \leq - 10(1-q) \frac p{1-pq}  \,.
\end{displaymath}
Again, as we control the strategy of \MinPl, 
we can fix $p = \frac{9}{10-q} < 1$ (since $q < 1$), and  
\begin{displaymath}
\E^{\minstrategy, \maxstrategy'}_{\loc_0, 0} 
\leq - 10(1-q) \frac 9{(10-q)(1-\frac{9}{10-q}q)} 
= - 10(1-q) \frac 9{10(1-q)}
= -9 \,.
\end{displaymath}
We deduce that $\mlowervalue(\loc_0, 0) \leq -9$. Thus, we conclude that 
$\mlowervalue(\loc_0, 0) \leq -9 < 0 = \muppervalue(\loc_0, 0)$ as expected.
\end{exa}

Moreover, even if we wanted to restrict our study to one of the memoryless values 
(upper or lower-values), the following example shows a case where memoryless lower 
and upper-values are equal, but different from the deterministic value.

\begin{figure}
	\begin{tikzpicture}
	\node[PlayerMax, label={above:$\loc_0$}] at (0,0) (s0) {$\mathbf{-1}$};
	\node[PlayerMin, label={above:$\loc_1$}] at (5,0) (s1) {$\mathbf{0}$};
	\node[target] at (8,0) (s2) {\LARGE \smiley}; 
	
	\draw[->]
	(s1) edge[bend left=10] node[below] {$\trans_0, x \leq 1$} (s0)
	(s0) edge[bend left=10] node[above] {$\trans_1, 0 < x, x := 0$} (s1)
	(s1) edge node[above] {$\trans_2, x \leq 1$} (s2)
	;
	\end{tikzpicture}
	\caption{One-clock \WTG where $\dValue_{\loc_1, 0} \neq \muppervalue_{\loc_1, 0} = \mlowervalue_{\loc_1, 0}$.
	}
	\label{fig:1clock}
\end{figure}

\begin{exa}
	\label{ex:1clock}
	We consider the one-clock \WTG depicted in \figurename{~\ref{fig:1clock}}. We explain why $\dValue_{\loc_1, 0} \neq \muppervalue_{\loc_1, 0} = \mlowervalue_{\loc_1, 0}$. Since
	$\MinPl$ has a way to obtain the weight $0$ by using transition $\trans_2$ (and thus with a memoryless and deterministic strategy),
	all values $\dValue_{\loc_1, 0}$, $\muppervalue_{\loc_1, 0}$, $\mlowervalue_{\loc_1, 0}$ are non-positive. 

	We consider the deterministic strategy \detmaxstrategy  
	for \MaxPl defined, for all plays \play with $(\loc_0, \val)$ as 
	last configuration, by 
	\begin{displaymath}
	\detmaxstrategy(\play) = 
	\begin{cases}
	(\trans_1, \frac{\varepsilon}{2^{|\play|}}) & \text{if $\val = 0$} \\
	(\trans_1, 0) & \text{if $\val > 0$} 
	\end{cases}
	\end{displaymath}
	Since each iteration of the cycle gives a cumulated weight of at most
	$-\frac{\varepsilon}{2^{|\play|}}$, then
	$\dValue^{\detmaxstrategy}_{\loc_1, 0} \geq -\varepsilon$ for all
	$\varepsilon > 0$. In particular, $\dValue_{\loc_1, 0}$ is also non-negative, which implies that
	$\dValue_{\loc_1, 0} = 0$. Notice that the
	strategy $\detmaxstrategy$ uses memory, to compute the length $|\play|$ of
	the current play. 
	
	Now, we consider the memoryless strategy for \MinPl such that, for $p\in (0,1)$, chooses transition $\trans_0$ with delay 0 with probability
	$1-p$ and transition $\trans_2$ with delay 0 with probability $p$. When $\MaxPl$ follows a
	memoryless strategy, each time we pass through the configuration ($\loc_0, 0)$, the same choice is always made that is to delay for a small amount of time before taking transition $\trans_1$. In particular, the two memoryless strategies combined allows one to expect $1/(1-p)$ turns in the cycle, each turn having a fixed negative weight. Thus, \MinPl can secure a weight as little as possible, by making $p$ approach $1$. Thus $\muppervalue_{\loc_1, 0} = -\infty$. And since $\mlowervalue_{\loc_1, 0} \leq \muppervalue_{\loc_1, 0}$, we also have $\mlowervalue_{\loc_1, 0} = -\infty$.
\end{exa}

We will be able to obtain a result comparing the memoryless values and the 
deterministic one, in a fragment of \WTG{s} that we present in the next section.

\section{Divergent weighted timed games}
\label{sec:divergent}

Interesting fragments of \WTG{s} have been designed, in order to regain 
decidability of the problem of determining whether the value of a \WTG is below a 
certain threshold. One such fragment is obtained by enforcing a semantical 
property of divergence (originally called \emph{strictly non-Zeno cost} when only
dealing with non-negative weights~\cite{BouyerCassezFleuryLarsen-04}):
it asks that every play following a cycle in the region automaton has
a weight far from $0$. We will consider this restriction in the
following, since it allows for a large class of decidable \WTG{s}, with
no limitations on the number of clocks.  Formally, a cyclic region
path~\rpath of~\rgame is said to be a positive cycle (resp.~a negative
cycle) if every finite play~\play following~\rpath satisfies
$\weightC(\play)\geq 1$ (resp.~$\weightC(\play)\leq -1$).

\begin{defiC}[\cite{BusattoGastonMonmegeReynier-17}]\label{def:divergent}
	A \WTG is \emph{divergent} if every cyclic region path is positive or
	negative.
\end{defiC}

In~\cite{BusattoGastonMonmegeReynier-17}, it is shown that this
definition is equivalent to requiring that for all strongly connected
components (SCC) $S$ of the graph of~\rgame, either every
cycle~$\rpath$ inside $S$ is positive (we say that the SCC is
positive), or every cycle~$\rpath$ inside $S$ is negative (we say that
the SCC is negative). The best computability result in this setting
is:

\begin{thmC}[\cite{BusattoGastonMonmegeReynier-17}]
	\label{thm:divergent-computation}
	The deterministic value of a divergent \WTG can be computed in
	triply-exponential-time.
\end{thmC}
\noindent 
We explain how to recover from Theorem~\ref{thm:divergent-computation}
the needed shape of $\varepsilon$-optimal strategies, since this is
one of the new technical ingredients we need afterwards.

\subsection{Switching strategies for \MinPl.}
\label{subsec:div-switching}

Theorem~\ref{thm:divergent-computation} is obtained in
\cite{BusattoGastonMonmegeReynier-17} by using a \emph{value
	iteration algorithm} (originally described in \cite{AlurBernadskyMadhusudan-04} for
acyclic timed automata). If $V$ represents a value function, i.e.~a
mapping $\Locs\times\Rplus^\Cl\to\Rbar$, 
we denote by $V_{\loc,\val}$
the image $V(\loc,\val)$, for better readability. One step of the game
is summarised in the following operator~$\mathcal{F}$ mapping each
value function $V$ to the value function defined for all
$(\loc,\val)\in \Locs\times\Rplus^\Cl$~by
\begin{displaymath}
\mathcal F(V)_{\loc,\val}= 
\begin{cases}
	0 & \text{if $\loc\in\LocsT$} \\
	\sup_{(\loc,\val) \moveto{\trans, t} (\loc',\val')}
	\big[t\times\weight(\loc)+\weight(\trans)+V_{\loc',\val'}\big] 
	& \text{if $\loc\in\LocsMax$} \\
	\inf_{(\loc,\val) \moveto{\trans, t} (\loc',\val')}
	\big[t\times\weight(\loc)+\weight(\trans)+V_{\loc',\val'} \big] 
	& \text{if $\loc\in\LocsMin$}
\end{cases}
\end{displaymath} 
where $(\loc,\val)\xrightarrow{\trans, t}(\loc',\val')$ ranges over valid 
edges in \game. Then, starting from $V^0$ mapping
every configuration to $+\infty$, except for the targets mapped to
$0$, we let $V^i= \mathcal{F}(V^{i-1})$ for all $i>0$. The  
value function~$V^i$ 
is intuitively what \MinPl can guarantee when forced to reach the 
target in at most~$i$ steps.

The value computation of Theorem~\ref{thm:divergent-computation} is
then obtained in two steps. First, configurations~$(\loc,\val)$ of
value $\dValue_{\loc,\val}=-\infty$ are found by using a decomposition
of the region game~$\rgame$ into SCCs. Indeed, in divergent \WTG{s}, 
configurations of value~$-\infty$
are all the ones from which $\MinPl$ has a strategy to visit
infinitely many times configurations of a single location $(\ell,r)$
of $\rgame$ contained in a negative SCC. This is thus a Büchi
objective on the region game, that can easily be solved with some
attractor computations. Notice that if a configuration $(\loc,\val)$
has value $-\infty$, this implies that all
configurations~$(\loc,\val')$ with~$\val'$ in the same region as
$\val$ have value $-\infty$. As we explained at the end of
Section~\ref{sec:prelim} for the values $+\infty$, we can then remove
configurations of value $-\infty$ by strengthening the guards on
transitions, while letting unchanged other finite values.

Then, the (finite) deterministic value $\dValue$ is obtained as 
an iterate $V^H$ of the previous operator, with~$H$ polynomial in the size 
of the region game and the maximal weights of $\game$. This means that 
playing for only a bounded number of steps is equivalent to the original 
game. In particular, at horizon $H$, we have that
$\mathcal F(V^H)=V^{H+1}= \dValue$ so that $\dValue$ is a fixpoint of
$\mathcal F$. As a side effect, this allows one to decompose the
clock space $\Rplus^{\Cl}$ into a finite number $\alpha$ of
\emph{cells} (a refinement of the classical regions) such that
$\dValue$ is affine on each cell. 

\begin{lemC}[\cite{BusattoGastonMonmegeReynier-17}]
	\label{lem:dVal_fixpoint-F}
	Let $(\loc, \val)$ be a configuration. Then $\dValue_{\loc, \val}$ is a 
	fixpoint of \F, i.e. $\F(\dValue_{\loc, \val}) = \dValue_{\loc, \val}$.
\end{lemC}
\noindent 
Based on this, we can construct good strategies for \MinPl that have a
special form, the so-called \emph{switching strategies} (introduced in
\cite{BrihayeGeeraertsHaddadMonmege-17} in the untimed setting, further 
extended in the timed
setting with only one-clock in \cite{BrihayeGeeraertHaddadLefaucheuxMonmege-15}).

\begin{defi}
	\label{def:switching}
	A \emph{switching strategy} $\detminstrategy$ is described by two
	deterministic memoryless strategies $\detminstrategy^1$ and
	$\detminstrategy^2$, as well as a switching threshold $K$. The
	strategy $\detminstrategy$ then consists in playing strategy
	$\detminstrategy^1$ until either we reach a target location, or the
	finite play has length at least $K$, in which case we switch to
	strategy $\detminstrategy^2$.
\end{defi}

Our new contribution is as follows:

\begin{thm}
	\label{thm:switching-divergent} 
	In a divergent \WTG, for all $\varepsilon>0$ and $N\in\N$, there exists a
	switching strategy $\detminstrategy \in \sdStratMinProper$ for \MinPl, for which the
	two components $\detminstrategy^1$ and $\detminstrategy^2$
	are smooth, such that for
	all configurations $(\loc, \val)$,
	$\dValue^{\detminstrategy}_{\loc, \val} \leq \max(-N,
	\dValue_{\loc, \val})+\varepsilon$.
\end{thm}
\noindent 
In particular, if all configurations have a finite deterministic
value, there exists an $\varepsilon$-optimal proper smooth switching
strategy wrt the deterministic value. In the presence
of a configuration with a deterministic value $-\infty$, we build from
Theorem~\ref{thm:switching-divergent} a family of proper smooth switching strategies
(indexed by the parameter $N$) whose value tends to
$-\infty$.

The proof of Theorem~\ref{thm:switching-divergent} requires to build
both (smooth) strategies $\detminstrategy^1$ and $\detminstrategy^2$, as well
as a switching threshold $K$. The second strategy $\detminstrategy^2$
only consists in reaching the target and is thus obtained as a
deterministic memoryless strategy from a classical attractor
computation in the region game \rgame. It is easy to choose
$\detminstrategy^2$ smooth (so that it fulfils
Hypothesis~\ref{hyp:measurability}). 
In contrast, the first strategy $\detminstrategy^1$ requires more
care. We build it so that it fulfils two properties, that we summarise
in saying that $\detminstrategy^1$ is
\emph{fake-$\varepsilon$-optimal} wrt the deterministic value:
\begin{enumerate}
	\item each finite play conforming to $\detminstrategy^1$
	from $(\loc,\val)$ and reaching the target has a cumulated weight
	at most $\dValue_{\loc,\val}+|\play|\,\varepsilon$ (in particular,
	if $\dValue_{\loc,\val}=-\infty$, no such plays should exist);
	\item each finite play conforming to $\detminstrategy^1$ following a
	\emph{long enough} cycle in the region game $\rgame$ has a cumulated
	weight at most~$-1$.
\end{enumerate}
\noindent 
Here, ``fake'' means that $\detminstrategy^1$ is not obliged to
guarantee that the target is reached, but if it does so, it must do it with a
cumulated weight close to $\dValue_{\loc,\val}$, the error term
depending linearly on the size of the play.
The second property ensures that playing long enough
$\detminstrategy^1$ without reaching the target results in diminishing
the cumulated weight. Then, if the switch happens at horizon $K$ big
enough,
($K=(\wmax^e |\rgame| (|\Locs|\alpha+2) +N)(|\rgame|(|\Locs|\alpha
+1)+1)$ suffices for instance), \MinPl is sure that the cumulated
weight so far is low enough so that the rest of the play to reach a
target location (following $\detminstrategy^2$ only) will not make the
weight increase too much. In the absence of values $-\infty$ in
$\dValue$, the first property allows one to obtain a
$K\varepsilon$-optimal strategy even in the case where the switch does
not occur (because we reach the target prematurely). The construction
of a fake-$\varepsilon/K$-optimal strategy $\detminstrategy^1$ (the
linear dependency on the length of the play in the first property of
fake-optimality is thus taken care by a division by $K$ here) relies
on the fact that $\mathcal F(\dValue)=\dValue$ 
(by Lemma~\ref{lem:dVal_fixpoint-F}) to play almost-optimally at horizon 1. 
More formally:
\begin{itemize}
	\item For all configurations of value $-\infty$, $\detminstrategy^1$
	is built as a winning strategy for the Büchi objective ``visit
	infinitely often configurations of a location $(\loc,r)$ of $\rgame$
	contained in a negative SCC''. By definition, all cyclic paths
	following $\detminstrategy^1$ will be inside a negative SCC, and
	thus be of cumulated weight at most $-1$, by divergence of the
	WTG. Moreover, no plays conforming to $\detminstrategy^1$ from such
	a configuration of value $-\infty$ will reach a target location,
	since the chosen negative SCC is a trap controlled by $\MinPl$. It
	is easy to choose $\detminstrategy^1$ smooth (so that it
	fulfils Hypothesis~\ref{hyp:measurability}).
	
	\item For the remaining configurations of finite deterministic value, 
	we rely upon operator $\mathcal F$, letting $\detminstrategy^1$ choose 
	a decision that minimises the deterministic value at horizon $1$. However, 
	because of the guards on clocks, infimum/supremum operators in $\mathcal F$ 
	are not necessarily minima/maxima, and we thus need to allow for a small
	error at each step of the strategy: this is the main difference with
	the untimed setting, which by the way explains why our definition of
	switching strategy needed to be adapted. We will use the
	$\arginfepsilon$ operator defined for all mappings
	$f\colon A \to \R$ and $B\subseteq A$ by
	$\arginfepsilon_B f = \{a\in B\mid f(a) \leq \inf_B f +
	\varepsilon\}$. Then, for all configurations
	$(\loc, \val)\in \LocsMin\times \Rplus^\Cl$, we choose
	$\detminstrategy^1(\loc, \val)$ as a pair $(\trans, t)$ such that 
	\begin{equation}
	\label{eq:div-sigma1}
	\detminstrategy^1(\loc, \val) \in 
	\arginfepsilon[\varepsilon/K]_{(\loc,\val) \moveto{\trans, t}(\loc',\val')} 
	(t \, \weight(\loc) + \weight(\trans) + \dValue_{\loc',\val'})
	\end{equation}
	This set is non empty since $\dValue$ is a fixpoint of
	operator~$\mathcal F$ in this case. Moreover, knowing that the
	mapping $\dValue_\loc$ is piecewise affine by the results shown in
	\cite{BusattoGastonMonmegeReynier-17}, it is possible to choose
	$\detminstrategy^1$ so that it fulfils the measurability (even
	piecewise continuity) conditions of
	Hypothesis~\ref{hyp:measurability}. More precisely, we can consider
	it to take the same kind of decision for all configurations of a
	same cell: same transition, and either no delay or a delay jumping
	to the same border of cell.
\end{itemize}
\noindent 
The strategy $\detminstrategy^1$ thus built makes a small error wrt
the optimal at each step. But once again strongly relying on the
divergence of the WTG, we can nevertheless show that
$\detminstrategy^1$ is fake-$\varepsilon/K$-optimal wrt the
deterministic value.

\begin{lem}
	\label{lem:fake-optimally}
	The strategy $\detminstrategy^1$ of $\MinPl$ 
	is fake-$\varepsilon/K$-optimal
	wrt the deterministic value, i.e.~
	\begin{enumerate}
		\item each finite play \play conforming to $\detminstrategy^1$
		from $(\loc,\val)$ and reaching the target has a cumulated weight
		at most $\dValue_{\loc,\val}+|\play|\,\varepsilon/K$: in
		particular, if $\dValue_{\loc,\val}=-\infty$, no such play exists;
		
		\item each finite play conforming to $\detminstrategy^1$ following a
		cycle in the region game $\rgame$ of length at least
		$|\Locs|\alpha+1$ has a cumulated weight at most~$-1$.
	\end{enumerate}
\end{lem}
\begin{proof}
	We show independently the two properties.
	\begin{enumerate}
		\item If $\dValue_{\loc,\val}= -\infty$, we have already seen before
		that no such play exists. We thus restrict ourselves to
		considering configurations of value different from $-\infty$ in
		the following. We then reason by induction on the length of the
		plays \play. If $|\play| = 0$, $\loc \in \LocsT$ and
		\begin{displaymath}
		\weight(\play) = 0 = \dValue_{\loc, \val} \leq 
		\dValue_{\loc, \val} + 0 \times \frac{\varepsilon}{K}.
		\end{displaymath} 
		If $|\play| = n > 0$, we write $\play = (\loc,\val) \moveto{\trans, t} 
		(\loc', \val') \dots$, and let $\play'$ be the play 
		following \play from $(\loc', \val')$.  By definition, we have
		$\weight(\play) = t \,\weight(\loc) + \weight(\trans) +
		\weight(\play')$. By induction hypothesis, we obtain
		\begin{equation}
		\label{eq:weight1}
		\weight(\play) \leq t\, \weight(\loc) + \weight(\trans) +
		\dValue_{\loc', \val'} + (n-1)\frac{\varepsilon}{K}
		\end{equation}
		First, we suppose that $\loc \in \LocsMax$, then we can 
		rewrite~\eqref{eq:weight1} (by applying the supremum) as
		\begin{displaymath}
		\weight(\play)
		\leq \sup_{(\loc,\val) \moveto{\trans,t} (\loc',\val')} 
		\big(t\, \weight(\loc) + \weight(\trans) +
		\dValue_{\loc', \val'}\big) + (n-1)\frac{\varepsilon}{K}.
		\end{displaymath}
		Moreover, since $\dValue$ is a fixpoint of $\F$ (by 
		Lemma~\ref{lem:dVal_fixpoint-F}), then 
		\begin{displaymath}
		\weight(\play)
		\leq \dValue_{\loc, \val} + (n-1)\frac{\varepsilon}{K} 
		\leq \dValue_{\loc, \val} + n\frac{\varepsilon}{K}.
		\end{displaymath}
		Otherwise, we suppose that $\loc \in \LocsMin$, then, using the 
		definition of $\detminstrategy^1$ given in~\eqref{eq:div-sigma1}, we can 
		rewrite~\eqref{eq:weight1} as
		\begin{displaymath}
		\weight(\play) \leq 
		\inf_{(\loc,\val) \moveto{\trans,t} (\loc',\val')} 
		\big(t\, \weight(\loc) + \weight(\trans) + 
		\dValue_{\loc', \val'}\big) + \frac{\varepsilon}{K} 
		+ (n-1)\frac{\varepsilon}{K}
		\end{displaymath}
		We conclude, since \dValue is a fixpoint of \F (by 
		Lemma~\ref{lem:dVal_fixpoint-F}), that 
		$\weight(\play) \leq  \dValue_{\loc, \val} + n\frac{\varepsilon}{K}$.
		
		\item The second property is trivial (and already discussed before)
		when the strategy $\detminstrategy^1$ is chosen according to the
		case of values $-\infty$. We therefore once again suppose that there
		are no remaining configurations of value $-\infty$ in this proof.
		
		We first show that, for all edges $(\loc,\val) \moveto{\trans,t} 
		(\loc',\val')$ conforming to $\detminstrategy^1$, we have
		\begin{equation}
		\dValue_{\loc,\val} \geq t \, \weight(\loc) + 
		\weight(\trans) + \dValue_{\loc', \val'} - 
		\frac{\varepsilon}{K}
		\label{eq:dVal}
		\end{equation}
		If $\loc\in\LocsMin$, since $\dValue$ is a fixpoint of $\F$ (by 
		Lemma~\ref{lem:dVal_fixpoint-F}), we have
		\[\dValue_{\loc, \val} = 
		\inf_{(\loc,\val) \moveto{\trans_1,t_1} (\loc_1,\val_1)}
		\big(t_1 \, \weight(\loc) + \weight(\trans_1) + 
		\dValue_{\loc_1, \val_1}\big).\]
		Then, by definition of $(\trans,t)$ chosen by $\detminstrategy^1$,
		\[\dValue_{\loc, \val} \geq
		t \, \weight(\loc) + \weight(\trans) + \dValue_{\loc', \val'}-
		\frac{\varepsilon}{K}.\] 
		If $\loc\in \LocsMax$, since $\dValue$ is a fixpoint of $\F$ (by 
		Lemma~\ref{lem:dVal_fixpoint-F}), we have
		\begin{displaymath}
		\dValue_{\loc, \val} = 
		\sup_{(\loc,\val) \moveto{\trans_1,t_1} (\loc_1,\val_1)}
		\big(t_1 \, \weight(\loc) + \weight(\trans_1) + 
		\dValue_{\loc_1, \val_1}\big).
		\end{displaymath}
		In particular, for the pair $(\trans, t)$ chosen in \play, we
		have
		\[
		\dValue_{\loc, \val} \geq t \, \weight(\loc) +
		\weight(\trans) + \dValue_{\loc', \val'} \geq t \,
		\weight(\loc) + \weight(\trans) + \dValue_{\loc',
			\val'} - \frac{\varepsilon}{K}.
		\]
		
		Then, let $\rpath$ be a cyclic path in $\rg$ of length at least
		$|\Locs|\alpha+1$, and \play a finite play following $\rpath$ and
		conforming to $\detminstrategy^1$. It goes through only states of a
		single SCC of $\rg$, that is either positive or negative, by
		divergence of the WTG $\game$. We show that this SCC is necessarily
		negative, therefore proving $\weightC(\play)\leq -1$. Suppose, in
		the contrary, that the SCC is positive.
		
		First, since $\rpath$ has length at least $|\Locs|\alpha+1$, it must
		go through the same pair of location and cell of $\dValue$ (remember
		that $\dValue$ is piecewise affine, and we have called cell each
		piece where it is affine) at least twice. Consider the cycle
		$\rpath'$ (of length at most $|\Locs|\alpha$) obtained by
		considering the part of $\rpath$ between these two occurrences, and
		$\play'$ a corresponding finite play starting in a configuration
		$(\loc,\val)$, ending in $(\loc,\val')$ (with $\val$ and $\val'$ in
		the same cell). Since the SCC is positive, $\play'$ has weight at
		least $1$. By the definition of $\detminstrategy^1$ taking a uniform
		decision on each cell, we can mimic this loop once again from
		$(\loc,\val')$, for $2\lceil \beta\rceil$ times, where $\beta$ is
		the maximum over the cells $c$ (where $\dValue$ is not $-\infty$) of
		the differences $\sup_c \dValue - \inf_c \dValue$, that is finite
		since each cell is compact. The obtained play $\play''$ has weight
		at least $2\lceil \beta\rceil$ (since it is decomposed as
		$2\lceil \beta\rceil$ cycles in a positive SCC), and length at most
		$2\lceil \beta\rceil|\Locs|\alpha$.
		
		By summing up all the inequalities given by~\eqref{eq:dVal} along
		the play $\play''$ (from $(\loc,\val)$ to $(\loc,\val'')$), and by
		simplifying the result, we obtain that
		\[\weightC(\play'') \leq 
		\dValue_{\loc,\val}-\dValue_{\loc,\val''} +
		\frac{\varepsilon}{K}|\play''|\leq
		\dValue_{\loc,\val}-\dValue_{\loc,\val''} +
		\frac{\varepsilon}{K}2\lceil \beta\rceil|\Locs|\alpha.\] 
		However, by definition of $\beta$ (the maximum difference of values in a
		cell), we obtain
		\[\weightC(\play'') \leq 
		\lceil\beta\rceil\Big(1+\frac{\varepsilon}{K}2|\Locs|\alpha\Big)
		\leq \lceil\beta\rceil(1+2\varepsilon) < 2 \lceil\beta\rceil\]
		since $K\geq 2|\Locs|\alpha$ and $\varepsilon<1/2$. This contradicts
		the fact that $\play''$ has weight at least $2\lceil \beta\rceil$,
		and concludes the proof of the second item. \qedhere
	\end{enumerate}
\end{proof}
\noindent 
Now, we conclude the proof of Theorem~\ref{thm:switching-divergent}.

\begin{proof}[Proof of Theorem~\ref{thm:switching-divergent}]
	We start by showing that the switching strategy $\detminstrategy$ defined above by 
	the triplet $(\detminstrategy^1, \detminstrategy^2, K$) is a proper smooth 
	deterministic strategy.
	First, $\detminstrategy$ is a smooth deterministic strategy since 
	$\detminstrategy^1$ and $\detminstrategy^2$ satisfy Hypothesis~\ref{hyp:measurability}
	by construction. Now, $\detminstrategy$ is also proper (by Proposition~\ref{prop:proper}) 
	since it satisfies the second condition of Hypothesis~\ref{hyp:proper}: by letting $m = K + |\rgame| +1$, the 
	probability to reach a target location in at most $m$ steps is $1$. 
	
	Now, we prove that for all
	configurations $(\loc, \val)$, we have
	\begin{displaymath}
	\dValue^{\detminstrategy}_{\loc, \val} 
	\leq \max(-N, \dValue_{\loc, \val}) + \varepsilon \,.
	\end{displaymath}
	Let \play be a play conforming to $\detminstrategy$ from 
	$(\loc, \val)$ that reaches the target $(\loc_t, \val_t)$. We distinguish 
	two cases according to the length of \play. First, we suppose that 
	$|\play| \leq K$, then \play is only conforming to~$\detminstrategy^1$ which is 
	a fake-$\varepsilon/K$-optimal strategy. Moreover, as \play reaches the target, 
	$\dValue_{\loc, \val}$ is finite. Indeed if $\dValue_{\loc, \val} = -\infty$, 
	$\detminstrategy^1$ is built with a Büchi objective and cannot reach the target. 
	In particular, $\weightC(\play) \leq \dValue_{\loc, \val} + 
	|\play|\frac{\varepsilon}{K}$, and we conclude since $|\play| \leq K$
	\begin{displaymath}
	\weightC(\play) 
	\leq \dValue_{\loc, \val} + \varepsilon
	\leq \max(-N, \dValue_{\loc, \val}) + \varepsilon.
	\end{displaymath}
		
	Otherwise, we suppose that $|\play| \geq K$, then we can decompose \play as 
	$\play_1$ conforming to $\detminstrategy^1$, and followed by $\play_2$ 
	conforming to $\detminstrategy^2$ with $|\play_1| = K$. As $\play_2$ is 
	conforming to $\detminstrategy^2$, an attractor strategy on the region game 
	$\rgame$, it is acyclic in $\rgame$, so $|\play_2| \leq |\rg|$. In particular, 
	\begin{displaymath}
	\weightC(\play_2) \leq \wmax^e|\rgame|.
	\end{displaymath}
	As $|\play_1| = K$, by definition of
	$K= (\wmax^e |\rgame| (|\Locs|\alpha+2)
	+N)(|\rgame|(|\Locs|\alpha +1)+1)$, $\play_1$ goes at least 
	$(\wmax^e |\rgame| (|\Locs|\alpha+2) +N)(|\Locs|\alpha +1)$
	times through the same pair of location and region, and thus contains
	at least $\wmax^e |\rgame| (|\Locs|\alpha+2) +N$ region cycles
	of length at least $|\Locs|\alpha +1$. Moreover, all of them
	are conforming to $\detminstrategy^1$ that is a
	fake-$\varepsilon/K$-optimal strategy. In particular, each
	cycle has a weight at most $-1$, and the weight of all cycles
	of $\play_1$ is therefore at most
	$-\wmax^e |\rgame| (|\Locs|\alpha+2) -N$. Moreover, $\play_1$
	has at most $|\rgame|(|\Locs|\alpha +1)$ remaining edges
	(otherwise it would contain a region cycle of length at least
	$|\Locs|\alpha +1$), once the region cycles are removed. The
	weight of one of these edges is at most $\wmax^e$, so the
	weight of this set of edges is at most
	$\wmax^e |\rgame|(|\Locs|\alpha +1)$. We can deduce that
	\begin{displaymath}
	\weightC(\play_1) \leq -\wmax^e |\rgame| (|\Locs|\alpha+2) -N +
	\wmax^e|\rgame|(|\Locs|\alpha +1)
	= -\wmax^e|\rg| - N.
	\end{displaymath}
	So, we can deduce that
	\begin{displaymath}
	\weightC(\play) 
	= \weightC(\play_1) + \weightC(\play_2)
	\leq -\wmax^e|\rg| - N + \wmax^e|\rg| = -N.
	\end{displaymath}
	In particular, we obtain that 
	$\weightC(\play) \leq \max(-N, \dValue_{\loc, \val}) + \varepsilon$. 
	To conclude the proof, we remark that \play is a play reaching the target, 
	so $\weight(\play) = \weightC(\play) \leq \max(-N, \dValue_{\loc, \val}) + 
	\varepsilon$.
\end{proof}

\subsection{Memoryless strategies for \MaxPl.}
\label{subsec:div_memless-Max}
In a divergent \WTG, we can prove that \MaxPl has an $\varepsilon$-optimal 
memoryless smooth deterministic strategy for all $\varepsilon > 0$. To make this proof, 
we take the point of view of $\MaxPl$; looking for good strategies for this other player. 
This is possible since all \WTG{s} are determined with respect to the deterministic 
value. We may thus associate a deterministic value with any deterministic strategy 
\detmaxstrategy of \MaxPl:
\begin{displaymath}
\dValue^\detmaxstrategy_{\loc,\val} = \inf_{\detminstrategy\in\dStratMin}
\weight(\outcomes((\loc,\val),\detminstrategy,\detmaxstrategy)) \,.
\end{displaymath} 
Then, the deterministic strategy \detmaxstrategy is $\varepsilon$-optimal wrt 
the deterministic value if $\dValue^\detmaxstrategy_{\loc,\val} \geq
\dValue_{\loc,\val}-\varepsilon$ for all configurations $(\loc,\val)$.

As \MaxPl does not wish to go to the target, we show that no switch is
necessary to play $\varepsilon$-optimally: memoryless strategies are
sufficient to guarantee a value as close as wanted to the
deterministic value. For a configuration with a value
equal to $-\infty$, all the deterministic strategies for \MaxPl are
equivalent, that is, they are all equally bad. Without loss of generality,
we can therefore suppose that there are no configurations in \game
with a value equal to $-\infty$. Then, it is shown in
\cite{BusattoGastonMonmegeReynier-17} that remaining values are bounded in 
absolute value by $\wmax^e|\rgame|$, since \emph{optimal plays} have no cycles.
We use that fact to build a memoryless deterministic strategy
$\detmaxstrategy$ analogous to strategy $\detminstrategy^1$ before:

\begin{thm}
	\label{thm:optimalMax-divergent}
	In a divergent \WTG, there exists a memoryless smooth deterministic
	$\varepsilon$-optimal strategy for player~\MaxPl wrt the
	deterministic value.
\end{thm}
\noindent 
Once again, $\MaxPl$ will use the fact that $\dValue$ is a fixpoint of
operator $\mathcal F$ (once removed configurations of value
$-\infty$). We must still accomodate small errors at each step,
because of the guards on clocks. Similarly to the switching parameter
$K$ for the switching strategy (with $N=0$), we let
$K=\wmax^e|\rgame|(|L|\alpha+2) (|\rgame|(|L|\alpha+1)+1)$. We use the
$\argsupepsilon$ operator, analogously to $\arginfepsilon$, defined
for all mappings $f\colon A\to \R$ and $B\subseteq A$ by
$\argsupepsilon_B f = \{a\in B\mid f(a) \geq \sup_B f -
\varepsilon\}$. Then, for all configurations
$(\loc, \val)\in \LocsMax\times \Rplus^\Cl$, we choose
$\detmaxstrategy(\loc, \val)$ as a pair $(\trans,t)$ such that 
\begin{equation}
\label{eq:def:tau_eps}
\detmaxstrategy(\loc, \val) \in 
\argsupepsilon[\varepsilon/K]_{(\loc,\val)\moveto{\trans, t}(\loc',\val')}
(t \, \weight(\loc) + \weight(\trans) + \dValue_{\loc',\val'})
\end{equation}
with similar smooth assumptions as for $\detminstrategy^1$ before. 
In a completely symmetrical way as for Lemma~\ref{lem:fake-optimally},
we show that $\detmaxstrategy$ satisfies the following useful
properties:

\begin{lem}
	\label{lem:Max-fakeopt}
	The strategy $\detmaxstrategy$ satisfies the following properties:
	\begin{enumerate}
		\item\label{item:Max-fakeopt_1} each finite play \play conforming to
		$\detmaxstrategy$ from $(\loc,\val)$ and reaching the target has a
		cumulated weight at least
		$\dValue_{\loc,\val} - |\play|\varepsilon/K$;
		\item\label{item:Max-fakeopt_2} each finite play conforming to
		$\detmaxstrategy$, that follows a cycle in the region game
		$\rgame$ of length at least $|L|\alpha+1$, has a cumulated weight
		at least $1$.
	\end{enumerate}
\end{lem}

\noindent 
Now, we have tools to prove that $\detmaxstrategy$ is an 
$\varepsilon$-optimal strategy for \MaxPl with respect to the deterministic 
value from $(\loc, \val)$. In other words, we show that 
$\dValue^{\detmaxstrategy}_{\loc, \val} \geq \dValue_{\loc, \val} - \varepsilon$.

\begin{proof}[Proof of Theorem~\ref{thm:optimalMax-divergent}]
	Let \play be a play conforming to $\detmaxstrategy$ from
	$(\loc,\val)$.  We distinguish two cases according to the length of
	\play. First, we suppose that $|\play| \leq K$, then, by 
	Lemma~\ref{lem:Max-fakeopt}.\ref{item:Max-fakeopt_1}, we
	have 
	\begin{displaymath}
	\weight(\play) \geq \dValue_{\loc, \nu} -
	\frac{\varepsilon}{K}|\play| \geq \dValue_{\loc, \nu} - \varepsilon. 
	\end{displaymath}
	Otherwise, we suppose that $|\play|>K$, then by definition of
	$K=\wmax^e|\rgame|(|L|\alpha+2) (|\rgame|(|L|\alpha+1)+1)$,
	\play contains at least $\wmax^e|\rgame|(|L|\alpha+2)$
	cycles of length at least $|L|\alpha+1$ in $\rgame$. By
	Lemma~\ref{lem:Max-fakeopt}.\eqref{item:Max-fakeopt_2}, all of these cycles 
	have a cumulated weight at least $1$. So, the cumulated weight of
	\play induced by the cycles in $\rgame$ is at least
	$\wmax^e|\rgame|(|L|\alpha+2)$. Moreover, \play has at most
	$|\rgame|(|L|\alpha+1)$ remaining edges, once the region
	cycles are removed. The weight of each of these edges is at least
	$-\wmax^e$, so the weight of this set of edges is at least
	$-\wmax^e |\rgame|(|L|\alpha+1)$. So, we can deduce that
	\begin{displaymath}
	\weight(\play) 
	\geq \wmax^e|\rgame|(|L|\alpha+2) - \wmax^e |\rgame|(|L|\alpha+1) 
	= \wmax^e|\rgame| \geq \dValue_{\loc, \val}.
	\end{displaymath} 
	In particular, we obtain that $\weight(\play) \geq \dValue_{\loc, \val} - 
	\varepsilon$. 
\end{proof}
\noindent 
With the existence of the (proper) smooth deterministic $\varepsilon$-optimal
strategies for both players with respect to the deterministic value, we can deduce that 
divergent \WTG{s} are determined with respect to the smooth deterministic value and 
this value is equal to the deterministic one. Finally, 
we will deduce that divergent \WTG{s} are determined with respect to the stochastic 
value, and the stochastic and (smooth) deterministic values are equal:

\begin{cor}\label{cor:sdVal-div}
	In all divergent \WTG{s}, for all $(\loc, \val)$, 
	\begin{displaymath}
	\dValue_{\loc, \val} = \sdlowervalue_{\loc, \val} = \sduppervalue_{\loc, \val} = 
	\lowervalue_{\loc, \val} = \uppervalue_{\loc, \val}
	\,.
	\end{displaymath}
\end{cor}
\begin{proof}
	Let $\varepsilon > 0$, there exists $\detminstrategy \in \sdStratMinProper$ and 
	$\detmaxstrategy \in \sdStratMax$ such that 
	\begin{displaymath}
	\dValue^{\detminstrategy}_{\loc, \val} - \varepsilon \leq \dValue \leq 
	\dValue^{\detmaxstrategy}_{\loc, \val} + \varepsilon
	\end{displaymath}
	by Theorem~\ref{thm:switching-divergent} and 
	Theorem~\ref{thm:optimalMax-divergent}. First, 
	since $\sdStratMax \subseteq \dStratMax$, we deduce that 
	\begin{displaymath}
		\sduppervalue_{\loc, \val} \leq 
		\sup_{\detmaxstrategy \in \sdStratMax} 
		\weight(\outcomes((\loc, \val), \detminstrategy, \detmaxstrategy)) \leq 
		\sup_{\detmaxstrategy \in \dStratMax} 
		\weight(\outcomes((\loc, \val), \detminstrategy, \detmaxstrategy)) = 
		\dValue^{\detminstrategy}_{\loc, \val} - \varepsilon \,.
	\end{displaymath}
	Next, with the same reasoning, since $\sdStratMinProper \subseteq \dStratMin$, we deduce that 
	\begin{displaymath}
	\dValue^{\detmaxstrategy}_{\loc, \val} - \varepsilon = 
	\inf_{\detminstrategy \in \dStratMin} 
	\weight(\outcomes((\loc, \val), \detminstrategy, \detmaxstrategy)) \leq 
	\inf_{\detminstrategy \in \sdStratMinProper} 
	\weight(\outcomes((\loc, \val), \detminstrategy, \detmaxstrategy))  
	\leq \sdlowervalue_{\loc, \val}\,.
	\end{displaymath}
	Thus, $\sdlowervalue_{\loc, \val} = \sduppervalue_{\loc, \val}$, since 
	$\sdlowervalue_{\loc, \val} \leq \sduppervalue_{\loc, \val}$. 
	Finally, we conclude the proof by applying Theorem~\ref{thm:detVal=Val}: 
	\begin{displaymath}
	\sdlowervalue_{\loc, \val} \leq \lowervalue_{\loc, \val} \leq \uppervalue_{\loc, \val} \leq 
	 \sduppervalue_{\loc, \val} = \sdlowervalue_{\loc, \val}\,. \qedhere
	\end{displaymath}
\end{proof}

\section{Memoryless value in divergent weighted timed games}
\label{sec:trading}

We finally study the memoryless values of divergent \WTG{s} showing that memory can be fully
emulated with stochastic choices, and vice versa. Moreover, we show that divergent \WTG{s} are determined with respect to the memoryless value (as we obtained previously for the stochastic value).
\begin{thm}
	\label{thm:divergent-equalvalue}
	In all divergent \WTG{s}, for all 
	$(\loc,\val)$,
	$\dValue_{\loc,\val} =
	\muppervalue_{\loc,\val} = \mlowervalue_{\loc,\val}$.
\end{thm}
\noindent 
The proof is decomposed into two inequalities:
\begin{itemize}
\item we show in Lemma~\ref{lem:divergent-detVal<lowerVal} that $\dValue_{\loc,\val} \leq \mlowervalue_{\loc,\val}$, relying on the existence of a $\varepsilon$-optimal deterministic and memoryless strategy of $\MaxPl$ against the deterministic value (Theorem~\ref{thm:optimalMax-divergent});
\item  we show in Lemma~\ref{lem:partition} that
memoryless stochastic strategies can emulate deterministic ones: $\muppervalue_{\loc,\val} \leq \dValue_{\loc,\val}$. 
\end{itemize}\noindent 
Combined with the fact that $\mlowervalue_{\loc,\val} \leq \muppervalue_{\loc,\val}$, we indeed obtain the desired equalities.

As shown in Example~\ref{ex:1clock}, the theorem does not hold for all \WTG{s}, and the divergence property is therefore useful (though it might be possible to weaken it).

We start by showing that the deterministic value is at most equal to the memoryless lower-value: 
\begin{lem}
	\label{lem:divergent-detVal<lowerVal}
	In all divergent \WTG{s} \game, for all locations $\loc$ and valuations $\val$,
	$\dValue_{\loc,\val} \leq \mlowervalue_{\loc,\val}$. 
\end{lem}
\begin{proof} Let $\varepsilon > 0$. By
Theorem~\ref{thm:optimalMax-divergent}, \MaxPl has a memoryless 
deterministic $\varepsilon$-optimal strategy \detmaxstrategy for the deterministic value. Against any deterministic strategy of
$\MinPl$, from $(\loc,\val)$, it guarantees a weight at least $\dValue_{\loc,
\val} - \varepsilon$. 

Let \minstrategy be any memoryless proper strategy of \MinPl. Every play
conforming to $\minstrategy$ and $\detmaxstrategy$ has a weight at least
$\dValue_{\loc, \val} - \varepsilon$ (since it is conforming to
$\detmaxstrategy$), so that, by definition of the expectation,
\[\E^{\minstrategy, \detmaxstrategy}_{\loc, \val} = \int_{\play}
\weight(\play)\mathrm{d}\Proba_{\loc,\val}^{\minstrategy,\detmaxstrategy}(\play)
\geq \dValue_{\loc, \val} - \varepsilon\,\] Since this inequality holds for all
proper strategies of \MinPl, we deduce that 
	\begin{displaymath}
	\inf_{\minstrategy \in \mStratMinProper} 
	\E^{\minstrategy, \detmaxstrategy}_{\loc, \val} \geq \dValue_{\loc, \val} - 
	\varepsilon\,.
	\end{displaymath} 
	By taking the supremum over all memoryless strategies of
	$\MaxPl$ (that contain the memoryless and deterministic strategy
	\detmaxstrategy), we have
	\begin{displaymath}
	\sup_{\maxstrategy \in \mStratMax} \inf_{\minstrategy \in \mStratMinProper} 
	\E^{\minstrategy, \maxstrategy}_{\loc, \val} \geq \dValue_{\loc, \val} - 
	\varepsilon\,.
	\end{displaymath} 
	In particular, $\mlowervalue_{\loc, \val} \geq \dValue_{\loc,\val} - 
	\varepsilon$.
	To conclude, we remark that this inequality holds for all 
	$\varepsilon > 0$, and thus also when $\varepsilon$ is set to $0$. 
\end{proof}
\noindent 
The most technical inequality is the remaining one, showing that we can simulate deterministic strategies with memoryless
strategies:
\begin{lem}\label{lem:partition}
In all divergent \WTG{s} \game, for all locations $\loc$ and valuations $\val$,
$\muppervalue_{\loc, \val} \leq \dValue_{\loc, \val}$. 
\end{lem}
\noindent 
We prove this result in the rest of this section. To do so, we build a
memoryless strategy of $\MinPl$ at least as good as a deterministic
strategy. By Theorem~\ref{thm:switching-divergent}, we can start from
a switching strategy for $\MinPl$.  For $N \in \N$ and
$\varepsilon > 0$, we thus consider a switching strategy
$\detminstrategy = (\detminstrategy^1, \detminstrategy^2, K)$ of value
$\dValue^{\detminstrategy}_{\loc, \val} \leq \max(-N, \dValue_{\loc,
	\val}) +\varepsilon$, and simulate it with a memoryless strategy for
\MinPl, denoted $\minstrategy^p$, with a probability parameter
$p \in (0, 1)$. This new strategy is a probabilistic superposition of
the two memoryless deterministic strategies $\detminstrategy^1$ and
$\detminstrategy^2$.

\paragraph{Definition of $\minstrategy^p$}
The definition of $\minstrategy^p(\loc, \val)$, with $\loc\in\LocsMin$, depends
on the sign of the SCC containing the location $(\loc,r)$, with $r$ the region
of $\val$, of the region game $\rgame$. In a positive SCC, \MinPl always
chooses to play $\detminstrategy^1$, thus looking for a negative cycle in the
next SCCs (in topological order) if any, i.e.~we let $\minstrategy^p(\loc,
\val) = \Dirac{\detminstrategy^1(\loc, \val)}$. In a negative SCC, \MinPl
chooses to play $\detminstrategy^1$ with probability $p$, and
$\detminstrategy^2$ with probability $1-p$, i.e.~we let $\minstrategy^p(\loc,
\val) = p\times \Dirac{\detminstrategy^1(\loc, \val)} + (1-p)\times \Dirac{\detminstrategy^2(\loc, \val)}$.

These choices can be decomposed as first choosing a transition and then a delay as follows: 
letting
$(\trans_1, t_1) = \detminstrategy^1(\loc, \val)$, and 
$(\trans_2, t_2) = \detminstrategy^2(\loc, \val)$, we define
\begin{equation*}
	\label{eq:def_eta-p_trans}
	\minstrategytrans^p(\loc, \val) =
	\begin{cases}
		\Dirac{\trans_1} & \text{if $\loc$ is in a positive SCC} \\
		p\times\Dirac{\trans_1} + (1-p)\times\Dirac{\trans_2} &
		\text{otherwise}
	\end{cases}
\end{equation*}
and
\begin{equation*}
	\label{eq:def_eta-p_delay}
	\minstrategydelay^p(\loc, \val)(\trans) =
	\begin{cases}
		\Dirac{t_1} & \text{if $\loc$ is in a positive SCC} \\
		p\times\Dirac{t_1} + (1-p)\times\Dirac{t_2}
		& \text{if $\loc$ is in a negative SCC, and $\trans_1 = \trans_2$} \\
		\Dirac{t_1} & \text{if $\loc$ is in a negative SCC,
			and $\trans = \trans_1$} \\
		\Dirac{t_2} & \text{if $\loc$ is in a negative SCC,
			and $\trans = \trans_2$}
	\end{cases}
\end{equation*}
Theorem~\ref{thm:switching-divergent} ensuring that strategies
$\detminstrategy^1$ and $\detminstrategy^2$ are smooth, the superposition
$\minstrategy^p$ is also smooth. Moreover, it is also proper:

\begin{lem}
	For all $p \in (0, 1)$, the strategy $\minstrategy^p$ is proper.
	\label{lem:eta_properties}
\end{lem}
\begin{proof}
	We prove that $\minstrategy^p$ satisfies the second item of Hypothesis~\ref{hyp:proper}, with
    $m = |\rgame|$ and $\alpha = (1-p)^{m+1}$. Let 
	$\maxstrategy \in \StratMax$, and \playBis a finite play. We prove that 
	$\Proba^{\minstrategy^p, \maxstrategy}_{\playBis}(\bigcup_{n \leq m} 
	\TPlays_{\playBis}^n) \geq \alpha$. Not reaching the target in at most 
	$m$ steps is equivalent to having a prefix of length $m+1$, so that 
	$\Proba^{\minstrategy^p, \maxstrategy}_{\playBis}(\bigcup_{n \leq m} 
	\TPlays_{\playBis}^n) = 1-
	\Proba^{\minstrategy^p, \maxstrategy}_{\playBis}(\bigcup_{\ppath \mid |\ppath| 
		= m+1} \Cyl_{\playBis}(\ppath))$. We thus prove that 
	$\Proba^{\minstrategy^p, \maxstrategy}_{\playBis}(\bigcup_{\ppath \mid 
		|\ppath| = m+1} \Cyl_{\playBis}(\ppath)) \leq 1-\alpha$.

	Since $m=|\rgame|$, a play is in $\bigcup_{\ppath \mid |\ppath| = m+1} 
	\Cyl_{\playBis}(\ppath)$ if and only if it is of the form 
	$\playBis \play\play'$ with $|\play|=m+1$: in particular, $\play$ contains a 
	cycle in the region game, so cannot be conforming to the attractor strategy 
	$\detminstrategy_2$. As a consequence, $\play$ goes through a transition 
	played by $\detminstrategy_1$, chosen with probability $p$ in 
	$\minstrategy^p$. Given $k \geq 1$, calling $S_{\playBis}^{k}$ the set 
	of plays $\playBis \play\play'$ with $\play$ of length $k$ containing a 
	transition of probability $p$ chosen by $\minstrategy^p$, we have 
	\[\Proba^{\minstrategy^p, \maxstrategy}_{\playBis}(\bigcup_{\ppath \mid 
		|\ppath| = m+1} \Cyl_{\playBis}(\ppath)) = 
	\Proba^{\minstrategy^p, \maxstrategy}_{\playBis}(S_{\playBis}^{m+1}) \,.\]

	We conclude by showing that 
	$\Proba^{\minstrategy^p, \maxstrategy}_{\playBis}(S_{\playBis}^{k}) \leq 
	1-(1-p)^k$ holds for all $k\geq 1$ by induction. We let $\ppathBis$ be the 
	path followed by $\playBis$.

	When $k=1$, the transition of every play in $S_{\playBis}^{1}$ after the 
	common prefix $\playBis$ is chosen by $\MinPl$ with a probability $p$. Thus, 
	$\playBis\in \FPlaysMin$, $\minstrategy^p(\playBis) = p 
	\Dirac{(\trans_0, \delay_0)} + (1-p) \Dirac{(\trans'_0, \delay'_0)}$ with $(\trans_0,\delay_0) \neq (\trans'_0,\delay'_0)$ and the transition of every 
	play in $S_{\playBis}^{1}$ after the common prefix $\playBis$ is 
	$(\trans_0, \delay_0)$. Indeed, $S_{\playBis}^{1} = 
	\Cyl_{\playBis}(\ppathBis\trans_0, \Rplus^{|\ppathBis|} \times \{\delay_0\})$. 
	As a consequence, 
	\begin{align*}
		\Proba^{\minstrategy^p, \maxstrategy}_{\playBis}(S_{\playBis}^{1}) 
		&= \Proba^{\minstrategy^p, 
			\maxstrategy}_{\playBis}(\Cyl_{\playBis}(\ppathBis\trans_0, 
		\Rplus^{|\ppathBis|} \times \{\delay_0\})) \\ 
		&= \minstrategy^p(\playBis)(\trans_0, \delay_0) \times 
		\Proba^{\minstrategy^p, 
			\maxstrategy}_{\playBis\extendto{\trans_0,\delay_0}}
		(\Cyl_{\playBis\extendto{\trans_0,\delay_0}}(\ppathBis\trans_0, 
		\Rplus^{|\ppathBis|} \times \{\delay_0\})) \\
		&= p \times 1 = 1- (1-p)^1
	\end{align*}
\noindent 
	When $k\geq 2$, we distinguish three cases. 

	\begin{figure}
		\centering
		\begin{tikzpicture}
		\draw (-1,0) -- (2,0);
		\draw (2,0) -- (3,1);
		\draw (2,0) -- (3,-1);
		\draw (3,1) -- (6,0.25) -- (6,1.75) -- (3,1);
		\draw[fill=blue!20!white, opacity=0.5] (3,-1) -- (6,-0.25) -- 
		(6,-1.75) -- (3,-1);
		
		\node[draw=none, rectangle] at (0.5, .25) {$\playBis$};
		\node[draw=none, rectangle] at (2, .8) {$\trans_0, \delay_0$};
		\node[draw=none, rectangle] at (2.7, .45) {$p$};
		\node[draw=none, rectangle] at (2, -.8) {$\trans'_0, \delay'_0$};
		\node[draw=none, rectangle] at (3, -.45) {$1-p$};
		\node[draw=none, rectangle] at (7, -1) {
			\textcolor{blue}{$S^{k-1}_{\playBis\extendto{\trans'_0, \delay'_0}}$}};
		\node[draw=none, rectangle] at (8.3, 1) {
			$\Cyl_{\playBis}(\ppathBis\trans_0, 
				\Rplus^{|\ppathBis|}\times \{\delay_0\})$};
		\end{tikzpicture}
		\caption{The partition of $S^k_{\playBis}$ when 
			$\playBis \in \FPlaysMin$.}
		\label{fig:Sk-partition}
	\end{figure}
	
	\begin{itemize}
		\item If $\playBis \in\FPlaysMin$ and $\minstrategy^p(\playBis) = p 
		\Dirac{(\trans_0, \delay_0)} + (1-p) \Dirac{(\trans'_0, \delay'_0)}$ with 
		$(\trans_0,\delay_0) \neq (\trans'_0,\delay'_0)$, we~have 
		(as illustrated in \figurename{~\ref{fig:Sk-partition}})
		\[S_{\playBis}^{k} = \Cyl_{\playBis}(\ppathBis\trans_0, 
		\Rplus^{|\ppathBis|}\times \{\delay_0\}) \uplus 
		S^{k-1}_{\playBis\extendto{\trans'_0,\delay'_0}} \,.\]
		As before, we have $\Proba^{\minstrategy^p, 
			\maxstrategy}_{\playBis}(\Cyl_{\playBis}(\ppathBis\trans_0, 
		\Rplus^{|\ppathBis|}\times \{\delay_0\})) = p$. We show that 
		\begin{equation}
			\label{eq:jolijoli}
			\Proba^{\minstrategy^p, \maxstrategy}_{\playBis}(S^{k-1}_{\playBis 
				\extendto{\trans'_0,\delay'_0}}) = 
			(1-p) \Proba^{\minstrategy^p, \maxstrategy}_{\playBis 
				\extendto{\trans'_0,\delay'_0}}(S^{k-1}_{\playBis 
				\extendto{\trans'_0,\delay'_0}}) 
		\end{equation}
		and, by induction hypothesis, we deduce that 
		$\Proba^{\minstrategy^p, \maxstrategy}_{\playBis 
			\extendto{\trans'_0,\delay'_0}}(S^{k-1}_{\playBis 
			\extendto{\trans'_0,\delay'_0}}) \leq 1-(1-p)^{k-1}$.
		In conclusion, we obtain 
		\begin{align*}
		\Proba^{\minstrategy^p, \maxstrategy}_{\playBis}(S_{\playBis}^{k}) &= 
		p + (1-p) \Proba^{\minstrategy^p, \maxstrategy}_{\playBis 
			\extendto{\trans'_0,\delay'_0}}(S^{k-1}_{\playBis 
			\extendto{\trans'_0,\delay'_0}}) \\
		&\leq p + (1-p)(1-(1-p)^{k-1}) = 1-(1-p)^k \,.
		\end{align*}
		
		Now, we show~\eqref{eq:jolijoli}: to do it, we must come back to the 
		definition of the probabilities with respect to cylinders. First, 
		$S^{k-1}_{\playBis\extendto{\trans'_0,\delay'_0}}$ can be decomposed a 
		union of cylinders over some paths $\ppathBis\trans'_0\ppath$ with 
		$|\ppath|=k-1$. For such a path $\ppath$, let 
		$C_\ppath$ be the set of sequences of delays of plays appearing in 
		$S^{k-1}_{\playBis\extendto{\trans'_0,\delay'_0}}$ so that 
		\[S^{k-1}_{\playBis\extendto{\trans'_0,\delay'_0}} =
		\bigcup_{|\ppath|=k-1} \Cyl_{\playBis}(\ppathBis\trans'_0\ppath, C_\ppath)=
		\bigcup_{|\ppath|=k-1} \Cyl_{\playBis 
			\extendto{\trans'_0,\delay'_0}}(\ppathBis\trans'_0\ppath, C_\ppath) 
		\,. \] 
		The set $C_\ppath$ is Lebesgue-measurable since it can be obtained as a
		finite unions and intersections of the constraints appearing in
		$\minstrategydelay^p$ (and thus in $\detminstrategy_1$ or
		$\detminstrategy_2$) or $\maxstrategydelay$, that are all
		Lebesgue-measurable by Hypothesis~\ref{hyp:measurability} 
		(the explanation of this decomposition is given in 
		Appendix~\ref{app:eta-proper_cylinders}). Thus,
		\begin{align*}
			\Proba^{\minstrategy^p, \maxstrategy}_{\playBis}(S^{k-1}_{\playBis 
				\extendto{\trans'_0,\delay'_0}}) 
			&= \sum_{|\ppath|=k-1}\Proba^{\minstrategy^p, 
				\maxstrategy}_{\playBis}(\Cyl_{\playBis}(\ppathBis 
			\trans'_0\ppath, C_\ppath)) \\
			&= \sum_{|\ppath|=k-1} \minstrategy^p(\playBis)(\trans'_0, \delay'_0) 
			\times \Proba^{\minstrategy^p, \maxstrategy}_{\playBis 
				\extendto{\trans'_0,\delay'_0}}(\Cyl_{\playBis 
				\extendto{\trans'_0,\delay'_0}}(\ppathBis\trans'_0\ppath, 
			C_\ppath)) \\
			&= (1-p) \sum_{|\ppath|=k-1} \Proba^{\minstrategy^p, 
				\maxstrategy}_{\playBis 
				\extendto{\trans'_0,\delay'_0}}(\Cyl_{\playBis 
				\extendto{\trans'_0,\delay'_0}}(\ppathBis\trans'_0\ppath, 
			C_\ppath)) \\
			&= (1-p) \Proba^{\minstrategy^p, 
				\maxstrategy}_{\playBis 
				\extendto{\trans'_0,\delay'_0}}(S^{k-1}_{\playBis 
				\extendto{\trans'_0,\delay'_0}}) 
		\end{align*}

		\item If $\playBis \in\FPlaysMin$ and $\minstrategy^p(\playBis) = 
		\Dirac{\trans_0,\delay_0}$, then $S_{\playBis}^{k} = 
		S^{k-1}_{\playBis\extendto{\trans_0,\delay_0}}$. As 
		in~\eqref{eq:jolijoli}, we can obtain that 
		\[\Proba^{\minstrategy^p, \maxstrategy}_{\playBis}(S^{k-1}_{\playBis 
			\extendto{\trans_0,\delay_0}}) = 
		\minstrategy^p(\playBis)(\trans_0, \delay_0) \times 
		\Proba^{\minstrategy^p, \maxstrategy}_{\playBis 
			\extendto{\trans_0,\delay_0}}(S^{k-1}_{\playBis 
			\extendto{\trans_0,\delay_0}}) \,. \]
		Then, we have directly 
		\[\Proba^{\minstrategy^p, \maxstrategy}_{\playBis}(S_{\playBis}^{k}) = 
		\Proba^{\minstrategy^p, \maxstrategy}_{\playBis 
			\extendto{\trans_0,\delay_0}}(S^{k-1}_{\playBis 
			\extendto{\trans_0,\delay_0}}) \leq 1-(1-p)^{k-1} \leq 1-(1-p)^k 
		\,.\]

		\item If $\playBis\in\FPlaysMax$, then 
	 	\[S_{\playBis}^{k} = 
	 	\bigcup_{(\trans_0,\delay_0)\in\support{\maxstrategy(\playBis)}} 
	 	S^{k-1}_{\playBis\extendto{\trans_0,\delay_0}}\,.\] 
	 	Again by decomposing on cylinders, we have that 
		\[\Proba^{\minstrategy^p, \maxstrategy}_{\playBis}(S_{\playBis}^{k})  = 
		\sum_{\trans_0}\int_{t_0} \maxstrategytrans(\playBis)(\trans_0) 
		\Proba^{\minstrategy^p, \maxstrategy}_{\playBis 
			\extendto{\trans_0,\delay_0}}(S^{k-1}_{\playBis 
			\extendto{\trans_0,\delay_0}}) 
		\mathrm{d}\maxstrategydelay(\playBis, \trans_0)(\delay_0) \,.\]
		By induction hypothesis, we obtain
		\begin{align*}
		\Proba^{\minstrategy^p, \maxstrategy}_{\playBis}(S_{\playBis}^{k}) &\leq 
		(1 - (1-p)^{k-1})\sum_{\trans_0}\int_{t_0} 
		\maxstrategytrans(\playBis)(\trans_0) 
		\mathrm{d}\maxstrategydelay(\playBis, \trans_0)(\delay_0) \\
		&\leq 1 - (1-p)^{k-1}\leq 1-(1-p)^k \,. \qedhere
		\end{align*}
	\end{itemize}
\end{proof}
\noindent 
To show that $\muppervalue_{\loc,\val}\leq \dValue_{\loc,\val}$ for all
$(\loc,\val)$, we prove that $\mValue^{\minstrategy^p}_{\loc,\val} \leq
\max(-N, \dValue_{\loc, \nu}) + 3\varepsilon$ for all $(\loc,\val)$, for $p$
\emph{close enough to} $1$: we conclude by taking the limit when $N$ tends to
$+\infty$ and $\varepsilon$ tends to $0$.  We get that inequality by showing
the following result, paired with the fact that
$\dValue^{\detminstrategy}_{\ell, \nu} \leq \max(-N, \dValue_{\ell, \nu}) +
\varepsilon$.

\begin{prop}
	\label{prop:partition}
	For all configurations $(\loc, \val)$ and $\varepsilon > 0$ small
	enough, there exists $\tilde p\in (0,1)$ so that for all
	$p\in [\tilde p,1)$,
	\begin{displaymath}
		\mValue^{\minstrategy^p}_{\ell, \nu}\leq
		\dValue^{\detminstrategy}_{\ell, \nu} + 2\varepsilon.
	\end{displaymath}
\end{prop}
\begin{proof}
	Since $\minstrategy^p$ is memoryless and proper (by
	Lemma~\ref{lem:eta_properties}), we can use
	Lemma~\ref{lem:best-response} (with $\varepsilon/2$), to obtain the
	existence of a (smooth) deterministic strategy $\detmaxstrategy\in\sdStratMax$ such
	that for all configurations $(\loc,\val)$,
	$\E^{\minstrategy^p,\detmaxstrategy}_{\loc,\val} \geq
	\mValue^{\minstrategy^p}_{\loc,\val}-\varepsilon/2$. The proof then
	consists in computing  the bound $\tilde p$ on the probabilities such that
	for all $p\in [\tilde p,1)$, $\E^{\minstrategy^p, \detmaxstrategy}_{\loc,
	\val} \leq \dValue^{\detminstrategy}_{\loc, \val} + 3\varepsilon/2$. We
	conclude by combining both inequalities. 

	Notice that $\minstrategy^p$ and $\detmaxstrategy$ are both discrete
	strategies so that we can apply Lemma~\ref{lem:expectation_discrete}: for
	all $(\loc, \val)$, 
	\begin{displaymath}
	\E^{\minstrategy^p, \detmaxstrategy}_{\loc, \val} = 
	\sum_{\play \in \TPlays^{\minstrategy^p, \detmaxstrategy}_{\loc, \val}} 
	\weight(\play) \times 
	\Proba^{\minstrategy^p, \detmaxstrategy}_{\loc, \val}(\play)
	\end{displaymath}
	where $\TPlays^{\minstrategy^p, \detmaxstrategy}_{\loc, \val}$ is the
	$\Proba^{\minstrategy^p,\detmaxstrategy}_{\loc, \val}$-measurable set of
	plays starting in $(\loc, \val)$, conforming to $\minstrategy^p$ and
	$\maxstrategy$ and ending in the target.

	The case where the whole region game only contains positive SCCs is
	easy, since then $\minstrategy^p$ chooses the transition and delay
	given by $\detminstrategy^1$ with probability 1. By divergence,
	$\game$ then contains no negative cycles. A play conforming to
	$\minstrategy^p$ is also conforming to the deterministic strategy
	$\detminstrategy^1$, so it must be acyclic. In particular, there
	exists only one play \play conforming to $\minstrategy^p$ and
	$\detmaxstrategy$. This one is also conforming to $\detminstrategy$
	and thus reaches the target with a cumulated weight
	$\weightC(\play)=\E^{\minstrategy^p, \detmaxstrategy}_{\loc, \val}
	\leq \dValue^\detminstrategy_{\loc, \val}$ as expected.
	
	Now, suppose that the region graph contains at least a negative
	SCC. Thus, we let $c > 0$ be the maximal
	size of an elementary cycle of the region game (that visits each pair
	$(\loc, r)$ at most once) and $w^- > 0$ be the opposite of the maximal
	cumulated weight of an elementary negative cycle in $\rgame$
	(necessarily bounded by $\wmax^e\, |\rgame|$).
	
	We partition the set $\TPlays^{\minstrategy^p, \detmaxstrategy}_{\loc,
	\val}$ into subsets $\Pi_{i, j}$ of plays according to the number $i$ of edges $(\loc, \val) \moveto{\trans, \delay} (\loc',
	\val')$ chosen by $\MinPl$ with a probability $\minstrategy^p(\loc, \val)(\trans, \delay) = 1-p$, and their
	length $j$ (we always have $i \leq j$). The partition is depicted in
	\figurename{~\ref{fig:partition}}:
	\begin{itemize}
		\item $\Pi_{\N, \geq K}$, depicted in blue, contains all plays with a
		length greater than $K$ (the switching threshold)
		\item $\Pi_{0, \leq K}$, depicted in yellow, contains all plays
		without edges of probability $1-p$, with a length at most $K$;
		\item $\widetilde\Pi$, depicted in red, contains the rest of the
		plays.
	\end{itemize}
	
	\begin{figure}[tbp]
		\centering
		\begin{tikzpicture}[xscale=.7,yscale=.3]
		\draw[fill=blue!60!black,draw=none] (0,0)--(9.6, 9.6)--(0,9.6)--(0,0);
		
		\draw[fill=red, draw=none] (0,0)--(7,7)--(0,7)--(0,0);
		
		\draw[fill=yellow!70!orange,rounded corners,draw=none] 
			(-.15,0) rectangle (.15,7);
		\draw[fill=white,draw=none] (-.2,9.6) rectangle (.2,10.6);
		
		\draw[fill=white,draw=none] (8,8) rectangle (10,10);
		
		\draw[->,very thick] (0,0) -- (8.5,0) node[right](){$i$};
		
		\draw[->,very thick] (0,0) -- (0,11) node[above](){$j$};
		\node at (-.5,7)(){$K$};
		
		\node at (-.9,3) () {$\Pi_{0,\leq K}$};
		\node[white] at (4.25,8) () {$\Pi_{\N,\geq K}$};
		\node[white] at (2.5,5) () {$\widetilde\Pi$};
		\end{tikzpicture}
		\caption{Partition of the plays $\TPlays^{\minstrategy^p,
		\detmaxstrategy}_{\loc, \val}$}
		\label{fig:partition}
	\end{figure}
	\noindent 
	We have
	\begin{equation}
	\label{eq:decomposition-expectation}
	\E_{\loc, \val}^{\minstrategy^p, \detmaxstrategy} 
	= \underbrace{\sum_{\play \in \Pi_{0, \leq K}} \weight(\play)
		\Proba^{\minstrategy^p, \detmaxstrategy}_{\loc,
			\val}(\play)}_{\gamma_{0, \leq K}} + \underbrace{\sum_{\play \in
			\Pi_{\N, \geq K}} \weight(\play) \Proba^{\minstrategy^p,
			\detmaxstrategy}_{\loc, \val}(\play)}_{\gamma_{\N, \geq K} }
	+\underbrace{\sum_{\play \in \widetilde{\Pi}} \weight(\play)
		\Proba^{\minstrategy^p, \detmaxstrategy}_{\loc,
			\val}(\play)}_{\widetilde{\gamma}}
	\end{equation}
	We now compute and bound the three expectations $\gamma_{0, \leq K}$,
	$\gamma_{\N, \geq K}$ and $\widetilde{\gamma}$. 
	
	\paragraph*{\texorpdfstring{The red zone is such that $\widetilde{\gamma}\leq \varepsilon/4$}{Red zone}}
	All plays in $\widetilde{\Pi}$ have a length at most $K$: so 
	the cumulated weight of all such plays is at most $K\wmax^e$. 
	So, we have
	\begin{displaymath}
	\widetilde{\gamma} 
	= \sum_{\play \in \widetilde{\Pi}} \weight(\play) 
	\Proba_{\loc, \val}^{\minstrategy^p, \detmaxstrategy}(\play) 
	\leq \sum_{\play \in \widetilde{\Pi}} K\wmax^e 
	\Proba_{\loc, \val}^{\minstrategy^p, \detmaxstrategy}(\play) 
	= K\wmax^e 
	\Proba_{\loc, \val}^{\minstrategy^p, \detmaxstrategy}(\widetilde{\Pi}).
	\end{displaymath}
	But all plays $\play\in \bigcup_{j\leq K} \Pi_{i, j}$ with $i \leq K$ take
	$i$ edges of probability $1-p$. In particular, by bounding all other
	probabilities by 1, and since there are at most $2^K$ plays in
	$\bigcup_{j\leq K} \Pi_{i, j}$ (since $\detmaxstrategy$ is deterministic
	and the support of all $\minstrategy^p(\loc,\val)$ contains at most two
	edges), we obtain (using that $1 - (1-p)^K \leq 1$)
	\begin{align}
	\Proba^{\minstrategy^p, \detmaxstrategy}_{\loc, \val}(\widetilde{\Pi})
	&\leq 2^K\sum_{i=1}^{K-1} (1-p)^i = 2^K\frac{(1-p)(1-(1-p)^K)}p \leq
	2^K \frac{1-p}{p}      \longrightarrow_{p\to 1} 0
	\label{eq:proba_red}
	\end{align}
	If we suppose that
	\begin{displaymath}
	p \geq \frac{1}{1+\varepsilon/(4\times 2^K K\wmax^e)}
	\end{displaymath}
	we obtain as desired
	\begin{displaymath}
	\widetilde{\gamma} \leq 2^K K\wmax^e\frac{1-p}p \leq \frac{\varepsilon}{4}.
	\end{displaymath}
	
	\paragraph*{\texorpdfstring{The yellow and blue zones are such that
			$\gamma_{0, \leq K} + \gamma_{\N, \geq K} \leq
			\dValue^{\detminstrategy}_{\ell, \nu} + 5\varepsilon/4$}{Yellow and blue zones}}
	
	All plays in $\Pi_{0, \leq K}$ reach the target without taking any edge of
	probability $1 - p$, so they are conforming to $\detminstrategy^1$. In the
	case where $\dValue_{\loc, \val} = - \infty$, $\Pi_{0, \leq K} = \emptyset$
	and $\gamma_{0, \leq K} = 0$, since no play conforming to
	$\detminstrategy^1$ from $(\loc, \val)$ reaches the target. In this case,
	\MinPl can stay in a cycle with a negative cumulated weight as long as 
	wanted. Now, if $\dValue_{\loc, \val}$ is finite, by
	Lemma~\ref{lem:fake-optimally}, the cumulated weight of a play in $\Pi_{0,
	\leq K}$ is at most $\dValue_{\loc, \val} + K\varepsilon/K = \dValue_{\loc,
	\val} + \varepsilon$. As $\dValue_{\loc, \val} = \inf_{\detminstrategy \in
	\dStratMin} \dValue^{\detminstrategy}_{\loc, \val} \leq
	\dValue^{\detminstrategy}_{\loc, \val}$, in both cases, we can
	write
	\[\gamma_{0, \leq K} 
	\leq \left(\dValue^{\detminstrategy}_{\ell, \nu} +
	\varepsilon \right)
	\Proba^{\minstrategy^p,\detmaxstrategy}_{\loc, \val}(\Pi_{0, \leq
		K}).\]
	
	Let \play be a play in $\Pi_{i, j}$ with $0 \leq i$ and $j \geq K$.
	Since $\minstrategy^p$ only allows cycles in negative SCCs, all region
	cycles in \play have a cumulated weight at most $-1$. By definition
	of $K$ and the proof of Theorem~\ref{thm:switching-divergent}, 
	$\weight(\play) \leq \dValue^{\detminstrategy}_{\ell, \nu}\leq
	\dValue^{\detminstrategy}_{\ell, \nu} + \varepsilon$. 
	
	By summing up the contribution of yellow and blue zones, we get
	\begin{equation}
	\label{eq:expectation_yelow-blue}
	\gamma_{0, \leq K} + \gamma_{\N, \geq K} 
	\leq \left(\dValue^{\detminstrategy}_{\ell, \nu} + \varepsilon \right) 
	\Proba^{\minstrategy^p,\detmaxstrategy}_{\loc, \val}(\Pi_{0, \leq K} \cup 
	\Pi_{\N, \geq K}).
	\end{equation}
	To conclude, we distinguish two cases. First, we suppose that  
	$\dValue^{\detminstrategy}_{\loc, \val} \geq -5\varepsilon/4$. By having
	$\Proba^{\minstrategy^p, \detmaxstrategy}_{\loc, \val}(\Pi_{0, \leq K} \cup 
	\Pi_{\N, \geq K}) \leq 1$, we get
	\begin{displaymath}
	\gamma_{0, \leq K} + \gamma_{\N, \geq K} \leq 
	\left(\dValue^{\detminstrategy}_{\ell, \nu} + \frac{5\varepsilon}{4} \right) 
	\Proba^{\minstrategy^p,\detmaxstrategy}_{\loc, \val}(\Pi_{0, \leq K}\cup 
	\Pi_{\N, \geq K})\leq \dValue^{\detminstrategy}_{\loc, \val} + \frac{5\varepsilon}{4}.
	\end{displaymath}
	Otherwise $\dValue^{\detminstrategy}_{\loc, \val} < 
	-5\varepsilon/4$, and then, by~\eqref{eq:proba_red}, 
	\[
	\Proba^{\minstrategy^p, \detmaxstrategy}_{\loc, \val}(\Pi_{0, \leq K} 
	\cup\Pi_{\N, \geq K}) 
	= 1 -
	\Proba^{\minstrategy^p, \detmaxstrategy}_{\loc,
		\val}(\widetilde{\Pi}) \geq 1 - 2^K\frac{1-p}{p}
	\longrightarrow_{p\to 1} 1.
	\]
	If we suppose that
	\begin{displaymath}
	p \geq \frac {2^K} {2^K+1-\frac{\dValue^{\detminstrategy}_{\ell, \nu} + 
			5\varepsilon/4}{\dValue^{\detminstrategy}_{\ell, \nu} + 
			\varepsilon}}\in(0,1)
	\end{displaymath}
	then
	\[
	\Proba^{\minstrategy^p, \detmaxstrategy}_{\loc, \val}(\Pi_{0, \leq
		K}\cup\Pi_{\N, \geq K}) \geq
	\frac{\dValue^{\detminstrategy}_{\ell, \nu} +
		5\varepsilon/4}{\dValue^{\detminstrategy}_{\ell, \nu} +
		\varepsilon}\] and, by negativity of
	$\dValue^{\detminstrategy}_{\loc, \val} +\varepsilon$, we can
	rewrite~\eqref{eq:expectation_yelow-blue} as
	\begin{displaymath}
	\gamma_{0, \leq K} + \gamma_{\N, \geq K} \leq 
	\left(\dValue^{\detminstrategy}_{\ell, \nu} + 
	\varepsilon \right) \frac{\dValue^{\detminstrategy}_{\ell, \nu} +
		5\varepsilon/4}{\dValue^{\detminstrategy}_{\ell, \nu} +
		\varepsilon} = \dValue^{\detminstrategy}_{\ell, \nu} +
	\frac{5\varepsilon}4.
	\end{displaymath}
	In all cases, we have $\gamma_{0, \leq K}+\gamma_{\N, \geq K} \leq 
	\dValue^{\detminstrategy}_{\loc, \val} + 5\varepsilon/4$.
	
	\paragraph*{Gathering all constraints on $p$}
	We gather all the lower bounds over $p$ that we need in the proof, letting
	\[\tilde p =
	\begin{cases}
	\max\left(\frac{1}{1+\varepsilon/(4\times 2^K K\wmax^e)}, \frac 1
	2\right) & \text{if } \dValue^{\detminstrategy}_{\loc,\val}\geq
	-5\varepsilon/4  \\
	\max\left(\frac{1}{1+\varepsilon/(4\times 2^K K\wmax^e)}, 
	\frac{1}{2},  
	\frac {2^K} {2^K+1-\frac{\dValue^{\detminstrategy}_{\ell, \nu} + 
			5\varepsilon/4}{\dValue^{\detminstrategy}_{\ell, \nu} + 
			\varepsilon}}
	\right)
	& \text{otherwise}
	\end{cases}\] Then, for $p\in(\tilde p,1)$, we have
	$\E^{\minstrategy^{p},\detmaxstrategy}_{\loc, \val} \leq
	\dValue^\detminstrategy_{\loc, \val} + 3\varepsilon/2$. Since
	$\tilde p$ does not depend on $\detmaxstrategy$, we conclude that for
	$p\in(\tilde p,1)$, we have
	$\mValue^{\minstrategy^{p}}_{\loc, \val} \leq
	\dValue^\detminstrategy_{\loc, \val} + 2\varepsilon$.
\end{proof}

\section{The untimed setting: shortest-path games} 
\label{sec:SPG}

In this section, we restrict ourselves to \WTG{s} that use no clocks. These are
generally called shortest-path games, and their
deterministic value have previously been
studied~\cite{BrihayeGeeraertsHaddadMonmege-17,BusattoGastonMonmegeReynier-17}:
both players have optimal strategies, $\MaxPl$ does not require memory, while
$\MinPl$ requires pseudo-polynomial size memory. Since there are no clocks, we
simply remove from transitions the component regarding the guards and clock
resets. We also only provide the weight of locations (usually called vertices
or states, in this setting), and not of transitions. To summarise, here is a
self-contained definition of the games we study in this section. 

\begin{defi}
	A \emph{shortest-path game} (SPG) is a tuple $\arenaEx$ where
	$\vertices:=\maxvertices\uplus \minvertices\uplus \finalvertices$ is a
	finite set of vertices partitioned into the sets $\maxvertices$ and
	$\minvertices$ of $\MaxPl$ and $\MinPl$ respectively, and a set
	$\finalvertices$ of target vertices,
	$\edges\subseteq \vertices\setminus\finalvertices\times \vertices$ is
	a set of \emph{directed edges}, and $\edgeweights\colon \edges \to \Z$
	is the \emph{weight function}, associating an integer weight with each
	edge.
\end{defi}

Since \SPG{s} are special cases of \WTG{s}, the definition of stochastic 
strategies and values from \WTG{s} can be adapted. In this context, all  
distributions over transitions are finite, and thus all strategies are smooth. Moreover, all strategies are
discrete, and if the strategies of \MinPl are supposed to be proper, 
the various values exist and
Lemma~\ref{lem:expectation_discrete} applies. 
In particular, smooth deterministic values are defined only by a restriction 
over strategies of \MinPl, i.e.~$\sdStratMax = \dStratMax$).


\subsection{Memoryless value}

In \SPG{s}, we are able to extend the result of
Theorem~\ref{thm:divergent-equalvalue} without imposing the divergence condition: 

\begin{thm}\label{thm:shortest-path}
	For all \SPG{s} $\game$, for all vertices $\loc$, we have 
	\begin{displaymath}
	\dValue_\loc = \muppervalue_\loc= \mlowervalue_\loc\,.	
	\end{displaymath} 
\end{thm}
\noindent 
The proof strategy follows the same path as in the divergent case: 
\begin{itemize}
\item we show in Lemma~\ref{lem:SPG-detVal<lowerVal} that $\dValue_{\loc} \leq \mlowervalue_{\loc}$, by adapting the proof of Lemma~\ref{lem:divergent-detVal<lowerVal} relying on the existence of an optimal deterministic and memoryless strategy of $\MaxPl$ against the deterministic value \cite{BrihayeGeeraertsHaddadMonmege-17};
\item  we show in Lemma~\ref{lem:partition-SPG} that
memoryless stochastic strategies can emulate deterministic ones: $\muppervalue_{\loc} \leq \dValue_{\loc}$ (with similar but more involved techniques as in Lemma~\ref{lem:partition}). 
\end{itemize}
Combined with the fact that $\mlowervalue_{\loc} \leq \muppervalue_{\loc}$, we indeed obtain the desired equalities. 

We start by adapting the proof of Lemma~\ref{lem:divergent-detVal<lowerVal}: 
\begin{lem}
	\label{lem:SPG-detVal<lowerVal}
	In all \SPG{s} \game, for all vertices $\loc$, $\dValue_{\loc} \leq \mlowervalue_{\loc}$. 
\end{lem}
\begin{proof} By \cite{BrihayeGeeraertsHaddadMonmege-17}, \MaxPl
has a memoryless and deterministic optimal strategy \detmaxstrategy for the deterministic value. 
Against any deterministic strategy of
$\MinPl$, from vertex $\loc$, it guarantees a weight at least $\dValue_{\loc}$. 

The rest of the proof is entirely similar, but reproduced here for ease. 
Let \minstrategy be any memoryless proper strategy of \MinPl. Every play
conforming to $\minstrategy$ and $\detmaxstrategy$ has a weight at least
$\dValue_{\loc, \val}$ (since it is conforming to
$\detmaxstrategy$), so that, by definition of the expectation,
\[\E^{\minstrategy, \detmaxstrategy}_{\loc, \val} = \int_{\play}
\weight(\play)\mathrm{d}\Proba_{\loc,\val}^{\minstrategy,\detmaxstrategy}(\play)
\geq \dValue_{\loc, \val}\,\] Notice that this integral is indeed a finite sum here, since the strategies are necessarily discrete. Since the inequality holds for all
proper strategies of \MinPl, we deduce that 
	\begin{displaymath}
	\inf_{\minstrategy \in \mStratMinProper} 
	\E^{\minstrategy, \detmaxstrategy}_{\loc} \geq \dValue_{\loc}\,.
	\end{displaymath} 
	By taking the supremum over all memoryless strategies of
	$\MaxPl$ (that contain the memoryless and deterministic strategy
	\detmaxstrategy), we have
	\begin{displaymath}
	\sup_{\maxstrategy \in \mStratMax} \inf_{\minstrategy \in \mStratMinProper} 
	\E^{\minstrategy, \maxstrategy}_{\loc} \geq \dValue_{\loc} \,.
	\end{displaymath} 
	In particular, $\mlowervalue_{\loc} \geq \dValue_{\loc}$, as expected. 
\end{proof}
\noindent 
The rest of this section is devoted to the proof of the final piece of the proof: 
\begin{lem}\label{lem:partition-SPG}
In all \SPG{s} \game, for all vertices $\loc$,
$\muppervalue_{\loc} \leq \dValue_{\loc}$. 
\end{lem}
\noindent 
In an SPG, once we fix a memoryless strategy $\minstrategy\in\mStrat_\MinPl$,
we obtain a \emph{Markov decision process} (MDP) where the other player must
still choose how to react. An MDP is a tuple $\langle \vertices, A, P\rangle$ 
where $\vertices$
is a set of vertices, $A$ is a set of actions, and $P\colon \vertices\times A\to \Distr
\vertices$ is a partial function mapping to some pair of vertices and actions a
distribution of probabilities over the successor vertices. In our context, we
let $\MDP$ be the MDP with the same set $\vertices$ of vertices as $\game$, actions
$A=\vertices\cup\{\bot\}$ being either successor vertices of the game or an additional
action $\bot$ denoting the random choice of $\minstrategy$, and a probability
distribution $P$ defined by:
\begin{itemize}
	\item if $\loc\in \maxvertices$, $P(\loc,\loc')$ is only defined if
	$(\loc,\loc')\in\edges$ in which case $P(\loc,\loc')=\Dirac {\loc'}$, and
	$P(\loc,\bot)$ is also undefined;
	\item if $\loc\in \minvertices$, $P(\loc,\bot)=\minstrategy(\loc)$, and
	$P(\loc,\loc')$ is undefined for all $\loc'\in\vertices$.
\end{itemize}
In drawings of MDPs (and also of Markov chains, later), we show
weights as trivially transferred from the game graph.

\begin{exa}
	In \figurename~\ref{fig:SP1}, an SPG is presented on
	the left, with the MDP in the middle obtained by picking as a
	memoryless strategy for $\MinPl$ the one choosing to go
	to~$\loc_\MaxPl$ with probability $p\in(0,1)$ and to the target vertex
	with probability $1-p$. Another more complex example is given in
	\figurename~\ref{fig:SP2} where the memoryless strategy for $\MinPl$
	consists, in vertex~$\loc_1$, to choose successor $\loc_0$ with
	probability $p\in(0,1)$ and successor $\loc_2$ with probability~$1-p$,
	and in vertex $\loc_3$, to choose successor $\loc_1$ with the same
	probability $p$ and the target vertex with probability $1-p$.
\end{exa}

In such an MDP, when player $\MaxPl$ has chosen a strategy, there
will remain no ``choices'' to make, and we will thus end up in a
\emph{Markov chain}. A Markov chain (\emph{MC}) is a tuple
$\mathcal M= \langle \vertices, P\rangle$ where $\vertices$ is a set of vertices, and
$P\colon \vertices\to \Distr \vertices$ associates to each vertex a distribution of
probabilities over the successor vertices. In our context, for all
memoryless strategies $\maxstrategy\in\mStrat_\MaxPl$, we
let $\MC$ the MC obtained from the MDP $\MDP$ by following strategy
$\maxstrategy$ and action $\bot$. Formally, it consists of the
same set $\vertices$ of vertices as $\game$, and mapping $P$ associating to a
vertex $\loc\in\minvertices$, $P(\loc)=\minstrategy(\loc)$ and to a
vertex $\loc\in\maxvertices$, $P(\loc)=\maxstrategy(\loc)$.

\begin{exa}
	On the right of \figurename~\ref{fig:SP1} is depicted the MC
	obtained when $\MaxPl$ decides to go to $\loc_\MinPl$ with probability
	$q\in[0,1]$ and to the target vertex with probability $1-q$.
\end{exa}

By \cite[Section~10.5.1]{BaierKatoen}, the value
	$\mValue^{\minstrategy}_\loc$ is finite if and only if strategy
	$\minstrategy$ ensures the reachability of a target vertex with probability
	1, no matter how the opponent plays. Thus, for memoryless strategies, Hypothesis~\ref{hyp:proper} can be restricted to only checking that for all vertices $\loc_0$ and strategies $\maxstrategy\in\StratMax$, $\Proba^{\minstrategy, \maxstrategy}_{\loc_0}(\TPlays) = 1$.
From Lemma~\ref{lem:Bellman}, we obtain a Bellman-like equation as follows, once  simplified since we restrict ourselves to memoryless strategies:
\begin{equation}
\label{eq:SPG-Bellman}
\E^{\minstrategy,\maxstrategy}_{\loc}  =
\begin{cases}
0 & \text{if }  \loc \in \finalvertices\\
\sum_{(\loc,\loc')\in \edges} \minstrategy(\loc,\loc') \times (\edgeweights(\loc,\loc') + 
\E^{\minstrategy,\maxstrategy}_{\loc'})
& \text{if } v \in\minvertices\setminus \finalvertices\\
\sum_{(\loc,\loc')\in \edges} \maxstrategy(\loc,\loc') \times (\edgeweights(\loc,\loc') + 
\E^{\minstrategy,\maxstrategy}_{\loc'})
& \text{if } \loc \in\maxvertices\setminus \finalvertices
\end{cases}
\end{equation}

\begin{exa}\label{example:for-intro}
	For the game of \figurename~\ref{fig:SP1}, we let $\minstrategy$ and
	$\maxstrategy$ the memoryless strategies that result in the MC on
	the right. Letting
	$x=\E_{\loc_\MinPl}^{\minstrategy,\maxstrategy}$ and
	$y=\E_{\loc_\MaxPl}^{\minstrategy,\maxstrategy}$, the
	system \eqref{eq:SPG-Bellman} rewrites~as
	$x = (1-p)\times 1 + p \times y$ and $y =
	q\times (-1 + x) + (1-q) \times (-10)$.
	We thus have $x=p(9q-10)/(1-pq)$. Two cases happen, depending on the
	value of $p$: if $p<9/10$, then $\MaxPl$ maximises $x$ by choosing
	$q=1$, while choosing $q=0$ when $p\geq 9/10$. In all cases,
	player $\MaxPl$ will therefore play deterministically: if $p<9/10$,
	the expected payoff from $\loc_\MinPl$ will then be
	$\mValue^{\minstrategy}(\loc_\MinPl)=-p/(1-p)$; if
	$p\geq 9/10$, it will be
	$\mValue^{\minstrategy}(\loc_\MinPl)=-10p$. This value is
	always greater than the optimum $-10$ that $\MinPl$ were able to
	achieve with memory, since we must keep $1-p>0$ to ensure reaching
	the target with probability $1$. We thus obtain
	$\muppervalue(\loc_\MinPl)= \muppervalue(\loc_\MaxPl)=-10$
	as before. There are no optimal strategies for $\MinPl$, but an
	$\varepsilon$-optimal one consisting in choosing probability
	$p\geq 1-\varepsilon/10$.
\end{exa}

The fact that $\MaxPl$ can play optimally with a deterministic
strategy in the MDP $\game^{\minstrategy}$ is not specific to
this example. Indeed, in an MDP $\game^{\minstrategy}$ such that
$\Proba^{\minstrategy,\maxstrategy}_{\loc} (\TPlays)=1$
for all $\maxstrategy$, $\MaxPl$ cannot avoid reaching the
target, and must then ensure the most expensive play
possible. Considering the MDP $\tilde{\game}^{\minstrategy}$
obtained by multiplying all the weights in the graph by $-1$, the
objective of $\MaxPl$ becomes a shortest-path objective. We can then
deduce from \cite{BertsekasTsitsiklis91} that \MaxPl has an optimal
deterministic memoryless strategy: the same applies in the original
MDP $\game^{\minstrategy}$.

\begin{prop}
	\label{prop:strat_opti_Eve}
	In the MDP $\game^{\minstrategy}$ such that
	$\Proba^{\minstrategy,\maxstrategy}_{\loc} (\TPlays)=1$
	for all $\maxstrategy$, $\MaxPl$ has an optimal deterministic
	memoryless strategy.
\end{prop}

\begin{figure}[tbp]
	\centering
	\begin{tikzpicture}[xscale=.8,yscale=.8]
	\node[PlayerMax] (s0){$\loc_0$}; 
	\node[PlayerMin] at (2, 0) (s1){$\loc_1$}; 
	\node[PlayerMax] at (4, -2.5) (s2){$\loc_2$};
	\node[PlayerMin] at (4, -1) (s3){$\loc_3$};
	\node[target] at (6, 0) (s4) {\LARGE $\smiley$};
	
	\draw[->]
	(s1)  edge [auto = left, bend left] node {$\mathbf{0}$} (s0)
	(s0) edge [auto = left] node {$\mathbf{-1}$} (s1)
	(s0) edge [auto = left, bend left=50,looseness=.7] node {$\mathbf{-10}$} (s4)
	(s1) edge [auto = right,bend right=10] node {$\mathbf{1}$} (s2)
	(s2) edge [auto = right] node {$\mathbf{1}$} (s3)
	(s3) edge [auto = right] node {$\mathbf{1}$} (s1)
	(s3) edge [auto = left] node {$\mathbf{0}$} (s4)
	(s2) edge [auto = right,bend right=10] node {$\mathbf{-15}$} (s4)
	;
	
	\begin{scope}[xshift=9cm]
	\node[PlayerMax] (s0){$\loc_0$}; 
	\node[PlayerMin] at (2, 0) (s1){$\loc_1$}; 
	\node[PlayerMax] at (4, -2.5) (s2){$\loc_2$};
	\node[PlayerMin] at (4, -1) (s3){$\loc_3$};
	\node[target] at (6, 0) (s4) {\LARGE $\smiley$};
	
	\node[proba] at (2, -1) (p1) {};
	\node[proba] at (4, 0) (p3) {};
	
	\draw[->]
	(s1)  edge (p1)
	(p1) edge [auto = left] node {$p,\mathbf{0}$} (s0)
	(s0) edge [auto = left] node {$\mathbf{-1}$} (s1)
	(s0) edge [auto = left, bend left=50,looseness=.7] node {$\mathbf{-10}$} (s4)
	(p1) edge [auto = right] node {$1{-}p,\mathbf{1}$} (s2)
	(s2) edge [auto = right] node {$\mathbf{1}$} (s3)
	(s3) edge (p3)
	(p3) edge [auto = right] node {$p,\mathbf{1}$} (s1)
	(p3) edge [auto = left] node[xshift=-1mm] {$1{-}p,\mathbf{0}$} (s4)
	(s2) edge [auto = right,bend right=10] node {$\mathbf{-15}$} (s4)
	;
	\end{scope}
	\end{tikzpicture}
	\caption{On the left, a more complex example of shortest-path
		game. On the right, the MDP associated with a randomised
		strategy of $\MinPl$ with a parametric probability
		$p\in (0,1)$.}
	\label{fig:SP2}
\end{figure}

\noindent 
The divergence of \WTG{s} was used to obtain an $\varepsilon$-optimal switching
strategy for \MinPl wrt the deterministic value. For \SPG{s}, divergence is not required to obtain it: 

\begin{propC}[\cite{BrihayeGeeraertsHaddadMonmege-17}]\label{prop:switching}
	There exists a switching strategy $\detminstrategy = \langle
	\detminstrategy^1,\detminstrategy^2,K\rangle$ with $\detminstrategy^1$ a
	fake-optimal strategy\footnote{Remember from \cite{BrihayeGeeraertsHaddadMonmege-17} that all plays conforming to a fake-optimal strategy and
reaching the target have a weight at most the deterministic value, and every cyclic play conforming to it has a negative cumulative weight.}, $\detminstrategy^2$ an attractor and
	$K=(2\wmax^{\Trans}(|V|-1)+n)|V|+1$ such that $\dValue^\detminstrategy_\loc \leq
	\max(-n,\dValue_\loc)$, for all initial vertices $\loc\in\vertices$ and $n\in \N$.
\end{propC}
\noindent 
In particular, if $\dValue_\loc$ is finite, for $n$ large enough,
the switching strategy is optimal. If $\dValue_\loc=-\infty$
however, the sequence $(\detminstrategy^n)_{n\in\N}$ of strategies, each
with a different parameter~$n$, has a value that tends to $-\infty$.

Now, we consider a switching strategy $\detminstrategy$ defined in 
Proposition~\ref{prop:switching}, and define the memoryless 
strategy $\minstrategy^p$, with $p\in (0,1)$ as follows: for $\loc \in \vertices$,
letting $\detminstrategy^1(\loc) = \edge_1$ and 
$\detminstrategy^2(\loc) = \edge_2$, we let 
\begin{displaymath}
\minstrategy^p(\loc) = 
\begin{cases}
\Dirac{\trans_1} & \text{if $\loc$ is in a positive SCC} \\
p\times\Dirac{\trans_1} + (1-p)\times \Dirac{\trans_2} & \text{otherwise} 
\end{cases}
\end{displaymath}

We conclude as before by showing the following result paired with the fact that
$\dValue^\detminstrategy_\loc \leq \max(-n,\dValue_\loc)$: 
\begin{prop}
	\label{prop:SPG_partition}
	For all vertices $\loc$ and $\varepsilon > 0$ small
	enough, there exists $\tilde p\in (0,1)$ so that for all
	$p\in [\tilde p,1)$,
	\begin{displaymath}
	\mValue^{\minstrategy^p}_{\loc}\leq
	\dValue^{\detminstrategy}_{\loc} + 2\varepsilon.
	\end{displaymath}
\end{prop}
\begin{proof}
Unlike in the proof of Proposition~\ref{prop:partition}, some SCCs of \SPG{s}
may contain positive and negative cycles (e.g.~the \SPG depicted on the left of
\figurename{~\ref{fig:SP2}}): we will thus refine our partition to compute the 
expected values. 

First, by Lemma~\ref{lem:eta_properties}, we still know
that $\minstrategy^p$ is proper, i.e.~$\Proba^{\minstrategy^{p},
\detmaxstrategy}_{\loc}(\TPlays)=1$. We can thus use again 
Lemma~\ref{lem:best-response} to obtain the existence of a deterministic 
strategy $\detmaxstrategy$ such that 
$\E^{\minstrategy^p,\detmaxstrategy}_{\loc} \geq
\mValue^{\minstrategy^p}_{\loc}-\varepsilon/2$.

We let $c>0$ be the maximal size of an elementary cycle (that visits a vertex
 at most once) in $\game$, $w^->0$ be the opposite of the maximal weight of an
 elementary negative cycle in $\game$, and $w^+\geq 0$ be the maximal weight of
 an elementary non-negative cycle in $\game$ (or $0$ if such cycle does not
 exist).

We now find the bound $\tilde p$ such that $\E^{\minstrategy^p,
\detmaxstrategy}_\loc\leq \dValue^{\detminstrategy}_{\loc} + 3\varepsilon/2$ to
conclude. To do so, we consider a new partition of the set $\Pi$ of plays
starting in $\loc$, conforming to $\minstrategy^{p}$ and $\detmaxstrategy$, and
reaching the target, as depicted in \figurename{~\ref{fig:SPG_partition}}: as
before the partition relies on the number $i$ of transitions of
probability~$1-p$ a play goes through, and its length $j$:
	\begin{itemize}
		\item $\Pi_{0, \N}$, depicted in yellow, contains all plays with no
		edges of probability $1-p$;
		\item $\Pi_{> 0, \geq J}$, depicted in blue, contains all plays with
		$i\geq 1$ edges of probability $1-p$, and a length
		at least
		\[
		J(i) = \lfloor ia + b\rfloor  \qquad \text{with} \quad 
		a = c\left(1 + \frac{w^+}{w^-}\right)
		\quad \text{and} \quad 
		b = \frac{|\dValue^{\detminstrategy}_{\loc}| + 
			|\vertices|\wmax^{\Trans} + w^-}{w^-}c + |\vertices|
		\]
		\item $\Pi_{> 0, < J}$, depicted in red, is the rest of the plays.
	\end{itemize}\noindent 
	
	\begin{figure}[tbp]
		\centering
		\begin{tikzpicture}[xscale=.7,yscale=.3]
		\draw[fill=blue!60!black,draw=none] 
		(0,0) -- (9.6, 9.6) -- (0,9.6) -- (0,0);
		
		\draw[fill=red, draw=none] (0,0) -- (0,3.5) -- (5.5,9.6) -- (9.6,9.6);
		
		\draw[fill=yellow!70!orange,rounded corners,draw=none] (-.15,0) 
		rectangle (.15,9.6);
		\draw[fill=white,draw=none] (-.2,9.6) rectangle (.2,10.6);

		\draw[fill=white,draw=none] (8,8) rectangle (10,10);
		
		\draw[->,thick] (0,0) -- (8.5,0) node[right](){$i$};
		
		\draw[->,thick] (0,0) -- (0,10.5) node[above](){$j$};
		\node at (6,10.25)(){$J(i)$};
		
		\node at (-.7,5) () {$\Pi_{0,\N}$};
		\node[white] at (2.25,8) () {$\Pi_{>0,\geq J}$};
		\node[white] at (4,5.6) () {$\Pi_{> 0, < J}$};
		\end{tikzpicture}
		\caption{A new partition of plays $\Pi$ in a \SPG.}
		\label{fig:SPG_partition}
	\end{figure}
	\noindent 
	We let $\gamma_{0, \N}$ (respectively, $\gamma_{> 0, \geq J}$ and
	$\gamma_{> 0, < J}$) be the expectation
	$\E_{\loc}^{\minstrategy^{p}, \detmaxstrategy}$ restricted to
	plays in $\Pi_{0, \N}$ (respectively, $\Pi_{> 0, \geq J}$ and 
	$\Pi_{> 0, < J}$). By Lemma~\ref{lem:expectation_discrete}, 
	\begin{equation}
	\label{eq:expectation}
	\mValue^{\minstrategy^{p},\detmaxstrategy}_{\loc}=
	\E_{\loc}^{\minstrategy^{p}, \detmaxstrategy} = 
	\sum_{\play} \weight(\play) \times 
	\Proba^{\minstrategy, \detmaxstrategy}_{\loc}(\play) = 
	\gamma_{0, \N} + \gamma_{> 0, \geq J} + \gamma_{> 0, < J}
	\end{equation}
	We thus analyse separately the three terms of \eqref{eq:expectation}. 
	
	First, we analyse the weight of a play depending on the
	number of edges $1-p$ it goes through.  Let $\play$ be a
	play in $\Pi_{i, j}$, with $1 \leq i$ and $j \geq i$: it
	goes through $i$ edges of probability $1-p$. 
	
	Each cycle conforming to $\minstrategy^p$ and
$\detmaxstrategy$ with a non-negative cumulative weight contains at least one transition
taken by $\minstrategy^p$ with probability $1-p$ (otherwise, all its edges would be of probability $p$ or $1$, and the cycle would be
conforming to the fake-optimal strategy $\detminstrategy^1$, implying that the cumulative weight would be negative). 

	In particular, $\play$ contains at most $i$ elementary cycles of
	non-negative cumulated weight (at most $w^+$). The total length of these
	cycles is at most $ic$. Once we have removed these cycles from the play, it
	remains a play of length at least $j-ic$. By a repeated pumping
	argument, it still contains at least
	$\left\lfloor\frac{j-ic-|\vertices|}{c}\right\rfloor$ elementary cycles, that
	all have a negative cumulated weight (at most $-w^-$). The remaining part,
	once removed the last negative cycles it contains, has length at most
	$|\vertices|$, and thus a cumulative weight at most $|\vertices|\wmax^{\Trans}$. In summary
	the cumulative weight of every play in $\Pi_{i, j}$ is at most
	\begin{equation}
	\label{eq:weight}
	iw^+ + \left\lfloor\frac{j-ic-|\vertices|}{c}\right\rfloor(-w^-) + 
	|\vertices|\wmax^{\Trans}
	\end{equation}
	
	\paragraph*{Red zone is such that $\gamma_{>0, <J} \leq \varepsilon/2$.}
	Let $\play$ be a play in $\Pi_{i, j}$, with $i\geq 1$ and
	$j < J(i)$. By~\eqref{eq:weight}, its cumulative weight is at
	most
	\begin{displaymath}
	iw^+ + \left\lfloor\frac{j-ic-|V|}{c}\right\rfloor(-w^-) + 
	|\vertices|\wmax^{\Trans} \leq iw^+ + |\vertices|\wmax^{\Trans} \,.
	\end{displaymath}
	So, we can decompose the expectation $\gamma_{>0, < J}$
	as follows:
	\begin{align}
	\gamma_{>0, < J} 
	&= \sum_{\play \in \Pi_{> 0, < J}} \weight(\play)
	\Proba_{\loc}^{\minstrategy^{p}, \detmaxstrategy}(\play) \leq
	\sum_{i = 1}^{+\infty} (iw^+ + |\vertices|\wmax^{\Trans})
	\Proba_{\loc}^{\minstrategy^{p}, \detmaxstrategy}
	(\Pi_{i, < J(i)}) 
	\label{eq:red_expectation1}
	\end{align}
	Moreover, the probability of a play in $\Pi_{i, < J(i)}$,
	given by the $i$ edges of probability $(1-p)$ and the
	$j -i$ edges with a probability bounded by $1$, is at most
	$(1-p)^i$. Since the number of plays in $\Pi_{i, < J(i)}$ is
	bounded by $2^{J(i)}$ (for each of the at most $J(i)$ steps,
	$\MinPl$ has at most 2 choices in its distribution, while
	$\MaxPl$ plays a deterministic strategy), we have
	\begin{equation}
	\label{eq:red_proba}
	\Proba_{\loc}^{\minstrategy^{p}, \detmaxstrategy}(\Pi_{i, < J(i)}) \leq 
	(1 - p)^i 2^{J(i)}
	\end{equation}
	We rewrite~\eqref{eq:red_expectation1} as 
	\begin{align*}
	\gamma_{>0, < J} 
	&\leq \sum_{i = 1}^{+\infty} (iw^++|\vertices|\wmax^{\Trans}) (1 - p)^i 2^{J(i)}\leq
	\sum_{i = 1}^{+\infty} (iw^++|\vertices|\wmax^{\Trans}) (1 - p)^i
	2^{ai+b} \\ 
	&\leq 
	w^+2^b \sum_{i = 1}^{+\infty} i ((1 - p)2^a)^i + |\vertices|\wmax^{\Trans}2^b \sum_{i = 1}^{+\infty} ((1 - p)2^a)^i
	\end{align*}
	these sums converging as soon as we consider $p \geq 1 -
	\frac{1}{2^a}$. We finally obtain
	\begin{displaymath}
	\gamma_{>0, < J} 
	\leq 
	 w^+2^b  \frac{2^a(1-p)}{(1 - 2^a(1-p))^2} + |\vertices|\wmax^{\Trans}2^b \frac{2^a(1-p)}{1 - 2^a(1-p)}\,.
	\end{displaymath}
	We consider a stronger assumption on $p$, namely that
	$p \geq 1 - \frac{1}{2^{a + 1}}$. Then, we know that $1 \leq
	\frac{1}{1 - 2^a(1 - p)} \leq 2$, so that 
	we rewrite the previous inequality as 
	\begin{displaymath}
	\gamma_{>0, < J} 
	\leq  w^+2^{b + a + 2} (1-p) +|\vertices|\wmax^{\Trans}2^{b + a + 1}(1-p) \,.
	\end{displaymath}
	By choosing $p$ such that
	\begin{displaymath}
	p \geq 1 - \frac{\varepsilon}{2^{b + a + 2}(|\vertices|\wmax^{\Trans} + 
		2 w^+)}
	\end{displaymath}
	we obtain as desired $\gamma_{>0, <J} \leq\varepsilon/2$. 
	
	\paragraph*{Yellow and blue zones are such that
		$\gamma_{0, \N} + \gamma_{>0, \geq J}\leq
		\dValue^{\detminstrategy}_{\loc}+ \varepsilon$.}
	We first upper-bound the cumulative weight of all plays of these
	two zones.  On the one hand, all plays of $\Pi_{0, \N}$ reach
	the target without edges of probability $1-p$, i.e.~by
	conforming to~$\detminstrategy^1$. By fake-optimality of
	$\detminstrategy^1$, their cumulative weight is upper-bounded by
	$\dValue^{\detminstrategy}_{\loc}$. 
	On the other hand, by~\eqref{eq:weight}, all plays $\play$ of
	$\Pi_{i, j}$, with $0 \leq i < I$ and $j \geq J(i)$, have a
	cumulative weight at most
	\begin{align*}
	\weight(\play) 
	& \leq iw^+ + \left\lfloor\frac{j-ic-|\vertices|}{c}\right\rfloor (-w^-) + 
	|\vertices|\wmax^{\Trans}\\
	&\leq iw^+ + \frac{J(i) - ic - |\vertices|}{c} (-w^-) + 
	|\vertices|\wmax^{\Trans} \\
	&= iw^+ + \frac{ai + \frac{|\dValue^{\detminstrategy}_{\loc}| + 
			|\vertices|\wmax^{\Trans}+w^-}{w^-}c + |\vertices| - ic - |\vertices|}{c} 
		(-w^-) + |\vertices|\wmax^{\Trans} \\
	&= iw^+ + \left(i\left(1 + \frac{w^+}{w^-}\right) + 
	\frac{|\dValue^{\detminstrategy}_{\loc}| + |\vertices|\wmax^{\Trans}}{w^-} - i\right) 
	(-w^-) + |\vertices|\wmax^{\Trans} \\
	&= iw^+ -iw^+ -|\dValue^{\detminstrategy}_{\loc}| -|\vertices|\wmax^{\Trans} + 
	|\vertices|\wmax^{\Trans} = -|\dValue^{\detminstrategy}_{\loc}|\leq \dValue^{\detminstrategy}_{\loc}
	\end{align*}
	Therefore, all plays in the yellow and blue zones have a
	cumulative weight bounded by $\dValue^{\detminstrategy}_{\loc}$. This implies
	\begin{align*}
	\gamma_{0, \N} + \gamma_{> 0, \geq J}
	&\leq \sum_{\play \in \Pi_{0, \N}} \dValue^{\detminstrategy}_{\loc}
	\Proba_{\loc}^{\minstrategy^{p}, \detmaxstrategy}(\play) +
	\sum_{\play \in \Pi_{> 0, \geq J}} \dValue^{\detminstrategy}_{\loc}
	\Proba_{\loc}^{\minstrategy^{p}, \detmaxstrategy}(\play)  \leq \dValue^{\detminstrategy}_{\loc}
	\Proba_{\loc}^{\minstrategy^{p}, \detmaxstrategy}\left(\Pi_{0, \N} \cup 
	\Pi_{> 0, \geq J}\right).
	\end{align*}
	Depending on the sign of $\dValue^{\detminstrategy}_{\loc}$, we can
	conclude:
	\begin{itemize}
		\item If $\dValue^{\detminstrategy}_{\loc}\geq 0$, then
		upper-bounding the probability
		$\Proba_{\loc}^{\minstrategy^{p}, \detmaxstrategy} 
		\left(\Pi_{0, \N} \cup \Pi_{> 0, \geq J}\right)$ by $1$, suffices to get
		$\gamma_{0, \N} + \gamma_{> 0, \geq J(i)} \leq
		\dValue^{\detminstrategy}_{\loc}$. 
		
		\item If $\dValue^{\detminstrategy}_{\loc}< 0$, then, by the 
		bound~\eqref{eq:red_proba} found for the red zone, we have
		\begin{align*}
		\Proba_{\loc}^{\minstrategy^{p}, \detmaxstrategy}\left(\Pi_{0, \N} \cup 
		\Pi_{> 0, \geq J}\right) 
		&= 1 - \Proba_{\loc}^{\minstrategy^{p}, \detmaxstrategy}
		\left(\Pi_{> 0, < J}\right)\\
		&\geq 1- \sum_{i=1}^{\infty}(1 - p)^i 2^{ai+b}\\
		&=1-\frac{2^{a+b}(1-p)}{1-(1-p)2^a}\\
		&\geq 1 - 2^{a + b+1}(1-p)  \qquad (\text{since }
		1/(1-(1-p)2^a) \leq 2).
		\end{align*}
		This allows us to obtain
		\begin{displaymath}
		\gamma_{0, \N} + \gamma_{> 0, \geq J} \leq 
		\dValue^{\detminstrategy}_{\loc} (1 - 2^{a + b + 1}(1-p)).
		\end{displaymath}
		In case we have moreover
		\begin{displaymath}
		p \geq 1 - \frac{\varepsilon}{2^{a + b + 1}| 
			\dValue^{\detminstrategy}_{\loc}|}
		\end{displaymath}
		we finally obtain
		$\gamma_{0, \N} + \gamma_{> 0, \geq J} \leq
		\dValue^{\detminstrategy}_{\loc}+ \varepsilon$ as expected.
	\end{itemize}
	
	\paragraph*{Lower bound over $p$}
	
	If we gather all the lower bounds over $p$ that we need in the proof,
	we get that:
	\begin{itemize}
		\item if
		$\dValue^{\detminstrategy}_{\loc}\geq 0$, we must have
		\[p\geq \max\left(1 - \frac{1}{2^{a + 1}}, 
		1 - \frac{\varepsilon}{2^{b + a + 2}(|\vertices|\wmax^{\Trans} + 
		2 w^+)} \right) \]
		\item if $\dValue^{\detminstrategy}_{\loc}< 0$, we must have
		\[p\geq \max\left(1 - \frac{1}{2^{a + 1}}, 
		1 - \frac{\varepsilon}{2^{b + a + 2}(|\vertices|\wmax^{\Trans} + 
		2 w^+)}, 
		1 - \frac{\varepsilon}{2^{a + b + 1}|\dValue^{\detminstrategy}_{\loc}|}
		\right)\]
	\end{itemize}
	with $\varepsilon$ small enough so that this bound is less than $1$.
\end{proof}
\noindent 
This ends the proof that for all vertices $\loc$,
$\muppervalue_\loc \leq \dValue_\loc$.

\paragraph{Discussion about smooth deterministic values in \SPG}
With the existence of a switching optimal strategy with respect to the 
deterministic value for \MinPl (by Proposition~\ref{prop:switching}), 
we obtain the existence of proper smooth deterministic optimal strategies 
with respect to the deterministic value for \MinPl in all \SPG{s}.
Finally, 
we can deduce that \SPG{s} are determined with respect to the stochastic 
value, and the stochastic and (smooth) deterministic values are equal:

\begin{cor}
	In all \SPG{s}, for all locations $\loc$, 
	\begin{displaymath}
	\dValue_{\loc} = \sdlowervalue_{\loc} = \sduppervalue_{\loc} = 
	\lowervalue_{\loc} = \uppervalue_{\loc}
	\,.
	\end{displaymath}
\end{cor}
\begin{proof}
	Since \MaxPl has an optimal deterministic strategy \cite{BrihayeGeeraertsHaddadMonmege-17}, 
	there exists $\detminstrategy \in \sdStratMinProper$ and $\detmaxstrategy \in \dStratMax$ 
	such that 
	\begin{displaymath}
	\dValue^{\minstrategy}_{\loc, \val} \leq \dValue \leq 
	\dValue^{\detmaxstrategy}_{\loc, \val} 
	\end{displaymath}
	by Proposition~\ref{prop:switching}. First, since $\sdStratMax = \dStratMax$, we deduce that 
	\begin{displaymath}
	\sduppervalue_{\loc, \val} \leq 
	\sup_{\detmaxstrategy \in \sdStratMax} 
	\weight(\outcomes((\loc, \val), \detminstrategy, \detmaxstrategy))  = 
	\dValue^{\detminstrategy}_{\loc, \val} \,.
	\end{displaymath}
	Next, since $\sdStratMinProper \subseteq \dStratMin$, we deduce that 
	\begin{displaymath}
	\dValue^{\detmaxstrategy}_{\loc, \val} = 
	\inf_{\detminstrategy \in \dStratMin} 
	\weight(\outcomes((\loc, \val), \detminstrategy, \detmaxstrategy)) \leq 
	\inf_{\detminstrategy \in \sdStratMinProper} 
	\weight(\outcomes((\loc, \val), \detminstrategy, \detmaxstrategy))  
	\leq \sdlowervalue_{\loc, \val}\,.
	\end{displaymath}
	Finally, we obtain $\sdlowervalue_{\loc, \val} = \sduppervalue_{\loc, \val}$
	and we conclude the proof by applying Theorem~\ref{thm:detVal=Val}.
\end{proof}

\subsection{Characterisation of optimality in weighted timed games without clocks}
\label{subsec:SPG-optimality}

All \SPG{s} admit an optimal deterministic strategy for both players: however,
as we have seen in \figurename{~\ref{fig:SP1}}, $\MinPl$ may require memory to
play optimally. In this case, $\MinPl$ does not have an optimal memoryless
(randomised) strategy, but only has $\varepsilon$-optimal ones, for all
$\varepsilon>0$. But some \SPG{s} indeed admit optimal memoryless strategies
for $\MinPl$: the strategy $\minstrategy^{p}$ described in the proof of Lemma~\ref{lem:partition-SPG} is indeed optimal in \SPG{s} not
containing negative cycles, for instance. We 
characterise here the \SPG{s} in which \MinPl admits an optimal memoryless strategy.
For sure, $\MinPl$ does not have an optimal strategy if there is some vertex
$\loc$ of value $\dValue_\loc =-\infty$. Thus, \textbf{we only consider in this
section \SPG{s} without vertices $\loc$ with a deterministic value $-\infty$}.

We first recall the computations performed in
\cite{BrihayeGeeraertsHaddadMonmege-17} to compute values $\dValue_\loc$. As
for divergent \WTG{s}, it consists of an iterated computation, called
\emph{value iteration} based on the operator $\F\colon
(\Z\cup\{+\infty\})^\vertices \to (\Z\cup\{+\infty\})^\vertices$ defined for
all $X=(X_\loc)_{\loc\in \vertices}\in (\Z\cup\{+\infty\})^\vertices$ and all
vertices $\loc\in \vertices$ by
\[\F(X)_\loc =
\begin{cases}
0 & \text{if } \loc\in \finalvertices\\
\min_{(\loc,\loc')\in \edges} (\edgeweights(\loc,\loc') + X_{\loc'}) 
& \text{if } \loc \in \minvertices\\
\max_{(\loc,\loc')\in \edges} (\edgeweights(\loc,\loc') + X_{\loc'}) 
& \text{if } \loc\in \maxvertices
\end{cases}\] 
We let $f^{(0)}_\loc=0$ if $\loc\in \finalvertices$ and
$+\infty$ otherwise. By monotony of $\F$, the sequence
$(f^{(i)}=\F^i(f^{(0)}))_{i\in \N}$ is non-increasing. It is proved to
be ultimately constant, and converging towards $(\dValue_\loc)_{\loc\in \vertices}$,
the smallest fixpoint of $\F$. The pseudo-polynomial complexity of
computing the values of \SPG{s} comes from the fact that this sequence may
become stationary after a pseudo-polynomial (and not polynomial)
number of steps: the game of \figurename{~\ref{fig:SP1}} is one of the
typical examples.

We introduce a new notion, being \emph{the most permissive strategy} of
$\MinPl$ at each step $i \geq 0$ of the computation. It maps each
vertex $\loc \in \minvertices$ to the set
\[\widetilde \edges^{(i)}(\loc) = 
\{(\loc, \loc')\in \edges \mid \edgeweights(\loc,\loc') + f^{(i-1)}_{\loc'} = f^{(i)}_\loc\}\] 
of vertices that $\MinPl$ can choose. For each such most permissive strategy
$\widetilde \edges^{(i)}$, we let $\widetilde\game^{(i)}$ be the \SPG 
where we remove all edges $(\loc,\loc')$ with $\loc\in \minvertices$ and
$(\loc,\loc') \notin \widetilde \edges^{(i)}(\loc)$. This allows us to state the
following result:

\begin{thm}
	\label{thm:caractOptiSansMem}
	The following assertions are equivalent:
	\begin{enumerate}
		\item\label{item:caractOptiSansMem-1} 
		$\MinPl$ has an optimal memoryless deterministic strategy in $\game$ 
		(for $\dValue$);
		\item\label{item:caractOptiSansMem-2} 
		$\MinPl$ has an optimal memoryless (randomised) strategy in $\game$ 
		(for $\muppervalue=\mlowervalue$);
		\item\label{item:caractOptiSansMem-3} 
		$f^{(|\vertices|-1)}_\loc=f^{(|\vertices|)}_\loc=\dValue_\loc$ for all vertices $\loc$ (this means 
		that the sequence $(f^{(i)})$ is stationary as soon as step $|\vertices|-1$), and 
		$\MinPl$ can guarantee that a vertex of $\finalvertices$ is reached from all vertices in the 
		game $\widetilde\game^{(|\vertices|-1)}$.
	\end{enumerate}
\end{thm}

\begin{rem}
This characterisation of the existence of optimal memoryless strategy
is testable in polynomial time since it is enough to compute vectors
$f^{(|\vertices|-1)}$ and $f^{(|\vertices|)}$, check their equality, compute the sets
$\widetilde \edges^{(|\vertices|-1)}(\loc)$ (this can be done while computing
$f^{(|\vertices|)}$) and check whether $\MinPl$ can guarantee that a
target is reached in $\widetilde\game^{(|\vertices|-1)}$ by an attractor computation.
\end{rem}

We prove Theorem~\ref{thm:caractOptiSansMem} in the rest of this section. Implication \underline{$\ref{item:caractOptiSansMem-1}\Rightarrow 
	\ref{item:caractOptiSansMem-2}$} is trivial by the result of 
	Theorem~\ref{thm:shortest-path}.
	
	\medskip
	
	For implication \underline{$\ref{item:caractOptiSansMem-3}\Rightarrow 
	\ref{item:caractOptiSansMem-1}$}, consider any memoryless deterministic 
	strategy $\detminstrategy$ that guarantees $\MinPl$ to reach 
	$\finalvertices$ from all vertices in the game graph 
	$\widetilde\game^{(|\vertices|-1)}$. Then, for all vertices $\loc$, we
	show by induction on $n$, that each play $\play$ from $\loc$ that reaches
	the target in at most $n$ steps, and conforming to $\detminstrategy$,
	has a cumulative weight $\weight(\play)\leq \dValue_\loc$. This is trivial for
	$n=0$. If~$\play= \loc \xrightarrow{\edge} \play'$ with $\play'$ starting in 
	$\loc'$, then
	\[\weight(\play) = \edgeweights(\edge) + \weight(\play') \leq 
	\edgeweights(\edge) + \dValue_{\loc'} = \edgeweights(\edge) + f^{(|\vertices|-1)}_\loc\,.\] 
	If $\loc \in \maxvertices$, we have
	\[\weight(\play) \leq \edgeweights(\edge) + f^{(|\vertices|-1)}_\loc \leq
	f^{(|\vertices|)}_\loc = \dValue_\loc\,.\] 
	If $\loc\in \minvertices$, since~$\loc' \in \widetilde \edges^{(|\vertices|-1)}(\loc)$,
	\[\weight(\play)= f^{(|\vertices|)}_\loc = \dValue_\loc\,.\]
	This ends the proof by induction. To conclude that 
	\ref{item:caractOptiSansMem-1} holds, since $\detminstrategy$ guarantees that the target is reached, all plays conforming to it reach the target in less than 
	$|\vertices|$ steps, which proves that~$\dValue^{\detminstrategy}_\loc \leq \dValue_\loc$,
	showing that~$\detminstrategy$ is optimal.
	
	\medskip
	
	For implication \underline{$\ref{item:caractOptiSansMem-1} \Rightarrow 
	\ref{item:caractOptiSansMem-3}$}, consider an optimal deterministic memoryless 
	strategy $\detminstrategy^*$, such
	that for all $\loc$, $\dValue^{\detminstrategy^*}_\loc= \dValue_\loc$.
	
	First, we show that $f^{(|\vertices|-1)}_\loc = \dValue_\loc$ for all vertices
	$\loc$. For that, consider the deterministic strategy $\detmaxstrategy$ of
	$\MaxPl$ defined for all finite plays $\play$ having $n\leq |\vertices|$
	vertices, ending in a vertex $\loc \in \maxvertices$, by
	$\detmaxstrategy(\play) = (\loc, \loc')$ such that
	$\edgeweights(\loc,\loc')+f_{\loc'}^{(|\vertices|-1-n)}=f_{\loc}^{(|\vertices|-n)}$. For longer
	finite plays, we define $\detmaxstrategy$ arbitrarily. Then, let $\play$
	be the play from $\loc$ conforming to $\detminstrategy^*$
	and~$\detmaxstrategy$. Since $\detminstrategy^*$ ensures reaching the
	target and is memoryless and deterministic, $\play$ reaches the target in
	at most $|\vertices|-1$ steps. Let $\play=\loc_0 \xrightarrow{\edge_0} \loc_1 \cdots 
	\xrightarrow{\edge_{k-1}} \loc_k$ with $k\leq |\vertices|$. Let us show that 
	$\weight(\play)\geq f_{\loc_0}^{(|\vertices|-1)}$. We prove
	by induction on $0\leq j\leq k$ that
	\[\sum_{i=j}^{k-1} \edgeweights(\edge_i) \geq f_{\loc_j}^{(|\vertices|-1-j)}\,.\] 
	When $j=k$, the result is trivial since
	the sum is \[0 = f_{\loc_k}^{(0)} \geq f_{\loc_k}^{(|\vertices|-1-k)}\]
	by using that $(f_{\loc_k}^{(i)})_i$ is non-increasing. Otherwise, by induction hypothesis
	\[\sum_{i=j}^{k-1} \edgeweights(\edge_i) \geq
	\edgeweights(\edge_j) + f_{\loc_{j+1}}^{(|\vertices|-1-(j+1))}\,.\] 
	If $\loc_j \in \maxvertices$, $\loc_{j+1}$ is chosen by $\detmaxstrategy$ so that
	\[\edgeweights(\edge_j) + f_{\loc_{j+1}}^{(|\vertices|-1-(j+1))} =
	f_{\loc_j}^{(|\vertices|-1-j)}.\] 
	If $\loc\in \minvertices$, by definition of $\F$,
	\[\edgeweights(\edge_j) + f_{\loc_{j+1}}^{(|\vertices|-1-(j+1))} \geq
	f_{\loc_j}^{(|\vertices|-1-j)}.\] 
	We can conclude in all cases, so that $f^{(|\vertices|-1)}_\loc = \dValue_\loc$ for all 
	vertices $\loc$.
	
	Then, we show that $\MinPl$ can guarantee to reach $\finalvertices$
	from all vertices in the \SPG~$\widetilde\game^{(|\vertices|-1)}$. Let us
	suppose that this is not the case. Then, there exists a set
	$\vertices'$ of vertices of $\widetilde\game^{(|\vertices|-1)}$ in which $\MaxPl$ can trap 
	$\MinPl$ forever: for all
	$\loc'\in\vertices'\cap\minvertices$,
	$\widetilde \edges^{(|\vertices|-1)}(\loc') \subseteq \vertices'$, and for all
	$\loc'\in\vertices'\cap\maxvertices$, 
	$\edges(\loc) \cap \edges'\neq \emptyset$. Since~$\detminstrategy^*$
	guarantees that a target is reached, there exists
	$\loc\in \vertices'\cap\minvertices$ such that
	$\detminstrategy^*(\loc)=\loc'\notin\vertices'$: then
	$\edgeweights(\loc,\loc')+\dValue_{\loc'} > \dValue_\loc$ (here we use that
	$\dValue_\loc = f_\loc^{(|\vertices|-1)} = f_\loc^{(|\vertices|)}$). Consider an optimal
	deterministic memoryless strategy $\detmaxstrategy^*$ of
	$\MaxPl$ in $\game$ (which we know exists by \cite{BrihayeGeeraertsHaddadMonmege-17}). Then, the play $\play$ from $\loc$ conforming to
	$\detminstrategy^*$ and $\detmaxstrategy^*$ starts by taking the edge
	$(\loc,\loc')$ and continues with a play $\play'$. By optimality, we know
	that $\weight(\play)=\dValue_\loc$ and $\weight(\play')=\dValue_{\loc'}$. However,
	\[\weight(\play)=\edgeweights(\loc,\loc')+\weight(\play') =
	\edgeweights(\loc,\loc')+\dValue_{\loc'} > \dValue_\loc\] which raises a
	contradiction.
		
		\medskip
		
	We finish the proof by showing \underline{$\ref{item:caractOptiSansMem-2} \Rightarrow 
	\ref{item:caractOptiSansMem-1}$}: the proof will be constructive and actually allows one to build an optimal memoryless deterministic strategy when it exists. Consider thus an
	optimal memoryless strategy~$\minstrategy^*$ for
	the memoryless value. We build a memoryless deterministic
	strategy $\detminstrategy^*$, and show that it is also optimal.
	The construction of $\detminstrategy^*$ is performed in two steps. 

	We first select
transitions that minimise the value of $\minstrategy^*$ at horizon $1$. Formally,
for all locations $\loc$, we define 
\begin{equation} 
\label{eq:Etilde}
\widetilde \Trans(\loc) = \argmin_{(\loc,\loc') \in \support{\minstrategy(\loc)}} 
\left[\weight(\loc,\loc')+\mValue^{\minstrategy}_{\loc'}\right]
\end{equation}
In a first approximation, we thus fix 
$\detminstrategy^*(\loc)$ to be any edge of $\widetilde \edges(\loc)$, for all vertices $\loc$. The next example shows that $\detminstrategy^*$ may not be optimal. 

\begin{figure}[tbp]
	\centering
	\begin{tikzpicture}[xscale=.8,yscale=.8,node distance=2cm]
	\node[PlayerMin] (s0){$\loc_0$};
	\node[PlayerMin,right of=s0] (s1){$\loc_1$};
	\node[target,right of=s1] (s4) {\LARGE $\smiley$};
	
	\draw[->]
	(s0)  edge [auto = left, bend left=10] node {$\mathbf{-1}$} (s1)
	(s1) edge [auto = left, bend left=10] node {$\mathbf{1}$} (s0)
	(s0) edge [auto = left, bend left=50,looseness=.7] node {$\mathbf{0}$} (s4)
	(s1) edge [auto = right] node {$\mathbf{-1}$} (s4)
	;
	
	\begin{scope}[xshift=8cm]
	\node[PlayerMin] (s0){$\loc_0$};
	\node[PlayerMin,right of=s0] (s1){$\loc_1$};
	\node[target,right of=s1] (s4) {\LARGE $\smiley$};
	
	\node[proba] at (1.3, -1) (p1) {};
	
	\draw[->]
	(s0)  edge [auto = left, bend left=10] node {$\mathbf{-1}$} (s1)
	(s1) edge (p1)
	(p1) edge [auto = left] node[xshift=2mm] {$1-p,\mathbf{1}$} (s0)
	(p1) edge [auto = right,bend right=5] node {$p,\mathbf{-1}$} (s4)
	;
	\end{scope}
	\end{tikzpicture}
	\caption{An \SPG (on the left) and the Markov chain obtained by fixing an optimal memoryless strategy, with $p\in(0,1)$.}
	\label{fig:SP3}
\end{figure}

\begin{exa}
	Consider the game of \figurename{~\ref{fig:SP3}}, 
	with the memoryless strategy $\minstrategy^*$ defined by
	$\minstrategy^*(\loc_0) = \Dirac{(\loc_0, \loc_1)}$ and
	$\minstrategy^*(\loc_1)=(1-p)\Dirac{(\loc_1, \loc_0)} + 
	p\Dirac{(\loc_1, \text{\large\smiley})}$, where $p\in(0,1)$.
	Then, we can
	check that $\mValue^{\minstrategy^*}(\loc_0)=-2$ and
	$\mValue^{\minstrategy^*}(\loc_1)=-1$, which implies that $\minstrategy^*$ is optimal. We have 
	$\widetilde \edges(\loc_0)=\{(\loc_0,\loc_1)\}$ and
	$\widetilde \edges(\loc_1)=\{(\loc_1,\loc_0),(\loc_1,\text{\Large\smiley})\}$. However, considering the strategy $\detminstrategy^*$ defined by $\detminstrategy^*(\loc_0) = (\loc_0,\loc_1)$ and $\detminstrategy^*(\loc_1) = (\loc_1,\loc_0)$, results in a value $\dValue^{\detminstrategy^*}(\loc_0)=\dValue^{\detminstrategy^*}(\loc_1)=+\infty$: thus $\detminstrategy^*$ is not optimal. 
\end{exa}

We employ techniques already used in \cite{ChatterjeeAlfaroHenzinger-04} to refine the definition of $\minstrategy^*$. For each vertex $\loc$ in the game, we let $\dist(\loc)$ be 
the attractor distance of $\loc$ to the target in the finite MDP $\MDP$, i.e. the 
smallest number of steps such that $\MinPl$ has a path of this length reaching the 
target, no matter how is playing \MaxPl (notice that this may be different from 
the attractor distance given in the whole game, since some edges are taken
with probability~$0$ in~$\minstrategy$). For all vertices $\loc$, $\dist(\loc)<+\infty$ 
since otherwise ($\minstrategy$ being a memoryless strategy) the 
finite MDP $\MDP$ 
would contain a bottom strongly connected component without target states, 
which would contradict the fact that $\minstrategy$ ensures 
to reach $\finalvertices$ with probability $1$.
We then let, for all vertices $\loc\in \minvertices$,
$\detminstrategy^*(\loc)$ be any edge in $\argmin_{(\loc, \loc')\in \widetilde \edges(\play)} \dist(\loc')$.

\begin{exa}
On the same example as above, using the fact
that $d(\text{\large\smiley})=0$, $d(\loc_1)=1$ and $d(\loc_0)=2$ (since
the edge $(\loc_0,\text{\large\smiley})$ is not present in
$\game^{\minstrategy}$), we have no choice but to let $\detminstrategy^*$ be defined by
$\detminstrategy^1(\loc_0)=(\loc_0,\loc_1)$ and
$\detminstrategy^1(\loc_1)=(\loc_1, \text{\large\smiley})$, which is indeed an optimal deterministic (and memoryless) strategy.
\end{exa}

\begin{lem}\label{lem:optimal-}
The memoryless deterministic strategy $\detminstrategy^*$ defined above is optimal.
\end{lem}
\begin{proof}
We show this by proving independently that: 
\begin{enumerate}
\item Each finite play \play conforming to $\detminstrategy^*$ from $\loc$ and reaching the target has a cumulated weight at most $\mValue^{\minstrategy}_{\loc}$.
\item No play conforming to $\detminstrategy^*$ can contain a cycle.
\end{enumerate}
\noindent 
From the second item, we are thus certain that all plays conforming to $\detminstrategy^*$ reach the target. The first item then allows us to conclude that $\detminstrategy^*$ is optimal. 

	\begin{enumerate}
		\item We prove the property by induction on the length of finite plays 
		$\play$ conforming to $\detminstrategy^*$ that reaches the target, for 
		all initial vertices $\loc$. If $\play$ has length $0$, this means that 
		$\loc\in \finalvertices$, in which case 
		$\weight(\play) = 0 = \mValue^{\minstrategy^*}_\loc$. Consider then a
		play $\play = \loc\xrightarrow{\trans} \play'$ such that $|\play| \geq 1$ 
		and $\play'$ starting by the vertex $\loc'$, so that 
		$\weight(\play) = \edgeweights(\trans) + \weight(\play')$. By 
		induction hypothesis, $\weight(\play') \leq \mValue^{\minstrategy^*}_{\loc'}$, 
		so that $\weight(\play) \leq \edgeweights(\trans) + 
		\mValue^{\minstrategy^*}_{\loc'}$.
		
		Suppose first that $\loc \in \maxvertices$. By Proposition~\ref{prop:strat_opti_Eve}, consider a deterministic best-response strategy \detmaxstrategy of \MaxPl against $\minstrategy^*$. Then
		$\E^{\minstrategy^*,\detmaxstrategy}_u = \mValue^{\minstrategy^*}_u$ for 
		all $u \in \maxvertices$, and by~\eqref{eq:SPG-Bellman}, letting $(u, u')=\detmaxstrategy(u)$,
		$\E^{\minstrategy^*,\detmaxstrategy}_u = \edgeweights(u, u') +
		\E^{\minstrategy^*,\detmaxstrategy}_{u'}$. We thus know that
		$\edgeweights(u, u') + \E^{\minstrategy^*,\detmaxstrategy}_{u'} = 
		\mValue^{\minstrategy^*}_u$. In particular, for all vertices $u\in \maxvertices$ and edge $(u,u')\in \edges$, 
		\begin{equation}
		\label{eq:inequality-Max}
		\edgeweights(u, u') + \mValue^{\minstrategy^*}_{u'} \leq 
		\mValue^{\minstrategy^*}_u 
		\end{equation}
		In particular,
		$\weight(\play) \leq \edgeweights(\loc, \loc') +
		\mValue^{\minstrategy^*}_{\loc'} \leq \mValue^{\minstrategy^*}_\loc$.
		
		If $\loc \in \minvertices$, then $\edge = (\loc, \loc') \in \widetilde \edges(\loc)$ 
		so that $\edgeweights(\edge)+\mValue^{\minstrategy^*}_{\loc'}$ is minimum
		over all possible 
		$(\loc,\loc') \in \support{\minstrategy^*(\loc)}$. The system
		\eqref{eq:SPG-Bellman} implies that 
		\begin{displaymath}
		\E^{\minstrategy^*,\maxstrategy}_\loc = 
		\sum_{(\loc, \loc'') \in \edges} \minstrategy^*(\loc)(\loc,\loc'') \times
		\big(\edgeweights(\loc,\loc'') + \E^{\minstrategy^*,\maxstrategy}_{\loc''}\big).
		\end{displaymath}
		for all memoryless strategies~$\maxstrategy$ of $\MaxPl$. In particular, for the best-response deterministic and memoryless strategy $\maxstrategy$ given by Proposition~\ref{prop:strat_opti_Eve}, 
		\begin{equation} 
		\label{eq:final-inequality}
		\hspace*{0.5\leftmargini}\mValue^{\minstrategy^*}_\loc = \E^{\minstrategy^*,\maxstrategy}_\loc =\quad\sum_{\mathclap{(\loc,\loc'') \in \support{\minstrategy^*(\loc)}}}\quad\; \minstrategy^*(\loc)(\loc,\loc'') \times
		(\edgeweights(\loc,\loc'') + \mValue^{\minstrategy^*}_{\loc''}) \geq
		\edgeweights(\loc,\loc') + \mValue^{\minstrategy^*}_{\loc'}
		\end{equation}
		so that we also get $\weight(\play) \leq \mValue^{\minstrategy^*}_\loc$.
		
		\item Suppose that a cycle
	$\loc_1\loc_2\cdots \loc_k\loc_1$ conforms to $\detminstrategy^*$, with $\loc_1$ a vertex of minimal distance $d(\loc_1)$. 
	We can choose $\loc_1$ such that it belongs to $\MinPl$, since the distance is computed by an attractor for $\MinPl$ (and thus all successors of a vertex of $\MaxPl$ must have a smaller distance). By minimality of $\dist(\loc_1)$ among the vertices of the
	cycle, $\dist(\loc_2) \geq \dist(\loc_1)$. Moreover, by the attractor
	computation, there exists $(\loc_1,\loc')\in \edges(\loc_1)$ such that
	$\dist(\loc')=\dist(\loc_1)-1 < \dist(\loc_1)$. By definition of~$\detminstrategy^*$, 
	we know that $(\loc_1, \loc') \notin \widetilde \edges(\loc_1)$, so that
	\[\edgeweights(\loc_1, \loc') + \mValue^{\minstrategy^*}_{\loc'} >
	\edgeweights (\loc_1,\loc_2) + \mValue^{\minstrategy^*}_{\loc_2}.\]
	By~\eqref{eq:final-inequality}, we know that in this case
	\[\mValue^{\minstrategy^*}_{\loc_1}> \edgeweights(\loc_1,\loc_2)
	+\mValue^{\minstrategy^*}_{\loc_2}.\] 
	By optimality of~$\minstrategy^*$, this rewrites in
	\[\muppervalue_{\loc_1} > \edgeweights(\loc_1,\loc_2) + \muppervalue_{\loc_2}\,.\] 
	By Theorem~\ref{thm:shortest-path}, this also rewrites in
	\[\dValue_{\loc_1} > \edgeweights(\loc_1,\loc_2) + \dValue_{\loc_2} \geq
	\F\big((\dValue_\loc)_{\loc\in \vertices}\big)(\loc_1)\,.\] 
	(since $\loc_1 \in \minvertices$): this contradicts the fact that the vector
	$(\dValue_\loc)_{\loc\in \vertices}$ is a fixpoint of~$\F$.
	\end{enumerate}
\end{proof}

\section{Discussion}
\label{sec:conclusion}

This article studies the trade-off between memoryless and deterministic
strategies, showing that $\MinPl$ guarantees the same value when
restricted to these two kinds of strategies, or when allowed to play with both memory and randomisation. This result holds both in divergent \WTG{s}, or \SPG{s}.   

We have studied the notions of deterministic values, (stochastic) values, and memoryless values. From a controller synthesis perspective, it could be more meaningful to study the value obtained by $\MinPl$ when being forced to play with a memoryless strategy, while not enforcing any properties on the strategy of $\MaxPl$. 
The result of Lemma~\ref{lem:best-response}, on the best response of \MaxPl, ensures that this value is identical to the memoryless value.

We aim at extending our study to more general \WTG{s}, though we have studied counter-examples (Examples~\ref{ex:mVal} and \ref{ex:1clock}) to the determinacy result with respect to the memoryless value, and thus of the equality between memoryless and deterministic values. We could still hope to weaken slightly the necessary condition to obtain our results, for instance considering the class of almost-divergent \WTG{s}
(adding the possibility for an execution following a region cycle to
have weight \emph{exactly 0}), used in 
\cite{BouyerJaziriMarkey-15,BusattoGastonMonmegeReynier-18} to obtain an
approximation schema of the optimal value. We wonder if similar
$\varepsilon$-optimal switching strategies may exist also in this
context, one of the crucial arguments in order to extend our emulation
result.

Another question concerns the implementability of the randomised
strategies: even if they use no memory, they still need to know the
precise current clock valuation. In (non-weighted) timed games,
previous work~\cite{ChatterjeeHenzingerPrabhu-08} aimed at removing
this need for precision, by using stochastic strategies where the
delays are chosen with probability distributions that do not require
exact knowledge of the clocks' measurements. In our setting, we aim at
further studying the implementability of the randomised strategies of
\MinPl in \WTG{s}, e.g.~by requiring them to be robust against small
imprecisions.

\bibliographystyle{alphaurl}
\bibliography{biblio.bib}

\appendix

\section{Construction of the probability measure}
\label{app:measure}

\propProbaDistribution*
\noindent 
We follow a similar proof schema as in~\cite[Appendix
A]{BertrandBouyerBrihayeMenetBaierGroserJurdzinski-14}. We start by proving
that $\Proba^{\minstrategy,\maxstrategy}_{\playBis}$ is a probability measure
over the algebra $(\Plays_{\playBis}^{\minstrategy,
\maxstrategy},\A_{\playBis})$ generated by cylinders defined from paths 
(then closed by finite union and complement). Then, we use Carathéodory's
extension theorem (Theorem~\ref{thm:Carateodory}) to conclude that
$\Proba^{\minstrategy,\maxstrategy}_{\playBis}$ is a probability measure over
the $\sigma$-algebra generated by all maximal~paths.

To define the probability $\Proba^{\minstrategy,\maxstrategy}_{\playBis}$ over
elements of the algebra $\A_{\playBis}$, we consider an inductive definition of
$\A_{\playBis}$. We remark that $\A_{\playBis} = \bigcup_n \A^n_{\playBis}$
where $\A^n_{\playBis}$ is the algebra generated by all cylinders defined with
a path of length at most\footnote{In
\cite{BertrandBouyerBrihayeMenetBaierGroserJurdzinski-14} authors defined this
set by cylinders with a length equal to $n$. Here, we need to relax this
hypothesis since maximal paths may be finite of length smaller than $n$, and if we do not keep these shorter paths, we will
lose some mass of probability.} $n$ from \playBis. More precisely, we let
$\FPaths_{\playBis}^n$ be the set of paths from \playBis that are either
non-maximal and length $n$ or maximal and length at most $n$. Then,
$\A^n_{\playBis}$ is generated by all cylinders of the form
$\Cyl_{\playBis}(\ppath, \C)$ with $\ppath \in \FPaths_{\playBis}^n$ and $\C$ a
Lebesgue-measurable set of $\Rplus^{|\ppath|}$. Elements of $\A^n_{\playBis}$
are finite unions of such disjoint cylinders and we let their probability be 
the sum of the probability of each~cylinder. 

\begin{lem}
	\label{lem:proba_measure-n}
	Let $n \in \N$ and \playBis be a finite play, then 
	$\Proba^{\minstrategy,\maxstrategy}_{\playBis}$ is a measure\footnote{More precisely, we 
	can prove that $\Proba^{\minstrategy,\maxstrategy}_{\playBis}$ is a probability measure over 
	$(\Plays_{\playBis}^{\minstrategy, \maxstrategy}, \A_{\playBis}^n)$. However, since this 
	property is not useful in the following, we do not prove that 
	$\Proba^{\minstrategy,\maxstrategy}_\play(\Plays_{\playBis}^{\minstrategy, \maxstrategy}) 
	= 1$ at this step of the proof.}
	over the algebra~$(\Plays_{\playBis}^{\minstrategy, \maxstrategy}, \A_{\playBis}^n)$.
\end{lem}
\begin{proof}
	By definition and Lemma~\ref{lem:proba_defi}.\eqref{item:proba_defi-bound},
	$\Proba^{\minstrategy,\maxstrategy}_{\playBis}$ is additive, non-negative
	and finite over $(\Plays_{\playBis}^{\minstrategy, \maxstrategy},
	\A_{\playBis}^n)$. In particular, by \cite{KemenyGSK-76},
	$\Proba^{\minstrategy,\maxstrategy}_{\playBis}$ is $\sigma$-additive, if
	and only if for all sequences $(A_i)_i$ of elements of $\A_{\playBis}^n$
	such that $A_0 \subseteq A_1 \subseteq \cdots$ and $A = \bigcup_i A_i \in
	\A_{\playBis}^n$, we have $\Proba^{\minstrategy,\maxstrategy}_{\playBis}(A)
	= \lim_i \Proba^{\minstrategy,\maxstrategy}_{\playBis}(A_i)$. So, let
	$(A_i)_i$ be such a sequence. Without lost of generality, we can suppose
	that each $A_i$ is generated by the same path $\ppath=\trans_1 \cdots
	\trans_k$ with $k \leq n$ (since \ppath may be a maximal path of a length
	less than $n$), otherwise it suffices to intersect each $A_i$ with
	$\Cyl_{\playBis}(\ppath)$. In particular, letting $(\C_i^j)_j$ be the sequence of 
	disjoint sets of constraints that appear in cylinders of $A_i$, we have 
	$A_i = \bigsqcup_{j=0}^{m_i} \Cyl_{\playBis}(\ppath, \C_i^j)$ where $m_i$ is 
	the number of cylinders in $A_i$. Now, by definition of probabilities (and its 
	additivity on cylinders), we have 
	\begin{displaymath}
	\Proba^{\minstrategy,\maxstrategy}_{\playBis}(A_i) = 
	\sum_{j=0}^{m_i} \Proba^{\minstrategy,\maxstrategy}_{\playBis}(\ppath, \C_i^j) \,.
	\end{displaymath}
	In particular, $\Proba^{\minstrategy,\maxstrategy}_{\playBis}(A_i)$ is equal to 
	\begin{multline*}
	\sum_{j=0}^{m_i}  \int_{I(\playBis,\trans_1)} \cdots 
	\int_{I(\playBis \extendto{\trans_1,\delay_1} \cdots 
		\extendto{\trans_{k-1}, \delay_{k-1}}, \trans_k)} \\
	\strategytrans(\playBis)(\trans_1) \cdots 
	\strategytrans(\playBis \extendto{\trans_1,\delay_1} \cdots 
	\extendto{\trans_{k-1}, \delay_{k-1}}) (\trans_k) \times 
	\Proba^{\minstrategy,\maxstrategy}_{\playBis \extendto{\trans_1,\delay_1} \cdots
		\extendto{\trans_k,\delay_k}}(\ppath, \C_i^j)\\
	\mathrm d\strategydelay(\playBis \extendto{\trans_1,\delay_1} \cdots 
	\extendto{\trans_{k-1}, \delay_{k-1}}, \trans_k)(\delay_k) \cdots
	\;\mathrm d\strategydelay(\playBis,\trans_1)(\delay_1) \,.
	\end{multline*}
	Moreover, by letting $\charact_Y$ be the characteristic function of the set $Y$, 
	we remark that $\Proba^{\minstrategy,\maxstrategy}_{\playBis \extendto{\trans_1,\delay_1} \cdots
		\extendto{\trans_k,\delay_k}}(\ppath, \C_i^j) = 
	\charact_{\Cyl(\ppath, \C_i^j)}(\playBis \extendto{\trans_1,\delay_1} \cdots
		\extendto{\trans_k,\delay_k})$ (by definition of the probability) and 
	$\Proba^{\minstrategy,\maxstrategy}_{\playBis}(A_i)$ is equal to 
	\begin{multline*}
	\sum_{j=0}^{m_i}  \int_{I(\playBis,\trans_1)} \cdots 
	\int_{I(\playBis \extendto{\trans_1,\delay_1} \cdots 
		\extendto{\trans_{k-1}, \delay_{k-1}}, \trans_k)} \\
	\strategytrans(\playBis)(\trans_1) \cdots 
	\strategytrans(\playBis \extendto{\trans_1,\delay_1} \cdots 
	\extendto{\trans_{k-1}, \delay_{k-1}}) (\trans_k) \times 
	\charact_{\Cyl(\ppath, \C_i^j)}(\playBis \extendto{\trans_1,\delay_1} \cdots
		\extendto{\trans_k,\delay_k})\\
	\mathrm d\strategydelay(\playBis \extendto{\trans_1,\delay_1} \cdots 
	\extendto{\trans_{k-1}, \delay_{k-1}}, \trans_k)(\delay_k) \cdots
	\;\mathrm d\strategydelay(\playBis,\trans_1)(\delay_1) \,.
	\end{multline*}
	In particular by linearity of the integral, we obtain that 
	$\Proba^{\minstrategy,\maxstrategy}_{\playBis}(A_i)$ is equal to 
	\begin{multline*}
	\int_{I(\playBis,\trans_1)} \cdots 
	\int_{I(\playBis \extendto{\trans_1,\delay_1} \cdots 
		\extendto{\trans_{k-1}, \delay_{k-1}}, \trans_k)} \\
	\strategytrans(\playBis)(\trans_1) \cdots 
	\strategytrans(\playBis \extendto{\trans_1,\delay_1} \cdots 
	\extendto{\trans_{k-1}, \delay_{k-1}}) (\trans_k) \times 
	\charact_{A_i}(\playBis \extendto{\trans_1,\delay_1} \cdots
		\extendto{\trans_k,\delay_k})\\
	\mathrm d\strategydelay(\playBis \extendto{\trans_1,\delay_1} \cdots 
	\extendto{\trans_{k-1}, \delay_{k-1}}, \trans_k)(\delay_k) \cdots
	\;\mathrm d\strategydelay(\playBis,\trans_1)(\delay_1)
	\end{multline*}
	since $\sum_{j = 0}^{m_i} \charact_{\Cyl(\ppath, \C_i^j)} = \charact_{A_i}$ 
	(as $\C_i^j$ are disjoint sets). Moreover, since, for all $i$, $\charact_{A_i}$ is 
	bounded by $1$, by dominated convergence, we deduce that $\lim_i 
	\Proba^{\minstrategy,\maxstrategy}_{\playBis}(A_i)$ is equal~to 
	\begin{multline*}
	\int_{I(\playBis,\trans_1)} \cdots 
	\int_{I(\playBis \extendto{\trans_1,\delay_1} \cdots 
		\extendto{\trans_{k-1}, \delay_{k-1}}, \trans_k)} \\
	\strategytrans(\playBis)(\trans_1) \cdots 
	\strategytrans(\playBis \extendto{\trans_1,\delay_1} \cdots 
		\extendto{\trans_{k-1}, \delay_{k-1}})(\trans_k) \times 
	\big(\lim_i \charact_{A_i}(\playBis \extendto{\trans_1,\delay_1} \cdots
		\extendto{\trans_k,\delay_k})\big)\\
	\mathrm d\strategydelay(\playBis \extendto{\trans_1,\delay_1} \cdots 
		\extendto{\trans_{k-1}, \delay_{k-1}},\trans_k)(\delay_k) \cdots
	\;\mathrm d\strategydelay(\playBis,\trans_1)(\delay_1) \,.
	\end{multline*}
	Moreover, since $A = \bigcup_i A_i$, then $\lim_i \charact_{A_i} = \charact_A$. 
	Thus, we deduce that $\lim_i \Proba^{\minstrategy,\maxstrategy}_{\playBis}(A_i)$ 
	is equal~to 
	\begin{multline*}
	\int_{I(\playBis,\trans_1)} \cdots 
	\int_{I(\playBis \extendto{\trans_1, \delay_1} \cdots 
		\extendto{\trans_{k-1}, \delay_{k-1}}, \trans_k)} \\
	\strategytrans(\playBis)(\trans_1) \cdots 
	\strategytrans(\playBis \extendto{\trans_1,\delay_1} \cdots 
		\extendto{\trans_{k-1}, \delay_{k-1}})(\trans_k) \times 
	\charact_{A}(\playBis \extendto{\trans_1,\delay_1} \cdots
		\extendto{\trans_k,\delay_k})\\
	\mathrm d\strategydelay(\playBis \extendto{\trans_1,\delay_1} \cdots 
	\extendto{\trans_{k-1}, \delay_{k-1}}, \trans_k)(\delay_k) \cdots
	\;\mathrm d\strategydelay(\playBis,\trans_1)(\delay_1) \,.
	\end{multline*}
	Thus, we conclude that 
	$\lim_i \Proba^{\minstrategy,\maxstrategy}_{\playBis}(A_i) = 
	\Proba^{\minstrategy,\maxstrategy}_{\playBis}(A)$ by the same way than for 
	$\Proba^{\minstrategy,\maxstrategy}_{\playBis}(A_i)$, since $A$ is a finite union 
	of cylinders (as $A \in \A_{\playBis}^n$).
\end{proof}
\noindent 
To prove that $\Proba^{\minstrategy,\maxstrategy}_{\playBis}$ is a probability measure on the algebra 
$(\Plays^{\minstrategy, \maxstrategy}_{\playBis}, \A_{\playBis})$, we use another property 
over $\A^n_{\playBis}$: for all $n \in \N$, $\A^n_{\playBis} \subseteq 
\A^{n+1}_{\playBis}$. Let $A$ be an element of $\A^n_{\playBis}$: either $A$ is a union of 
cylinders generated by maximal paths (of length at most $n$) and $A \in \A^{n+1}_{\playBis}$, 
or there exists a cylinder of $A$ generated by a non maximal path \ppath of length $n$. 
Since \ppath is non-maximal, it can be extended by a transition \trans. In particular, for 
all constraints $\C \subseteq \Rplus^{n}$, $\Cyl_{\playBis}(\ppath, \C) = \bigcup_{\trans}
\Cyl_{\playBis}(\ppath\trans, \C \times \Rplus)$. Thus, we conclude that $A \in
\A^{n+1}_{\playBis}$ by definition of $\A^{n+1}_{\playBis}$.

\begin{lem}
	\label{lem:proba_measure}
	Let \playBis be a finite play, then $\Proba^{\minstrategy,\maxstrategy}_{\playBis}$ is 
	a probability measure over $(\Plays^{\minstrategy, \maxstrategy}_{\playBis}, \A_{\playBis})$.
\end{lem}
\begin{proof}
	We start by proving that $\Proba^{\minstrategy,\maxstrategy}_{\playBis}$ is
	a measure on $\A_{\playBis}$. Since
	$\Proba^{\minstrategy,\maxstrategy}_{\playBis}$ is additive (as  
	$\A^n_{\playBis} \subseteq \A^{n+1}_{\playBis}$), non-negative and finite
	over $\A^n_{\playBis}$ (by
	Lemma~\ref{lem:proba_defi}.\eqref{item:proba_defi-bound}), the
	$\sigma$-additivity can be obtained by showing that $\lim_n
	\Proba^{\minstrategy,\maxstrategy}_{\playBis}(B_n) = 0$ for all sequences
	$(B_n)_n$ of elements of $\A_{\playBis}$ such that $B_0 \supseteq B_1
	\supseteq \cdots$ and $\bigcap_n B_n = \emptyset$ (by~\cite{KemenyGSK-76}).
	Let $(B_n)_n$ be such a sequence. Without loss of generality, we suppose
	that for every $n$, $B_n \in \A^n_{\playBis}$ (otherwise, all $B_n$ are in
	a fixed $\A^{n_0}_{\playBis}$ and Lemma~\ref{lem:proba_measure-n} allows us
	to conclude). 
	
	To easily write the limit of probabilities of $B_n$, we begin by proving 
	that all $B_n$ are generated by an unique cylinder where paths follow 
	the same prefix. By~\cite[Lemma~3.3]{KemenyGSK-76}, we know that if 
	$\lim_n \Proba^{\minstrategy,\maxstrategy}_{\playBis}(B_n) > 0$, then there 
	exists a sequence of cylinders $(\Cyl_n)_n$ such that for all $n$,
	$\Cyl_n \subseteq B_n$ and $\Cyl_{n+1} \subseteq \Cyl_n$. In other words, we 
	can suppose, without loss of generality, that there exists a 
	maximal path \ppath such that $B_n = \bigsqcup_{j=0}^{m_n} \Cyl_{\playBis}(\ppath_n, \C_n^j)$ 
	where $\ppath_n = \trans_1\cdots\trans_n$ is the prefix of \ppath of length $n$, 
	$(\C_n^j)_j$ is a sequence of constraints for $\ppath_n$ and $m_n$ the finite number 
	of cylinders of $B_n$ (since $B_n \in \A_{\playBis}^n$). Otherwise when
	$\Proba^{\minstrategy,\maxstrategy}_{\playBis}$ is not a measure on $\A_{\playBis}$, 
    such a sequence suffices to prove it. In particular, $B_n = 
	\Cyl_{\playBis}(\ppath_n, \C_n)$ where $\C_n = \bigsqcup_{j=0}^{m_n} \C_n^j$ (which is a 
	Lebesgue-measurable set of $\R^n$ since each $\C_n^j$ is). Moreover,
	with the same reasoning than in previous lemma and by letting 
	$\charact_Y$ the characteristic function of the set $Y$, 
	$\Proba^{\minstrategy,\maxstrategy}_{\playBis}(B_n)$ is equal~to 
	\begin{multline*}
	\int_{I(\playBis,\trans_1)} \strategytrans(\playBis)(\trans_1) \times
	\int_{I(\playBis \extendto{\trans_1,\delay_1}, \trans_2)} 
	\strategytrans(\playBis \extendto{\trans_1, \delay_1})(\trans_2) \cdots \\
	\int_{I(\playBis \extendto{\trans_1,\delay_1} \cdots 
		\extendto{\trans_{n-1}, \delay_{n-1}}, \trans_n)}
	\strategytrans(\playBis \extendto{\trans_1,\delay_1} \cdots 
		\extendto{\trans_{n-1}, \delay_{n-1}})(\trans_n) \times 
	\charact_{B_n}(\playBis \extendto{\trans_1,\delay_1}\cdots
		\extendto{\trans_n,\delay_n}) \\
	\mathrm d\strategydelay(\playBis \extendto{\trans_1,\delay_1} \cdots 
		\extendto{\trans_{n-1}, \delay_{n-1}},\trans_n)(\delay_n) \cdots\;
	\mathrm d\strategydelay(\playBis \extendto{\trans_1,\delay_1}, \trans_2)(\delay_2)
	\;\mathrm d\strategydelay(\playBis,\trans_1)(\delay_1).
	\end{multline*}
	
	Now, to prove that the limit of probabilities is equal to $0$, we will prove
	that there exists $i \in \N$ such that $B_i = \emptyset$ (more precisely, 
	$\C_i = \emptyset$) from the hypothesis of $\bigcap_n B_n = \emptyset$. 
	For all $i > 0$, we let $p_i \colon \Rplus^{i+1} \mapsto \Rplus^i$ 
	be the continuous (for product topologies) projection over the $i$ 
	first components of a constraint in $\Rplus^{i+1}$. For all $i < n$, we can 
	define the constraint for the $i$ first steps by induction along $\C_n$ such 
	that $\C^i_n = p_i(\C^{i+1}_n)$. Moreover, since $B_{n+1} \subseteq B_n$ and 
	all sets follows the same prefix, we have for all $i \leq n$ that 
	$\C^i_{n+1} \subseteq \C^i_n$. In particular, for some fixed $i$, the 
	sequence $(\C^i_n)_n$ decreases, thus it converges to $\C^i \subseteq \Rplus^i$. 
	Now, we want to prove that there exists $i > 0$ such that $\C^i = \emptyset$. 
	To do so, we reason by contradiction and we suppose that for all $i$, 
	$\C^i \neq \emptyset$, and we want to define a play where its delays are in 
	$\cap_i \C^i$. To define this play, we define, by induction on $i$, a sequence 
	of delays that satisfies $\C^i$:  for all $i$ if we have a sequence of delays 
	$(\delay_i)_i$ that satisfies $\C^i$ (for all $i$), then since $\C^i = p_i(\C^{i+1})$ 
	(by continuity of $p_i$), there exists $\delay_{i+1}$ such that 
	$(\delay_1, \ldots, \delay_i, \delay_{i+1})$ satisfies $\C^{i+1}$ 
	(with $(\delay_1, \ldots, \delay_i)$ satisfying $\C^i$). This sequence of delays 
	define a play $\play \in \bigcap_i \Cyl_{\playBis}(\ppath_i, \C^i) = \bigcap_i B_i$. 
	However, by hypothesis on $(B_n)_n$, $\bigcap_i B_i = \emptyset$, thus, we obtain a 
	contradiction and there exists $i$ such that $\C^i = 0$, i.e. $B_i = \emptyset$.
	
	To conclude about the $\sigma$-additivity, we prove that 
	$\lim_n \Proba^{\minstrategy,\maxstrategy}_{\playBis}(B_n) = 0$. 
	We let $B_i$ be the subset such that $B_i = \emptyset$, and for all $n \geq i$, 
	by letting $\charact_Y$ the characteristic function of the set $Y$, 
	$\Proba^{\minstrategy,\maxstrategy}_{\playBis}(B_n)$ is equal~to 
	\begin{multline*}
	\int_{I(\playBis,\trans_1)} \strategytrans(\playBis)(\trans_1) \times
	\int_{I(\playBis \extendto{\trans_1,\delay_1}, \trans_2)} 
	\strategytrans(\playBis \extendto{\trans_1, \delay_1})(\trans_2) \cdots \\
	\int_{I(\playBis \extendto{\trans_1,\delay_1} \cdots 
		\extendto{\trans_{n-1}, \delay_{n-1}}, \trans_n)}
	\strategytrans(\playBis \extendto{\trans_1,\delay_1} \cdots 
	\extendto{\trans_{n-1}, \delay_{n-1}})(\trans_n) \times 
	\charact_{B_n}(\playBis \extendto{\trans_1,\delay_1}\cdots
	\extendto{\trans_n,\delay_n}) \\
	\mathrm d\strategydelay(\playBis \extendto{\trans_1,\delay_1} \cdots 
	\extendto{\trans_{n-1}, \delay_{n-1}},\trans_n)(\delay_n) \cdots\;
	\mathrm d\strategydelay(\playBis \extendto{\trans_1,\delay_1}, \trans_2)(\delay_2)
	\;\mathrm d\strategydelay(\playBis,\trans_1)(\delay_1).
	\end{multline*}
	In particular, since $i \leq n$, by restricting the probability to the $i$ first 
	components, $\Proba^{\minstrategy,\maxstrategy}_{\playBis}(B_n)$ is bound~by 
	\begin{multline*}
	\int_{I(\playBis,\trans_1)} \strategytrans(\playBis)(\trans_1) \times
	\int_{I(\playBis \extendto{\trans_1,\delay_1}, \trans_2)} 
	\strategytrans(\playBis \extendto{\trans_1, \delay_1})(\trans_2) \cdots \\
	\int_{I(\playBis \extendto{\trans_1,\delay_1} \cdots 
		\extendto{\trans_{i-1}, \delay_{i-1}}, \trans_i)}
	\strategytrans(\playBis \extendto{\trans_1,\delay_1} \cdots 
	\extendto{\trans_{i-1}, \delay_{i-1}})(\trans_i) \times 
	\charact_{B_i}(\playBis \extendto{\trans_1,\delay_1}\cdots
	\extendto{\trans_i,\delay_i}) \\
	\mathrm d\strategydelay(\playBis \extendto{\trans_1,\delay_1} \cdots 
	\extendto{\trans_{i-1}, \delay_{i-1}},\trans_i)(\delay_i) \cdots\;
	\mathrm d\strategydelay(\playBis \extendto{\trans_1,\delay_1}, \trans_2)(\delay_2)
	\;\mathrm d\strategydelay(\playBis,\trans_1)(\delay_1).
	\end{multline*}
	By dominated convergence (since $\charact_{B_i} \leq 1$), we obtain that 
	$\lim_n \Proba^{\minstrategy,\maxstrategy}_{\playBis}(B_n)$ is bound~by 
	\begin{multline*}
	\int_{I(\playBis,\trans_1)} \strategytrans(\playBis)(\trans_1) \times
	\int_{I(\playBis \extendto{\trans_1,\delay_1},\trans_2)} 
	\strategytrans(\playBis \extendto{\trans_1,\delay_1})(\trans_2) \cdots \\
	\int_{I(\playBis \extendto{\trans_1,\delay_1} \cdots 
		\extendto{\trans_{i-1}, \delay_{i-1}},\trans_n)}
	\strategytrans(\playBis \extendto{\trans_1,\delay_1}\cdots 
		\extendto{\trans_{i-1}, \delay_{i-1}})(\trans_i) \times 
	\lim_n \charact_{B_n}(\playBis \extendto{\trans_1,\delay_1}\cdots
		\extendto{\trans_i,\delay_i}) \\
	\mathrm d\strategydelay(\playBis \extendto{\trans_1,\delay_1} \cdots 
		\extendto{\trans_{i-1}, \delay_{i-1}},\trans_i)(\delay_i) \cdots\;
	\mathrm d\strategydelay(\playBis \extendto{\trans_1,\delay_1}, \trans_2)(\delay_2)
	\;\mathrm d\strategydelay(\playBis, \trans_1)(\delay_1).
	\end{multline*}
	Now, since the sequence $(B_n)_n$ decreases, we know that for all $n \geq i$, 
	$B_n = \emptyset$. Thus, $\lim_n \charact_{B_n}(\playBis \extendto{\trans_1,\delay_1}\cdots
	\extendto{\trans_i,\delay_i}) = 0$, and we deduce that
	$\lim_n \Proba^{\minstrategy,\maxstrategy}_{\playBis}(B_n) = 0$.
	
	Finally, we conclude the proof by proving that 
	$\Proba^{\minstrategy,\maxstrategy}_{\playBis}(\Plays^{\minstrategy, \maxstrategy}_{\playBis}) 
	= 1$. To do that, we remark that $\A_{\playBis}^0 = 
	(\Plays^{\minstrategy, \maxstrategy}_{\playBis}, \Cyl_{\playBis}(\ppathBis))$, i.e. 
	$\Plays^{\minstrategy, \maxstrategy}_{\playBis} = \Cyl_{\playBis}(\ppathBis)$ 
	where \playBis follows \ppathBis. Thus, since \playBis follows \ppathBis,  
	$\Cyl_{\playBis}(\ppathBis) \neq \emptyset$, and by definitions, 
	we obtain that
	\begin{displaymath}
	\Proba^{\minstrategy,\maxstrategy}_{\playBis}(\Plays^{\minstrategy, \maxstrategy}_{\playBis}) 
	= \Proba^{\minstrategy,\maxstrategy}_{\playBis}(\ppathBis) = 1. \qedhere
	\end{displaymath}
\end{proof}
\noindent 
Finally, we conclude the proof of Proposition~\ref{prop:proba_distribution} by 
using the following theorem.

\begin{thm}[Carathéodory's extension theorem]
	\label{thm:Carateodory}
	Let $S$ be a set, and $\nu$ be a $\sigma$-finite measure defined on an algebra 
	$A \subseteq 2^S$. Then, $\nu$ can be extended in a unique manner to the 
	$\sigma$-algebra generated by $A$.
\end{thm}

\begin{proof}[Proof of Proposition~\ref{prop:proba_distribution}]
	We apply Theorem~\ref{thm:Carateodory} to the set $S =
	\Plays_{\playBis}^{\minstrategy, \maxstrategy}$, $A = \A_{\playBis}$, and
	$\nu = \Proba^{\minstrategy,\maxstrategy}_{\playBis}$ which is a
	$\sigma$-finite measure on $\A_{\playBis}$ (by
	Lemma~\ref{lem:proba_measure}). Hence, there is a unique extension of
	$\Proba^{\minstrategy,\maxstrategy}_{\playBis}$ on the $\sigma$-algebra
	generated by the cylinders which is a probability measure over maximal plays.
\end{proof}
\noindent 
\section{A (non-proper) smooth deterministic strategy of \MinPl with an infinite expected payoff}
\label{ex:det-proper}

\begin{figure}[tbp]
	\centering
	\begin{tikzpicture}[xscale=.8,every label/.style={font=\scriptsize}]
	\node[PlayerMin, label=below:$\loc_0$] at (0, 0) (s0) {$\mathbf{0}$};
	\node[PlayerMax, label=below:$\loc_1$] at (4, 0) (s1) {$\mathbf{0}$};
	\node[PlayerMin, label=below:$\loc_2$] at (8, 0) (s2) {$\mathbf{0}$};
	\node[target] at (12, 0) (t) {\LARGE $\smiley$};
	
	\draw[->]
	(s0) edge node[above]{$\trans_0, x \leq 1$} (s1)
	(s1) edge node[above]{$\trans_1, 0 < x < 1$} (s2)
	(s2) edge[loop above] node[above]{$\trans_2, x \leq 1, \mathbf{1}$} (s2)
	(s2) edge node[above]{$\trans_3, x \leq 1$} (t)
	;
	\end{tikzpicture}
	\caption{A \WTG where $\MinPl$ has a non-proper 
		(smooth) deterministic strategy that reaches the target with probability 
		$1$ against all (stochastic) strategies for \MaxPl.}
	\label{Fig:determinsticProper}
\end{figure}

We consider the \WTG depicted in \figurename{~\ref{Fig:determinsticProper}}, with 
$(\loc_0, 0)$ as initial configuration. We define a smooth deterministic strategy $\detminstrategy$ that always chooses $(\trans_0, 0)$ from $\loc_0$, 
and chooses $i$ times $(\trans_2, 0)$ before $(\trans_3, 0)$ when 
$\loc_2$ is reached with a valuation 
$\val \in [1-\frac{1}{i}, 1 - \frac{1}{i+1})$. 
First, this strategy guarantees that all plays conforming to it reach a 
target location, even if the number of steps to do it depends on the 
first configuration reached in $\loc_2$.

Now, we consider $\maxstrategy \in \StratMax$ defined such that, 
for all configurations $(\loc_1, \val)$, 
\begin{displaymath}
\maxstrategytrans(\loc_1, \val) = \Dirac{\trans_1}
\qquad \text{and} \qquad 
\maxstrategydelay((\loc, \val), \trans_1) = \mathcal{U}(\val, 1)
\end{displaymath} 
where $\mathcal{U}(\val, 1)$ is the uniform distribution over the open interval $(\val, 1)$, and 
we compute $\E^{\detminstrategy, \maxstrategy}_{\loc_0, 0}$.
By Lemma~\ref{lem:expectation-split}, we have 
\begin{align*}
\E^{\detminstrategy, \maxstrategy}_{\loc_0, 0} &= 
\sum_{\ppath \in \TPaths_{\loc_0, 0}}
\E^{\detminstrategy,\maxstrategy}_{\loc_0, 0}(\ppath) 
= \sum_n \sum_{\ppath \in \TPaths^{n+2}_{\loc_0, 0}} 
\E^{\detminstrategy,\maxstrategy}_{\loc_0, 0}(\ppath)  \\
&=\sum_n \sum_{\ppath \in \TPaths^{n+2}_{\loc_0, 0}}
\E^{\detminstrategy,\maxstrategy}_{(\loc_0, 0)\extendto{\trans_0, 0}}(\ppath) 
\qquad \text{(by definition of $\detminstrategy$)} \\
&= \sum_n \sum_{\ppath \in \TPaths^{n+2}_{\loc_0, 0}}
\int_0^1 
\E^{\detminstrategy,\maxstrategy}_{(\loc_0, 0)\extendto{\trans_0, 0}
	\extendto{\trans_1, \delay}}(\ppath) 
\mathrm d\minstrategydelay((\loc_1, 0),\trans_1)(\delay)
\qquad \text{(by definition of $\maxstrategy$)} \,.
\end{align*}
Now, since $\detminstrategy$ is a deterministic strategy, for all $n$, there 
exists only one path $\ppath_n \in \TPaths^{n+2}_{\loc_0, 0}$. In particular, we obtain that 
\begin{displaymath}
\E^{\detminstrategy, \maxstrategy}_{\loc_0, 0} = 
\sum_n \int_0^1 
\E^{\detminstrategy,\maxstrategy}_{(\loc_0, 0)\extendto{\trans_0, 0}
	\extendto{\trans_1, \delay}}(\ppath_n) 
\mathrm d\minstrategydelay((\loc_1, 0),\trans_1)(\delay) \,.
\end{displaymath}
By definition of $\detminstrategy$, we know that $\ppath_n$ is conforming to 
$\detminstrategy$ only when $\delay \in [1-\frac{1}{n}, 1-\frac{1}{n+1})$, i.e. 
\begin{displaymath}
\E^{\detminstrategy,\maxstrategy}_{(\loc_0, 0)\extendto{\trans_0, 0}
	\extendto{\trans_1, \delay}}(\ppath_n) = 
n\times \charact_{[1-\frac{1}{n}, 1-\frac{1}{n+1})}(\delay) \,.
\end{displaymath} 
Since $\minstrategydelay((\loc_1, 0),\trans_1)$ follows a uniform distribution over $(0, 1)$, we deduce that 
\begin{displaymath}
\E^{\detminstrategy, \maxstrategy}_{\loc_0, 0} 
= \sum_n n \int_{\frac{n-1}{n}}^{\frac{n}{n+1}}  \mathrm d\delay 
= \sum_n \frac{n}{n(n+1)} = +\infty \,.
\end{displaymath}

\section{Probabilities under a proper strategy for \MinPl}
\label{app:proof_lem-proper-bound}

\lemProperBound*
\noindent 
We proceed by induction, showing first that for all $n \in \N$ and 
all finite plays $\playBis$ following~\ppathBis, 
\begin{displaymath}
\sum_{\ppath \in \TPaths_{\playBis}^{nm}} 
\Proba^{\minstrategy,\maxstrategy}_{\playBis}(\ppathBis\ppath) \leq (1-\alpha)^{n} \,. 
\end{displaymath}
For $n = 0$, the property holds since 
$(1 - \alpha)^0 = 1$. Otherwise, suppose that the property holds for $n$, 
and we prove it for $n+1$ by decomposing the sum as
\begin{equation}
\label{eq:sums}
\sum_{\ppath \in \TPaths_{\playBis}^{(n+1)m}} 
\Proba^{\minstrategy,\maxstrategy}_{\playBis}(\ppathBis\ppath) = 
\sum_{\trans_0} \sum_{\trans_1} \cdots \sum_{\trans_{m-1}}
\sum_{\ppath \in \TPaths_{\playBis}^{nm}} 
\Proba^{\minstrategy,\maxstrategy}_{\playBis}(\ppathBis 
	\trans_0 \trans_1\cdots\trans_{m-1}\ppath) \,.
\end{equation}
For a fixed $\trans_0 \cdots \trans_{m-1} \ppath \in
\TPaths_{\playBis}^{(n+1)m}$, letting $\ppath' = \ppathBis \trans_0
\trans_1\cdots\trans_{m-1} \ppath$, we can unravel the $m$ first steps of the
definition of $\Proba^{\minstrategy,\maxstrategy}_{\playBis}(\ppath')$. Letting
$\strategy_i$ denote the strategy $\minstrategy$ or $\maxstrategy$ depending
if the source location of the transition $\trans_i$ belongs to \MinPl or \MaxPl, we can write
$\Proba^{\minstrategy,\maxstrategy}_{\playBis}(\ppath')$ under the form:
\begin{multline*} 
\int_{I(\playBis,\trans_0)} \cdots 
\int_{I(\playBis \extendto{\trans_0,\delay_0} \cdots 
	\extendto{\trans_{m-1},\delay_{m-1}})}  \\
\strategytrans^0(\playBis)(\trans_0) \cdots  
\strategytrans^{m-2}\big(\playBis \extendto{\trans_0,\delay_0} \cdots 
	\extendto{\trans_{m-1},\delay_{m-2}}\big)(\trans_{m-1}) 
\Proba^{\minstrategy,\maxstrategy}_{\playBis \extendto{\trans_0,\delay_0} 
	\cdots \extendto{\trans_{m-1},\delay_{m-1}}}(\ppath') \\
\mathrm d\strategydelay^{m-1} \big(\playBis \extendto{\trans_0,\delay_0} \cdots 
	\extendto{\trans_{m-2},\delay_{m-2}}, \trans_{m-1}\big)(\delay_{m-1}) \cdots 
\,\mathrm d\strategydelay^0 (\playBis,\trans_0)(\delay_{0}) .
\end{multline*}
Thus,~\eqref{eq:sums} can be rewritten as
\begin{multline*}
\sum_{\trans_0} \sum_{\trans_1} \cdots \sum_{\trans_{m-1}} 
\int_{I(\playBis,\trans_0)} \int_{I(\playBis \extendto{\trans_0,\delay_0})} 
\cdots \int_{I(\playBis \extendto{\trans_0,\delay_0} \cdots 
	\extendto{\trans_{m-1},\delay_{m-1}})} \\
\strategytrans^0(\playBis)(\trans_0) \cdots  
\strategytrans^{m-1}\big(\playBis \extendto{\trans_0,\delay_0} \cdots
	\extendto{\trans_{m-2},\delay_{m-2}}\big)(\trans_{m-1}) \\
\sum_{\ppath \in \TPaths_{\playBis}^{nm}} 
\Proba^{\minstrategy,\maxstrategy}_{\playBis \extendto{\trans_0,\delay_0} 
	\cdots \extendto{\trans_{m-1},\delay_{m-1}}}(\ppathBis \trans_0
\trans_1\cdots\trans_{m-1} \ppath) \\
\mathrm d\strategydelay^{m-1} \big(\playBis \extendto{\trans_0,\delay_0} \cdots 
	\extendto{\trans_{m-2},\delay_{m-2}},\trans_{m-1}\big)(\delay_{m-1}) \cdots 
\mathrm d\strategydelay (\playBis,\trans_{0})(\delay_{0}).
\end{multline*}
By induction hypothesis,
\begin{displaymath}
\sum_{\ppath\in \TPaths_{\playBis}^{nm}}
\Proba^{\minstrategy,\maxstrategy}_{\playBis \extendto{\trans_0,\delay_0} 
	\cdots \extendto{\trans_{m-1},\delay_{m-1}}}(\ppathBis \trans_0
\trans_1\cdots\trans_{m-1} \ppath) \leq (1-\alpha)^n
\end{displaymath}
so that $\sum_{\ppath \in \TPaths_{\playBis}^{(n+1)m}} 
\Proba^{\minstrategy,\maxstrategy}_{\playBis}(\ppathBis \trans_0
\trans_1\cdots\trans_{m-1} \ppath)$
can be bounded by
\begin{multline*}
(1-\alpha)^n \sum_{\trans_0} \sum_{\trans_1} \cdots \sum_{\trans_{m-1}} 
\int_{I(\playBis,\trans_0)} \cdots 
\int_{I(\playBis \extendto{\trans_0,\delay_0} \cdots 
	\extendto{\trans_{m-1},\delay_{m-1}})} \\
\strategytrans^0(\playBis)(\trans_0) \cdots 
\strategytrans^{m-1}\big(\playBis \extendto{\trans_0,\delay_0} \cdots 
	\extendto{\trans_{m-2},\delay_{m-2}}\big)(\trans_{m-1})  \\
\mathrm d\strategydelay^{m-1} \big(\playBis \extendto{\trans_0,\delay_0} \cdots
	\extendto{\trans_{m-2},\delay_{m-2}},\trans_{m-1}\big)(\delay_{m-1}) \cdots
\,\mathrm d\strategydelay^0(\playBis,\trans_{0})(\delay_{0})
\end{multline*}
that is $(1-\alpha)^n$ multiplied by the probability of all paths that can
continue after \play for at least $m$ steps without having reached
the target set $\LocsT$,
i.e.~by the second item of Hypothesis~\ref{hyp:proper} 
\[(1-\alpha)^n\, \Proba^{\minstrategy,\maxstrategy}_{\playBis}
(\bigcup_{n \geq m}\TPlays_{\playBis}^n) \leq (1-\alpha)^n(1-\alpha) \,.\] 
This concludes the
induction proof. To conclude, we simply need to consider paths of length that 
is not a multiple of $m$: we thus unravel similarly the first $k \in (0,m)$ 
transitions, bounding all their probabilities by $1$.

\section{\texorpdfstring{Cylinders decomposition for the set of plays conforming to 
$\minstrategy^p$}{Cylinders decomposition for the set of plays conforming to 
the strategy of Min}}
\label{app:eta-proper_cylinders}

We want to justify why the set of delays that appears in a path $\ppath$ 
of $S^k_{\playBis}$, $C_{\ppath}$ is a constraint, i.e. is a finite union and 
intersection of the constraints in $\minstrategydelay^p$ and $\maxstrategydelay$. 

Let consider a finite path $\ppath = \playBis\trans_{0}\cdots\trans_{k}$ that 
appears in $S^k_{\playBis}$, i.e.~that contains a transition chosen by 
$\minstrategy^p$ with probability $p$. A play following $\ppath$ and 
conforming to $\minstrategy^p$ and $\maxstrategy$ must satisfy certain constraints 
over each transition. If $\trans_{k}$ belongs to \MinPl, then the set of delays
such that $\playBis\extendto{\trans_0, \delay_0}\cdots \extendto{\trans_{k-1}, \delay_{k-1}}$ 
satisfies a (Lebesgue-measurable) constraint $C_{\trans_{k}}$ given by 
Hypothesis~\ref{hyp:measurability} over $\minstrategy^p$ (as the inverse image 
of the strategy applying on $\playBis\extendto{\trans_0, \delay_0}\cdots 
\extendto{\trans_{k-1}, \delay_{k-1}}$). The case where $\trans_{k}$ belongs to \MaxPl 
is analogous. Now, let $\trans_i$ with $0 \leq i \leq k$ belongs to \MinPl with the constraint 
$C_{\trans_{i+1} \cdots \trans_{k}}$. The constraint $C_{\trans_{i} \cdots \trans_{k}}$ 
is given by the inverse image of $\minstrategy^p$ on 
$\playBis\extendto{\trans_0, \delay_0}\cdots \extendto{\trans_{i-i}, \delay_{i-1}}$ 
(that gives a measurable set of delay) and the intersection with 
$C_{\trans_{i+1} \cdots \trans_{k}}$ (we consider only delays that satisfying the 
next constraints). Thus, by induction, we can define the constraint $C_{\ppath}$ by a 
finite (since $\ppath$ is finite) intersection of the constraints in 
$\minstrategydelay^p$ and $\maxstrategydelay$ (that are measurable by 
Hypothesis~\ref{hyp:measurability}). 

To conclude, we remark that in $\ppath$, we need to choose the position of the 
transition $p$ (when $\detminstrategy_1$ and $\detminstrategy_2$ choose the same 
transition). In particular, there exists at most $k$ available possible positions, and 
$C_{\trans_{k}}$ is given by the finite union over all possible positions.

\end{document}